\documentclass[reqno]{amsart}
\usepackage{bbm}
\usepackage[lite]{amsrefs}
\usepackage{amssymb}
\usepackage[all,cmtip]{xy}
\usepackage{amsmath}
\usepackage{amsthm}
\usepackage{amsfonts}
\usepackage{hyperref}
\usepackage{enumerate}
\usepackage{physics}
\usepackage[titletoc]{appendix}
\usepackage{fancyhdr}
\usepackage[dvipsnames]{xcolor}
\usepackage{comment}
\usepackage{tikz-cd}
\usepackage{tikzpagenodes}
\usepackage{caption}
\usepackage{float}
\usetikzlibrary[patterns]
\usepackage[margin = 1in]{geometry}
\usepackage{setspace}

\spacing{1.2}

\allowdisplaybreaks

 \newtheorem{thm}{Theorem}[section]

 \newtheorem{lem}[thm]{Lemma}
 \newtheorem{prop}[thm]{Proposition}
 
 \newtheorem{defn}[thm]{Definition}
  
  \newtheorem{defn-thm}[thm]{Definition-Theorem}
   \newtheorem{ex}[thm]{Example}
   \newtheorem{example}[thm]{Example}
   \newtheorem{rem}[thm]{Remark}

   \renewcommand\[{\begin{equation}}
\renewcommand\]{\end{equation}}
\numberwithin{equation}{section}

\usepackage{amsxtra}
\usepackage{graphicx}
\usepackage{bbm,newlfont}
\usepackage{amscd}
\usepackage{hyperref}
\usepackage[german,english]{babel}
\usepackage{cancel}
\usepackage{amsmath,amsthm,amssymb}
\usepackage{enumerate}
\usepackage{color}
\usepackage{varwidth}
\usepackage{tasks}
\usepackage{mathrsfs}  
\usepackage{setspace}
\usepackage{collectbox}


\usepackage{graphicx}

\DeclareFontFamily{U}{mathx}{}
\DeclareFontShape{U}{mathx}{m}{n}{<-> mathx10}{}
\DeclareSymbolFont{mathx}{U}{mathx}{m}{n}
\DeclareMathAccent{\widehat}{0}{mathx}{"70}
\DeclareMathAccent{\widecheck}{0}{mathx}{"71}

\begin{document}

\title[sATO, W-superalgebra and integrable system]{ Integrable systems on rectangular $\mathcal{W}$-superalgebras\\
via \\
super Adler-type operators
}

\author{Sylvain Carpentier}
\address{QSMS 25dong-102ho, Seoul National University, 1 Gwanak-ro Gwanak-gu, Seoul 08826, Korea}
\email{sylcar@snu.ac.kr 
 }
\thanks{
S. Carpentier thanks the National Research Foundation of Korea(NRF) for the grant funded by the Korean government(MSIT) (No.2020R1A5A1016126). G.S. Lee and U.R. Suh thank to the NRF for the grant \#2022R1C1C1008698 and Seoul National University for the Creative-Pioneering Researchers Program.
}

\author[G.S.Lee]{Gahng Sahn Lee}
\address{Department of Mathematical Sciences and Research institute of Mathematics, Seoul National University, Gwanak-ro 1, Gwanak-gu, Seoul 08826, Korea}
\email{polyhedra@snu.ac.kr}

\author[U.R.Suh]{Uhi Rinn Suh}
\address{Department of Mathematical Sciences and Research institute of Mathematics, Seoul National University, Gwanak-ro 1, Gwanak-gu, Seoul 08826, Korea}
\email{uhrisu1@snu.ac.kr}

\begin{abstract}
In this paper, we introduce a class of super Adler-type operators associated with the Lie superalgebra $\mathfrak{gl}(m|n)$. We show that these operators generate Poisson vertex superalgebras which are isomorphic to the classical $\mathcal{W}$-superalgebras associated with $\mathfrak{gl}(m|n)$ and some rectangular nilpotent elements. We use this isomorphism to construct integrable hierarchies on these rectangular $\mathcal{W}$-superalgebras.
\end{abstract}
\maketitle

\setcounter{tocdepth}{-1}

\pagestyle{plain}
\section{Introduction}

The Korteweg–de Vries (KdV) equation is one of 
the most well-known integrable Hamiltonian systems and is closely related with the differential operator $L^D_{[2]}=\partial^2 +u$ which is a reduction of $L_{[2]}=\partial^2+u_1\partial +u_0$. 
More generally, for $N \geq 2$, one can construct an integrable system on the algebra of differential polynomials $\mathcal{P}_{[N]}$ generated by the coefficients of the differential operator
\begin{equation} \label{eq:degree N adler}
L_{[N]}= \partial^N + u_{N-1} \partial^{N-1} +\cdots + u_0.
\end{equation}
The hierarchy of pairwise commuting evolutionary derivations $\frac{d}{dt_k}$ for $k=\mathbb{Z}_{>0}$ is given by the Lax equations 
\begin{equation}\label{eq: hierarchy}
\frac{d}{dt_k}(L_{[N]})=[(L_{[N]}^{k/N})_+,L]. 
\end{equation}

The residues of fractional powers of $L_{[N]}$ are conserved modulo total derivatives for each of these derivations 
\begin{equation} \label{res}
  \frac{d}{dt_k}( \mathrm{Res} \,  L^{l/N}) \in \partial \mathcal{P}_{[N]} \, \, \, \text{for all } k,l \in \mathbb{Z}_+.
\end{equation} 
Gelfand and Dickey \cite{GD76} showed that these conserved quantities are in involution for each of the two independent Lie brackets
 \begin{equation} \label{eq:adler bracket}
 \begin{split}
       &\big[\smallint g  ,  \smallint h \big]_1  =  \, \mathrm{Res} \,\,   \int (L \frac{\delta g}{\delta L})_+L\frac{\delta h}{\delta L}-L(\frac{\delta g}{\delta L}L)_+ \frac{\delta h}{\delta L}, \quad \textup{(quadratic GD bracket)}
       \\ &\big[\smallint g  ,  \smallint h \big]_2  = \, \mathrm{Res} \,\, \int (L \frac{\delta g}{\delta L})_+\frac{\delta h}{\delta L}-(\frac{\delta g}{\delta L}L)_+ \frac{\delta h}{\delta L}, \, \, \, \, \, \,   \quad \textup{     (linear GD bracket)}
\end{split}
\end{equation} 
on $\mathcal{P}_{[N]}/\partial \mathcal{P}_{[N]}$, where $\frac{\delta g}{\delta L} := \sum_{i=0}^{N-1}\partial^{-i-1}\frac{\delta g}{\delta u_{i}}$ and $ \frac{\delta g}{\delta u_i}:= \sum_{n \geq 0}(-\partial)^n(\frac{\partial g}{\partial u_i^{(n)}})$ for $  u_i^{(n)} := \partial^n(u_i)$. Here  $\smallint g$ and $\smallint h $ are the projections of two elements $g, h \in \mathcal{P}_{[N]} $ onto the quotient space $\mathcal{P}_{[N]}/\partial \mathcal{P}_{[N]}$.

 Shortly afterwards, Drinfeld and Sokolov \cite{DS85} generalized this picture and connected the integrable systems with the theory of Kac-Moody algebras. Given a set of Chevalley generators $\{e_i,f_i,h_i \}_{i=1,...,r}$ of a simple Lie algebra $\mathfrak{g}$ of rank $r$ and an element $s$ of maximal principal grading such that $\sum_{i=1}^r f_i +zs\in \mathfrak{g}(\!(z^{-1})\!)$ is semisimple, they considered the  differential operator  
\begin{equation}\label{eq:matrix Lax}
\textstyle 
\mathcal{L}= \partial + \sum_{\alpha \in \Delta} E_{\alpha} \otimes q_{\alpha} + (\sum_{i=1}^r f_i +zs) \otimes 1  \, \, \, \, \, \, \in \mathbb{C}[\partial]\ltimes\left(\mathfrak{g}(\!(z^{-1})\!)\otimes  \mathcal{V}(\mathfrak{b})\right),
\end{equation}
where the elements $E_\alpha$ form a basis of the Borel subalgebra generated by $e_i$ and $h_i$ and $\mathcal{V}(\mathfrak{b})$ is the algebra of differential polynomials generated by the indeterminates $q_{\alpha}$. 
They constructed integrable hierarchies on the algebra $\mathcal{V}(\mathfrak{b})$ and showed that these are Hamiltonian for each of the following Lie brackets defined on the quotient spaces $\mathcal{V}(\mathfrak{b})/\partial \mathcal{V}(\mathfrak{b}) $
\begin{equation} \label{eq:DS brackets}
 \big[\smallint g, \smallint h \big]_1: = \int \big( \frac{\delta g }{\delta q} \big| \big[  \mathcal{L}, \frac{\delta h }{\delta q} \big] \big), \quad \big[\smallint g, \smallint h\big]_2: = \int \big( \frac{\delta g }{\delta q} \big| \big[ s, \frac{\delta h }{\delta q}  \big] \big), 
\end{equation}
where $\frac{\delta g }{\delta q} := \sum_{\alpha} E^{\alpha} \otimes \frac{\delta g}{\delta q_{\alpha}}$ and $\{E^{\alpha}\}_{\alpha \in \Delta}$ is a basis of $\mathfrak{b}$ which is dual to $\{E_{\alpha}\}_{\alpha \in \Delta}$ with respect to the Killing form $( \, \,  | \, \,  )$. Moreover, they proved that both brackets and the infinite family of derivations can be reduced to the algebra $\mathcal{W}(\mathfrak{g})$ consisting of elements in $\mathcal{V}(\mathfrak{b})$ that are invariant under some gauge action on $\mathcal{L}$ by the nilpotent algebra $\mathfrak{n}$. In the case of $\mathfrak{gl}_N$, these reduced algebras  are isomorphic to $\mathcal{P}_{[N]}$ endowed with the two Gelfand-Dickey brackets \eqref{eq:adler bracket}.\\

 Note that in \eqref{eq:matrix Lax} the nilpotent element $f= \sum_{i=1}^r f_i$ is principal. In the series of papers \cites{DSKV13,DSKV14,DSKV15,DSKV16b,DSKV18} De Sole, Kac, and Valeri generalized further the Drinfeld-Sokolov approach and managed to construct integrable Hamiltonian systems for any simple Lie algebra $\mathfrak{g}$ and any nilpotent element $f$. These systems are defined on the so-called classical $\mathcal{W}$-algebras $\mathcal{W}(\mathfrak{g},f)$ whose generators and brackets can be realized as quasi-determinants of certain operator valued matrices and 
 the matrix analogue of Gelfand-Dickey algebras and brackets. 
 \\

 They used a powerful language of Poisson vertex algebras (PVAs) which are differential algebras $(\mathcal{V}, \partial)$ together with a  $\lambda$-\textit{bracket} $\{\,{}_{\lambda}\,\}:\mathcal{V}\otimes \mathcal{V} \rightarrow \mathcal{V}[\lambda]$. This map must satisfy some axioms which are introduced in Section \ref{Sec:Preliminaries}. The $\lambda$-bracket induces a Lie bracket on the quotient space $\mathcal{V}/\partial \mathcal{V}$. The Gelfand-Dickey brackets \eqref{eq:adler bracket} can be generalized to the matrix case where $L$ is 
 a degree $N$ matrix differential operator of size $n \times n$, and $\mathcal{V}_L$ is the algebra of differential polynomials generated by its coefficients.
There is a unique PVA structure on the algebra $\mathcal{V}_{L}$ such that the induced Lie bracket on $\mathcal{V}_L/ \partial \mathcal{V}_L$ is the quadratic Gelfand-Dickey bracket. It is given by the so-called \textit{Adler identity}  (\cite{DSKV16b})
 \begin{equation} \label{eq:Adler_intro1}
\{L_{ij}(z){}_{\lambda}L_{hk}(w)\}=\Big(L_{hj}(w+\lambda+\partial)(z-w-\lambda-\partial)^{-1}(L_{ik})^{*}(\partial+\lambda-z)-L_{hj}(z)(z-w-\lambda-\partial)^{-1}L_{ik}(w) \Big)(1),
\end{equation}
 for all $1 \leq i,j,h,k \leq n $ and indeterminates $z$ and $w$. In \eqref{eq:Adler_intro1}, we denoted by $P(1)$ the constant coefficient of a pseudo-differential operator $P$ and  the \textit{adjoint} and \textit{symbol} of a pseudo-differential operator $ L(\partial)= \sum_{m\in \mathbb{Z}} a_m \partial^m$ are given by $L^*(\partial) = \sum_{m\in \mathbb{Z}} (-\partial)^m a_m$ and $L(z)= \sum_{m\in \mathbb{Z}} a_m z^m,$ respectively.
 \\

 The main objective of this paper is to build the foundations for our long term goal which is to construct integrable Hamiltonian hierarchies on classical $\mathcal{W}$-superalgebras associated with any simple Lie superalgebra $\mathfrak{g}$ and any even nilpotent element $f$. In the case where there exists an even element $s \in \mathfrak{g}$ of maximal Dynkin grading such that $f +z s$ is semisimple in $\mathfrak{g}[z^{\pm1}]$, one can follow the Drinfeld-Sokolov method \cites{DS85, DSKV13, Suh18}. However, such an element $s \in \mathfrak{g}$ does not always exist. In this paper, we realize Poisson vertex superalgebra (PVsA) structures of $\mathcal{W}$-superalgebras for rectangular nilpotent elements $f$ by introducing a super analogue of the $\lambda$-bracket \eqref{eq:Adler_intro1}. Even though rectangular nilpotents do admit an element $s$ making $f + zs$ semisimple, this new method will enable us in our next project to tackle the general case where $f$ is any nilpotent element, as this strategy was the main breakthrough in the nonsuper case.  
\\

Let $V$ be a finite dimensional vector superspace over $\mathbb{C}$ with a homogeneous  basis $\mathcal{B}:=\{e_i\}_{i \in I}$ and a positive integer $N$. We suppose that $I$ is an ordered set and write $\tilde{\imath} \in \{\bar{0},\bar{1}\}= \mathbb{Z}/2\mathbb{Z}$ for the parity of $e_i$. We denote the superalgebra of endomorphisms on  $V$ by $\mathfrak{gl}(I)= \text{Span}_\mathbb{C}\{e_{ij}|i,j\in I\}$, where $e_{ij}(e_k)=\delta_{jk} e_i$ for $i,j,k\in I$. If $m$ is the number of even elements in $\mathcal{B}$ and $n= |I|-m$, it is clear that $\mathfrak{gl}(I)$ is isomorphic to $\mathfrak{gl}(m|n)$. To this data, we attach the even matrix differential operator
\begin{equation} \label{eq:intro_L}
L(\partial)=\sum_{i\in I} e_{ii} \otimes \partial^N + \sum_{M=0}^{N-1} \sum_{i,j\in I} e_{ij}\otimes u_{M,ij}\partial^M \, \, \, \, \, \in \mathfrak{gl}(I) \otimes \mathcal{V}_I^{N},
\end{equation}
where the $u_{M,ij}$'s are generators of a superalgebra of differential polynomials $\mathcal{V}_I^N$, and the elements $e_{ij}$ form the natural basis of $\mathfrak{gl}(I)$. Note that both $e_{ij}$ and $u_{M,ij}$ have parity $\tilde{\imath}+\tilde{\jmath}$.
 We prove in Section \ref{subsec:Generic} that the following identity defines a unique PVsA structure on $\mathcal{V}_I^N$
\begin{equation} \label{eq:Adler_intro}
\begin{aligned}
 \{L_{ij}(z){}_{\lambda}L_{hk}(w)\}=&(-1)^{\tilde{\imath}\tilde{\jmath}+\tilde{\imath}\tilde{h}+\tilde{\jmath}\tilde{h}}L_{hj}(w+\lambda+\partial)(z-w-\lambda-\partial)^{-1}(L_{ik})^{*}(\lambda-z)(1) \\
&-(-1)^{\tilde{\imath}\tilde{\jmath}+\tilde{\imath}\tilde{h}+\tilde{\jmath}\tilde{h}}L_{hj}(z)(z-w-\lambda-\partial)^{-1}L_{ik}(w)(1),
\end{aligned}
\end{equation}
for $i,j,h,k\in I$.  We call this identity the \textit{super Adler identity} which is a super analogue of \cite{DSKV15}. 

We show in Proposition \ref{prop:GD} that the $\lambda$-bracket \eqref{eq:Adler_intro} on the PVsA $\mathcal{V}_I^{N}$ is connected to the following quadratic super Gelfand-Dickey Lie bracket 
\begin{equation} \label{sgdq}
     \smallint  \{f \, {}_{\lambda}  \, g \}_{\lambda=0}  =  \, \, \mathrm{Res} \,\,  \mathrm{str} \int (L \star \frac{\delta f}{\delta L})_+ \star L \star\frac{\delta g}{\delta L}-L \star(\frac{\delta f}{\delta L} \star L)_+  \star \frac{\delta g}{\delta L}
\end{equation}
on $\mathcal{V}_I^N/\partial \mathcal{V}_I^N$, for $L$ in \eqref{eq:intro_L} and $f,g\in \mathcal{V}_I^N.$  In this formula, the supertrace $\mathrm{str}$ is defined by $\mathrm{str}(e_{ii})=(-1)^{\tilde{\imath}}$ and the $\star$-product is given by 
\begin{equation} \label{eq:spro}
(a \otimes v) \star (b\otimes w):= (-1)^{\tilde{b}\tilde{v}+\tilde{a}\tilde{b}}  ab \otimes vw.
\end{equation}
Additionally, this $\lambda$-bracket can be deformed by adding constant multiples of the identity matrix to $L$. The deformed bracket $ [ \, \, {}_{\lambda} \, \, ] $ reduces on the quotient $\mathcal{V}_I^N/\partial \mathcal{V}_I^N$ to the linear super Gelfand-Dickey Lie bracket
\begin{equation}\label{sgdl}
      \smallint [ f \, {}_{\lambda}  \,  g]_{\lambda=0} =  \, \, \mathrm{Res} \,\,  \mathrm{str} \int \Big(L \star \frac{\delta f}{\delta L}-L \star \frac{\delta f}{\delta L}\Big)_+  \star\frac{\delta g}{\delta L}.
    \end{equation}

More generally, an \textit{Adler-type PVsA} is defined to be a PVsA which is generated as a differential superalgebra by the coefficients of some matrix pseudo-differential operator and whose $\lambda$-brackets between generators are given by the super Adler identity for that operator. We call such an operator $A(\partial)$ a \textit{super Adler-type operator} (sATO).
We prove in Theorem \ref{Thm_gl:Skew-Jacobi,scalar} that if  $\mathcal{V}$ is a differential superalgebra generated by the coefficients of some even matrix pseudo-differential operator $A(\partial)$ and if the super Adler identity \eqref{eq:Adler_intro} is well-defined, then the axioms of a PVsA are satisfied. \\

In Section \ref{sec:W-algebra}, we connect Adler-type PVsAs to the theory of classical $\mathcal{W}$-superalgebras. After noting that when $N=1$ the corresponding Adler-type PVsA is the affine PVsA $\mathcal{V}^{-1}(\mathfrak{gl}(m|n))$ of level $-1$, we prove the following theorem (Theorem \ref{thm:W-algebra and sATO-ver2} in Section \ref{sec:W-algebra}).
\begin{thm} \label{thm:intro-1}
    The differential superalgebra $\mathcal{V}_I^N$ endowed with the $\lambda$-bracket \eqref{eq:Adler_intro} is isomorphic to 
the classical $\mathcal{W}$-superalgebra $\mathcal{W}(\mathfrak{gl}(Nm|Nn),f)$ of level $-1$ associated with the so-called $N\times (m|n)$ rectangular nilpotent element $f$.
\end{thm}

To prove the Theorem \ref{thm:intro-1}, we consider a $(Nm|Nn) \times (Nm|Nn)$ matrix of the form \begin{equation} \label{eq:P(partial)}
P(\partial)= \left[
\begin{array}{ccccc}
 \mathbbm{1}_{(m|n)}\partial+ P_{[11]} & \mathbbm{1}_{(m|n)}& 0  & \cdots & 0 \\
P_{[12]} &\mathbbm{1}_{(m|n)}\partial + P_{[22]}& \mathbbm{1}_{(m|n)} & \cdots & 0 \\
\vdots & \vdots& \vdots  & \ddots  & \vdots  \\
 P_{[N-1\, 1]}&  P_{[N-1\, 2]} & P_{[N-1\, 3]}  & \cdots & \mathbbm{1}_{(m|n)} \\
P_{[N1]}& P_{[N2]}& P_{[N3]}  & \cdots & \mathbbm{1}_{(m|n)}\partial + P_{[NN
]} \\
\end{array}
\right],
\end{equation}
 where the $P_{[ij]}$'s are even matrices of size $(m|n) \times (m|n)$ whose coefficients are the generators of some superalgebra of differential polynomials. Next, we take the last $(m+n)$ rows and first $(m+n)$ columns of $P(\partial)$ and define the corresponding {\it quasi-determinant} matrix $\mathrm{qdet}(P(\partial))\in \mathfrak{gl}(m|n)\otimes \mathcal{V}(\partial).$  We emphasize that the quasi-determinant is defined via the $\star$-product \eqref{eq:spro}. Finally, we show in Theorem \ref{thm:W-algebra and sATO-ver1} that the generators of the $\mathcal{W}$-superalgebra $\mathcal{W}(\mathfrak{gl}(Nm|Nn),f)$ are given by the coefficients of this quasi-determinant.
\\

In Section \ref{Sec:integrable}, we construct  integrable hierarchies on these rectangular $\mathcal{W}$-superalgebras. By Theorem \ref{thm:intro-1}, it is equivalent to build an integrable hierarchy on an Adler-type PVsA $\mathcal{V}_I^N$ associated with $L(\partial)$ in \eqref{eq:intro_L}. The main ingredient in this section is the infinite family of
Hamiltonians 
\begin{equation} \label{eq:intro_Hamiltonian}
    h_k=\frac{N}{k}\mathrm{Res}\;\text{str}L^{\frac{k}{N}}(\partial)\in \mathcal{V}_I^{N},
\end{equation}
for a positive integer $k$. A key point is that the fractional power $L^{\frac{k}{N}}(\partial)$ is computed with respect to a different  product
\begin{equation} \label{cpro}
 (a \otimes v) \circ (b\otimes w):= (-1)^{\tilde{b}\tilde{v}} ab \otimes vw
\end{equation}
on $\mathfrak{gl}(I) \otimes \mathcal{V}_I^N$, while  the super Gelfand-Dickey brackets \eqref{sgdq} and \eqref{sgdl} can only be expressed using the $\star$-product \eqref{eq:spro}.
We check in Lemmas \ref{lem:LM} and \ref{lem:LM-initial} that 
$ \{h_k \, {}_\lambda  \, u\} \, |_{\lambda =0 }= \{ h_{k+N} \, {}_\lambda  \, u \} \, |_{\lambda =0 }$ for a positive integer $k$ and $\{h_{k'} \, {}_\lambda  \, u\} \, |_{\lambda =0 }=0$
for $k'=1,2,\cdots, N$. Finally, by the Lenard-Magri scheme, we obtain the following theorem.
\begin{thm}[Proposition  \ref{thm:integragle system} and Theorem \ref{thm:main-2}]
Let $f$ be the
$N\times (m|n)$ rectangular nilpotent in the Lie superalgebra $\mathfrak{gl}(Nm|Nn)$.
For positive integers $k$,  the family of equations
\begin{equation} \label{eq:intro_integrable hierarchy}
    \frac{du}{dt_k}= \{h_k\, {}_\lambda \, u\} \, |_{\lambda=0} \quad  \Leftrightarrow  \quad \frac{dL}{dt_{k}}=({L}^{\frac{k}{N}})_{+} \circ  L- L \circ ({L}^{\frac{k}{N}})_{+}
\end{equation}
is an integrable system on $\mathcal{V}_I^{N}\simeq \mathcal{W}(\mathfrak{gl}(Nm|Nn),f)$. In addition, the local functionals $\int h_{k'}$ for positive integers $k'$ are integrals of motion for any equation in \eqref{eq:intro_integrable hierarchy}. 
\end{thm}

This system is a specialization of noncommutative KdV \cite{OS98} to the algebra $(\mathfrak{gl}(I) \otimes \mathcal{V}_I^N, \, \circ)$. Finally, we stress that for the construction of the Hamiltonians and the Lax formulation of the hierarchy the $\circ$-product had to be used while  the super Gelfand-Dickey brackets \eqref{sgdq} and \eqref{sgdl} can only be expressed using the $\star$-product.

\newpage

\section{Preliminaries}\label{Sec:Preliminaries}

In this section, we present an overview of some basic notions used
throughout the paper. A  \textit{vector superspace} is a $\mathbb{Z}/2\mathbb{Z}$-graded
vector space $V=V_{\bar{0}}\oplus V_{\bar{1}}$. We only consider vector superspaces and tensor products over $\mathbb{C}$, the field of complex numbers. A nonzero element
$a\in V$ is called \textit{homogeneous} if it belongs either to $V_{\bar{0}}$
or $V_{\bar{1}}$ and each homogeneous element $a\in V_{\bar{i}}$
is assigned a \textit{parity} $\tilde{a}=i\in\{0,1\}$. The direct
summand $V_{\bar{0}}$ (resp. $V_{\bar{1}}$) of $V$ is called the
\textit{even} (resp. \textit{odd}) subspace of $V$, and an element
in $V_{\bar{0}}$ (resp. $V_{\bar{1}}$) is said to be \textit{even}
(resp. \textit{odd}). A \textit{superalgebra} $(\mathcal{A},\cdot)$
consists of a vector superspace $\mathcal{A}$ and a bilinear product on it satisfying $\mathcal{A}_{\bar{i}}\cdot\mathcal{A}_{\bar{j}}\subseteq\mathcal{A}_{\bar{i}+\bar{j}}$
for $i,j\in\{0,1\}$. For two vector superspaces $V$ and $W$, $\mathrm{Hom}(V,W)$ is also a vector superspace by letting $\phi\in \mathrm{Hom}(V,W)_{\bar{i}}$
if and only if $\phi(V_{\bar{j}})\subseteq W_{\bar{i}+\bar{j}}$. For the sake of simplicity, it is assumed that every
element is
homogeneous unless stated otherwise.

\subsection{Poisson vertex superalgebras} 
A \textit{Lie superalgebra} $\text{\ensuremath{\mathfrak{g}}}=\mathfrak{g}_{\bar{0}}\oplus\mathfrak{g}_{\bar{1}}$
is a vector superspace equipped with an even linear map $[ \ ,\ ]:\text{\ensuremath{\mathfrak{g}}}\otimes\text{\ensuremath{\mathfrak{g}}}\rightarrow\text{\ensuremath{\mathfrak{g}}}$
which satisfies 
\begin{enumerate}[]
\item \quad (skew-symmetry) $[a,b]=-(-1)^{\tilde{a}\tilde{b}}[b,a]$,
\item \quad (Jacobi identity) $[a,[b,c]]=[[a,b],c]+(-1)^{\tilde{b}\tilde{c}}[b,[a,c]]$,
\end{enumerate}
for $a,b,c\in\text{\ensuremath{\mathfrak{g}}}$. The space of endomorphisms $\textrm{End}(V)$ on a vector superspace
$V=V_{\bar{0}}\oplus V_{\bar{1}}$ is one of the fundamental
examples of Lie superalgebras whose bracket is defined by $[\phi,\psi]:=\phi\psi-(-1)^{\tilde{\phi}\tilde{\psi}}\psi\phi$
for $\phi,\psi\in \text{End}(V)$. It is called the \textit{general linear Lie superalgebra}
$\mathfrak{gl}(V)$. For any integers $m, n \geq 0$, the general linear Lie superalgebra  
$\mathfrak{gl}(m|n)$ is given by $\mathfrak{gl}(\mathbb{C}^{(m|n)})$, where  $\mathbb{C}^{(m|n)}$ is the vector superspace with an ordered basis whose first $m$ elements are even and last $n$ elements are odd.

Let $\mathcal{V}$ be 
a \textit{differential superalgebra}, i.e. a superalgebra endowed with an even derivation $\partial:\mathcal{V}\to \mathcal{V}$. We assume that $\mathcal{V}$ is associative and supercommutative. The derivation can be naturally extended to the algebra $\mathcal{V}[\lambda]$ of polynomials in the even indeterminate $\lambda$. 
An even linear map $\{\ {}_{\lambda}\ \}:\mathcal{V}\otimes\mathcal{V}\rightarrow\mathcal{V}[\lambda]$
satisfying the following conditions
\begin{enumerate}[]
\item \quad (sesquilinearity) $\{\partial a{}_{\lambda}b\}=-\lambda\{a{}_{\lambda}b\}$
and $\{a{}_{\lambda}\partial b\}=(\lambda+\partial)\{a{}_{\lambda}b\}$,
\item \quad (left Leibniz rule)$\{a{}_{\lambda}bc\}=\{a{}_{\lambda}b\}c+(-1)^{\tilde{a}\tilde{b}}b\{a{}_{\lambda}c\}$, 
\item \quad (right Leibniz rule) $\{ab{}_{\lambda}c\}=(-1)^{\tilde{b}\tilde{c}}\{a{}_{\lambda+\partial}c\}_{\rightarrow}b+(-1)^{\tilde{a}(\tilde{b}+\tilde{c})}\{b{}_{\lambda+\partial}c\}_{\rightarrow}a$,
\end{enumerate}
for $a,b,c\in\mathcal{V}$, is called a $\lambda$-\textit{bracket} on $\mathcal{V}$. We use the notation
$ \{a{}_\lambda b\}= \sum_{n\in \mathbb{Z}_+} \lambda^n a_{(n)}b$. In the right Leibniz rule, the right arrow means that the operator $\partial$ acts only on $b$. More precisely, 
$\{a{}_{\lambda+\partial}c\}_{\rightarrow}b:=\sum_{n\in \mathbb{Z}_+}{\displaystyle a_{(n)}c\,(\lambda+\partial)^{n}b}$.

\begin{defn}
A \textit{Poisson vertex superalgebra (PVsA)} is a differential superalgebra $(\mathcal{V}, \partial)$ with 
a $\lambda$-bracket $\{\ {}_{\lambda}\ \}$ satisfying the  following two additional axioms: 
\begin{enumerate}[]
\item \quad \textup{(skew-symmetry)} $\{a{}_{\lambda}b\}=-(-1)^{\tilde{a}\tilde{b}}\{b{}_{-\lambda-\partial}a\}$,
\item \quad \textup{(Jacobi identity)} $\{a{}_{\lambda}\{b{}_{\mu}c\}\}=\{\{a{}_{\lambda}b\}{}_{\lambda+\mu}c\}+(-1)^{\tilde{a}\tilde{b}}\{b{}_{\mu}\{a{}_{\lambda}c\}\}$,
\end{enumerate}
for $a,b,c\in\mathcal{V}$. In the skew-symmetry, $\{b{}_{-\lambda-\partial}a\}:=\sum_{n\in \mathbb{Z}_+}(-\lambda-\partial)^{n}\,b_{(n)}a$
and the Jacobi identity axiom holds in $\mathcal{V}[\lambda,\mu]$.
Note that the sesquilinearity axioms (resp. left and right Leibniz rules) are compatible
with the skew-symmetry axiom. 
\end{defn}


Now, let us present some useful results on $\lambda$-brackets.
Consider a vector superspace $U$. By the Leibniz rule, there is a unique derivation on the superalgebra $\mathcal{V}(U)= S(\mathbb{C}[\partial]\otimes U)$ extending the action of $\partial$ on the $\mathbb{C}[\partial]$-module $\mathbb{C}[\partial]\otimes U$.
Recall that
the supersymmetric algebra of a vector superspace $V$ is
$ 
S(V):=S(V_{\bar{0}})\otimes{\bigwedge}(V_{\bar{1}})
$, where $S(V_{\bar{0}})$ is the symmetric algebra of $V_{\bar{0}}$
and $\bigwedge(V_{\bar{1}})$ is the exterior algebra of $V_{\bar{1}}$. 
If we fix a homogeneous basis $\{u_{i}\}_{i\in I}$
of $U$, then as a differential superalgebra
\begin{equation} \label{eq:P differential polynomial}
 \mathcal{V}(U)=\mathbb{C}[u_{i}^{(n)}\big\vert i\in I,\,n\in\mathbb{Z}_{+}], 
\end{equation}
where the even derivation $\partial$ is given by $u_{i}^{(n)}=\partial^n u_{i}$ for $n\in \mathbb{Z}_+$. The partial derivatives $(\frac{\partial}{\partial u_{i}^{(m)}})_{i \in I, m \geq 0}$ are the derivations of $\mathcal{V}(U)$
   of the same parity as ${u}_{i}$ and are defined by $\frac{\partial}{\partial u_{i}^{(m)}}u_{j}^{(n)}=\delta_{ij}\delta_{mn}$, for $i\in I$ and $m \in \mathbb{Z}_+$.

\begin{thm}[\cites{BDSK09,Suh20}] \label{thm:PVA extension}Let $U$ be a vector superspace and $\mathcal{V}(U)$ be the superalgebra of differential polynomials given in \eqref{eq:P differential polynomial} endowed with a $\lambda$-bracket $\{\ {}_\lambda \ \}$. Then the following properties hold.

\begin{enumerate}[]
\item \textup{(a)} (master formula)
For $f,g\in \mathcal{V}(U)$, we have 
\begin{equation} 
\{f{}_{\lambda}g\}={\displaystyle \sum_{i,j\in I,m,n\in\mathbb{Z}_{\geq0}}}C_{i,j}^{f,g}\frac{\partial g}{\partial u_{j}^{(n)}}(\lambda+\partial)^{n}\{u_{i}{}_{\lambda+\partial}u_{j}\}_{\rightarrow}(-\lambda-\partial)^{m}\frac{\partial f}{\partial u_{i}^{(m)}}\label{eq:master formula},
\end{equation}
where $C_{i,j}^{f,g}=(-1)^{\tilde{f}\tilde{g}+\tilde{u}_{i}\tilde{u}_{j}+\tilde{g}\tilde{u}_{j}+\tilde{u}_{j}}$.
\item  \textup{(b)}  The $\lambda$-bracket 
satisfies the skew-symmetry axiom if and only if 
\begin{equation}
\{u_{i}{}_{\lambda}u_{j}\}=-(-1)^{\tilde{u}_{i}\tilde{u}_{j}}\{u_{j}{}_{-\lambda-\partial}u_{i}\}\label{eq:skew-sym}
\end{equation}
holds for any $i,j\in I$.

\item \textup{(c)} Assume that the $\lambda$-bracket satisfies the skew-symmetry axiom.
Then the differential superalgebra $\mathcal{V}(U)$ endowed with this $\lambda$-bracket is a PVsA, provided that
\begin{equation}
\{u_{i}{}_{\lambda}\{u_{j}{}_{\mu}u_{k}\}\}=\{\{u_{i}{}_{\lambda}u_{j}\}{}_{\lambda+\mu}u_{k}\}+(-1)^{\tilde{u}_{i}\tilde{u}_{j}}\{u_{j}{}_{\mu}\{u_{i}{}_{\lambda}u_{k}\}\}\label{eq:Jacobi}
\end{equation}
 holds for any $i,j,k\in I$.
\end{enumerate}
\end{thm}

\begin{proof}
For a detailed proof, refer to Theorem 1.15 in \cite{BDSK09} and Proposition 2.4 and 2.5 in \cite{Suh20}.
\end{proof}

In other words, Theorem \ref{thm:PVA extension} (a) says that a $\lambda$-bracket on a superalgebra of differential polynomial $\mathcal{P}$ is completely determined by its values on pairs of elements in a generating set $S$. Moreover,
in order to see if the $\lambda$-bracket on $\mathcal{P}$ is a PVsA $\lambda$-bracket, it is enough to check the skew-symmetry and Jacobi identity axioms between the elements of $S$. The following example is one of the most fundamental example of PVsA and will be widely used in this paper.

\begin{example}[Affine Poisson vertex superalgebra] \label{ex:affine}
Let $\mathfrak{g}$ be a Lie superalgebra with an even supersymmetric invariant
bilinear form $(\ |\ )$ and let $k\in \mathbb{C}.$ 
Consider the $\lambda$-bracket on $\mathcal{V}(\mathfrak{g})$ defined by 
\begin{equation} \label{eq: Affine PVA}
\{a{}_{\lambda}b\}=[a,b]+k\lambda(a|b)\quad\text{for} \quad a,b\in\mathfrak{g}
\end{equation}
and Theorem \ref{thm:PVA extension} (a). 
We can also check the skew-symmetry and Jacobi identity axioms by  Theorem \ref{thm:PVA extension} (b) and (c).
The differential superalgebra
$\mathcal{V}(\mathfrak{g})$ endowed with the bracket \eqref{eq: Affine PVA} is called the \textit{affine Poisson vertex
superalgebra} (affine PVsA) associated with $\mathfrak{\mathfrak{g}}$. To emphasize the role of $k$ in the definition of the bracket \eqref{eq: Affine PVA}, we sometimes denote the affine PVsA by $\mathcal{V}^k(\mathfrak{g})$ and call $k$ its \textit{level}. 
\end{example}

\subsection{Classical $\mathcal{W}$-superalgebras} \label{sec:2.2} Let $\mathfrak{g}$ be a basic classical Lie superalgebra and $f$ be an even nilpotent element in an $\mathfrak{sl}_2$-triple $(e,h,f)$ in $\mathfrak{g}$. Consider the nondegenerate invariant supersymmetric even bilinear form $( \ |\ )$ on $\mathfrak{g}$ normalized by $(e|f)= \frac{1}{2}(h|h)=1$ and the $\frac{\mathbb{Z}}{2}$-graded decomposition of $\mathfrak{g}:= \bigoplus_{i\in\frac{\mathbb{Z}}{2}}\mathfrak{g}_{i}$, where $\mathfrak{g}_{i} = \{\, a\in \mathfrak{g}\, | \, [h,a]=2i a\, \}.$ 
We often denote subspaces of $\mathfrak{g}$ by
$
 \mathfrak{g}_{\geq j}=\bigoplus_{i\geq j}\mathfrak{g}_{i}$, $
 \mathfrak{g}_{\leq j}=\bigoplus_{i\leq j}\mathfrak{g}_{i}$ and $\mathfrak{g}_{<j}=\bigoplus_{i<j}\mathfrak{g}_{i}.
$
In particular, we let 
\begin{alignat}{2} \label{eq:Lie subalgebra}
\mathfrak{n}:=\mathfrak{g}_{\geq\frac{1}{2}},\qquad & \mathfrak{\mathfrak{p}}:=\mathfrak{g}_{<1}.
\end{alignat}

Define the differential superalgebra homomorphism $\rho: \mathcal{V}(\mathfrak{g})\to \mathcal{V}(\mathfrak{p})$  determined by $a \mapsto \pi_{\mathfrak{p}}(a)-(f|a)$ for $a\in \mathfrak{g}$, where $\pi_{\mathfrak{p}}: \mathfrak{g} \to \mathfrak{p}$ is the canonical projection map. Extend the map $\rho$ to the linear map
$\rho: \mathcal{V}(\mathfrak{g})[\lambda]\to \mathcal{V}(\mathfrak{p})[\lambda]$ such that $\rho(A \lambda^n)=\rho(A)\lambda^n$ for $A\in \mathcal{V}(\mathfrak{g})$
and 
consider the subspace of $\mathcal{V}(\mathfrak{p})$: 
\begin{equation} \label{eq: W-bracket}
\mathcal{W}^k(\mathfrak{g},f)=\{a\in\mathcal{V}(\mathfrak{p})|\,\rho ( \{n{}_{\lambda}a\}_{\text{Aff}})=0\;\text{{for\,all}}\;n\in\mathfrak{n}\},
\end{equation}
where $\{\ {}_{\lambda}\ \}_{\text{Aff}}$ is the $\lambda$-bracket of the affine PVsA $\mathcal{V}^k(\mathfrak{g})$ in Example \ref{ex:affine}. One can show that $\mathcal{W}^k(\mathfrak{g},f)$ is a differential subalgebra of $\mathcal{V}(\mathfrak{p})$ and it can be given a PVsA structure for the $\lambda$-bracket defined by  
\begin{equation}\label{lambda bracket of W-superalgebra}
\{A{}_\lambda B\}:=\rho(\{A{}_\lambda B\}_{\text{Aff}})
\end{equation}
for $A,B\in \mathcal{W}^k(\mathfrak{g},f).$ The PVsA $\mathcal{W}^k(\mathfrak{g},f)$ is called the \textit{classical $\mathcal{W}$-superalgebra associated with $\mathfrak{g}$ and $f$.} For later usage, we note that if the images $\rho(A),\rho(B)$ of $A, B\in \mathcal{V}(\mathfrak{g})$ are both in $\mathcal{W}^k(\mathfrak{g},f)$, then 
\begin{equation} \label{eq:W-bracket;property}
    \{\rho(A){}_\lambda \rho(B)\}= \rho(\{\rho(A){}_\lambda \rho(B)\}_{\text{Aff}})=\rho (\{A {}_\lambda B\}_{\text{Aff}}).
\end{equation}
The first equality follows from \eqref{lambda bracket of W-superalgebra}, and the second equality follows from Corollary 3.3 \cite{DSKV13} and Proposition 3.7 \cite{Suh16}.

\begin{prop}[\cites{DSKV16c, Suh20}]\label{prop:free generating set of W-algebra}
  Let $\{v_i|i\in I\}$ be a basis of $\mathfrak{g}^f:= \mathrm{ ker}(\mathrm{ad} f)\subset \mathfrak{g}$. Then there is a free generating set $\{w_i|i\in I\}$ of the $\mathcal{W}$-superalgebra $\mathcal{W}^k(\mathfrak{g},f)$ as a differential superalgebra satisfying
   \begin{equation} \label{eq:property of generator of W-algebra}
       w_i-v_i \in \partial(\mathbb{C}[\partial]\otimes \mathfrak{p}) \oplus \bigoplus_{m\geq 2}(\mathbb{C}[\partial]\otimes \mathfrak{p})^{\otimes m}
   \end{equation}
   for $i\in I.$ Moreover, if a subset $\{w_i|i\in I\}\subset\mathcal{W}^k(\mathfrak{g},f)$ consists of elements which (i) satisfy the property \eqref{eq:property of generator of W-algebra} and (ii) are homogeneous with respect to the {conformal grading}, then it is a free generating set. The conformal grading $\Delta$ on the differential superalgebra $\mathcal{V}(\mathfrak{p})$ is defined inductively by 
   \begin{equation}\label{eq:conformal weight}
       \Delta_a = 1- j_a, \quad \Delta_{\partial A} = \Delta_A+1, \quad \Delta_{AB} = \Delta_A + \Delta_B,
   \end{equation}
   where $a\in \mathfrak{p} \cap \mathfrak{g}_{j_a}$ and $A,B \in \mathcal{V}(\mathfrak{p})$ are homogeneous elements for that grading.
\end{prop}
\begin{proof}
 We refer to Proposition 3.12 \cite{Suh16}, and Proposition 2.10 and Remark 2.11 \cite{Suh20}.
\end{proof}

There are several known methods to describe a generator subset of $\mathcal{W}^k(\mathfrak{g},f)$ explicitly. In particular, when $\mathfrak{g}$ is a classical finite simple Lie algebra, a generating set can be obtained by the so-called \textit{generalized quasi-determinants} and \textit{generalized Adler-type operators} \cite{DSKV18}. In \cite{DSKV15}, this method has proved crucial in constructing integrable systems on the corresponding $\mathcal{W}$-algebras as it connects the $\lambda$-brackets to the Gelfand-Dickey theory. As discussed in the introduction, our goal is to develop  an analogous method for $\mathcal{W}$-superalgebras. Let us give an example of the quasi-determinant method for $\mathcal{W}^k(\mathfrak{gl}_{3},f)$ when $f$ is the principal nilpotent element.

\begin{example}
    Let $\mathfrak{g}= \mathfrak{gl}_{3}$ and $f=e_{21}+e_{32}$. Consider the matrix with entries in $\mathbb{C}[\partial]\ltimes \mathcal{V}(\mathfrak{p})$:
\begin{equation}
\mathcal{L}= \left[\begin{array}{ccc} k\partial+q_{11} & q_{21} & q_{31} \\ -1 & k\partial+ q_{22} & q_{32} \\ 0 & -1 & k\partial+q_{33}\end{array} \right].
\end{equation}
Recall that $\mathcal{V}(\mathfrak{p})$ is the differential algebra of polynomials in the indeterminates $q_{ij}$.
Consider the quasi-determinant of $\mathcal{L}$ with respect to the top right corner $e_{13}$ and denote its coefficients by $w_1, w_2, w_3\in \mathcal{V}(\mathfrak{p})$:
\begin{equation}
 |\mathcal{L}|_{13}=q_{31}- \left[k\partial+ q_{11}\ q_{21}\right]
\left[\begin{array}{cc}  -1 & -k\partial- q_{22}  \\ 0 & -1  \end{array} \right] 
\left[ \begin{array}{c}  q_{32} \\ k\partial+q_{33}\end{array} \right]\\=: k^3 \partial^3+ w_1 k^2\partial^2 + w_2 k\partial + w_3.
\end{equation}
The set $\{w_1,w_2, w_3\}$ generates the classical $\mathcal{W}$-algebra $\mathcal{W}^k(\mathfrak{g},f).$ In particular, we have $w_1=q_{11}+q_{22}+q_{33}$ and $w_2=q_{21}+ q_{32}+q_{11}q_{22}+q_{11}q_{33}+q_{22}q_{33} + k \partial(q_{22})+2k\partial(q_{33}).$
\end{example}

The following example is the simplest classical $\mathcal{W}$-superalgebra with nontrivial odd part. In this case, we can find generators using generalized Drinfeld-Sokolov reduction (See Example 3.15 in \cite{Suh18}).
\begin{example} \label{ex:W(osp(1|2))}
   Let $\mathfrak{g}=\mathfrak{osp}(1|2)$ and $(e,h,f)$ be the unique $\mathfrak{sl}_2$-triple  in $\mathfrak{g}$. Additionally, there are two independent odd elements $f_{\text{od}}\in \mathfrak{g}_{-1/2}$ and $e_{\text{od}}\in \mathfrak{g}_{1/2}$. After a proper normalization, we have  $[f,e_{\text{od}}]=f_{\text{od}}$, $[e,f_{\text{od}}]=e_{\text{od}}$, $[f_{\text{od}},f_{\text{od}}]=-2 f$,  $[e_{\text{od}},e_{\text{od}}]=2 e$, and $[e_{\text{od}},f_{\text{od}}]=-h$. Then the two elements, 
   \begin{align*}
       & w_1 = f_{\text{od}}-\frac{1}{2} e_{\text{od}} h -k \partial(e_{\text{od}})\;\;\text{and} \\
       & w_2 = f+ \frac{1}{2} f_{\text{od}} e_{\text{od}}-\frac{1}{4} h^2 + \frac{1}{4} e_{\text{od}} \partial(e_{\text{od}}) - \frac{1}{2} k \partial(h)
   \end{align*}
   generate $\mathcal{W}^k(\mathfrak{g},f)$. Moreover,  $\mathcal{W}^k(\mathfrak{g},f)$ is known to be isomorphic to the Neveu-Schwartz PVA.
\end{example}

The last example in this section is one of the main ingredients of this paper: a classical $\mathcal{W}$-superalgebra associated with the algebra $\mathfrak{gl}(Nm|Nn)$ where $N,m, n \geq 2$ and its \textit{rectangular nilpotent} $f$. Explicitly, $f$ is 
the even nilpotent element in $\mathfrak{gl}(Nm|Nn)$ such that its corresponding Jordan blocks are parameterized by the partition $(\underbrace{N,\cdots, N}_{m-\text{copies}}|\underbrace{N,\cdots, N}_{n-\text{copies}})$.  We denote this partition by $N\times (m|n)$ (see Definition 3.4 \cite{Peng14}).

\begin{example} \label{ex: rectangular W-super}
Let $\mathfrak{g}= \mathfrak{gl}(Nm|Nn)$ for three positive integers $m,n, N$ such that $N\geq 2$. \\
 (i)  After  a proper change of basis, we get the following presentation of $\mathfrak{gl}(Nm|Nn)$ : 
 \begin{equation} \label{ex:gl(ml|nl)}
\mathfrak{gl}(Nm|Nn)= \left\{ \left. \left[
\begin{array}{cccc}
A_{[11]} & A_{[12]} & \cdots & A_{[1N]} \\
A_{[21]} & A_{[22]} & \cdots & A_{[2N]} \\
\vdots & \vdots & \ddots & \vdots  \\
A_{[N1]} & A_{[N2]} & \cdots & A_{[NN]} \\
\end{array}\right] \ \right| \ A_{[ij]}\in \mathfrak{gl}(m|n) \text{ for } i,j=1,2,\cdots,  N\  \right\}
 \end{equation}
(ii) Take $f\in \mathfrak{gl}(Nm|Nn)$ by letting $A_{[21]}= A_{[32]}= \cdots = A_{[N\, N-1]}=\mathbbm{1}_{(m|n)}$, where $\mathbbm{1}_{(m|n)}$ is the identity matrix in $\mathfrak{gl}(m|n)$, and $A_{ij}=0$ otherwise. Using the matrix presentation in \eqref{ex:gl(ml|nl)}, 
\begin{equation} \label{ex:gl(ml|nl)-f}
f=   \left[
\begin{array}{ccccc}
0 &0 &0  & \cdots & 0  \\
\mathbbm{1}_{(m|n)} & 0 & 0& \cdots & 0 \\
0  &  \mathbbm{1}_{(m|n)} & 0& \cdots  & 0 \\
\vdots & \vdots & & \ddots & \vdots  \\
0 & 0 & 0 & \cdots&0 \\
\end{array}\right].
 \end{equation}
Then the classical $\mathcal{W}$-superalgebra $\mathcal{W}^k(\mathfrak{g},f)$ is freely generated by $N\cdot (m+n)^2$ elements and it is called a {\textit (classical) rectangular $\mathcal{W}$-superalgebra.}  We will omit the term ``classical'' from now on.
\end{example}


%

%

\subsection{Integrable Hamiltonian systems}  
In this section, we briefly review some notions on integrable Hamiltonian systems associated with a PVsA. 


\begin{defn} Let $\mathcal{P}$ be a PVsA with a $\lambda$-bracket $\{\, {}_\lambda \, \}$.
  A {\it Hamiltonian equation} associated with $\mathcal{P}$ is an evolutionary equation 
\begin{equation} \label{eq:Hamiltonian eq}
\frac{d\phi}{dt}=\{\, h\, {}_{\lambda}\, \phi \, \}\Big\vert_{\lambda=0}
\end{equation}
for some $h\in\mathcal{P}_{\bar{0}}$. Note that by the Leibniz rule, the equation \eqref{eq:Hamiltonian eq} uniquely defines a derivation of $\mathcal{P}$.
\end{defn}

Consider the quotient space 
 $\int \mathcal{P}:= \mathcal{P}/\partial \mathcal{P}$. The image of $a\in \mathcal{P}$ in $\int \mathcal{P}$ is denoted by $\int a$ and called a {\it local functional}. By the skew-symmetry and the Jacobi identity of PVsA, the space $\mathcal{P}/\partial \mathcal{P}$ of local functionals is a Lie superalgebra for the bracket: 
\begin{equation} \label{eq:local bracket}
[ \, \smallint f\, ,\, \smallint g\, ]:=\smallint\, \{\, f\, {}_{\lambda} \, g\, \} \vert_{\lambda=0}\quad \text{ for} \quad f,g\in\mathcal{P}.
\end{equation}

 A local functional $\int f \in \int \mathcal{P}$ is 
an \textit{integral of motion} of the Hamiltonian equation \eqref{eq:Hamiltonian eq} if and only if
\[
[\smallint f,\smallint h]=0.
\]
Moreover by the PVsA axioms, we have 
 \begin{equation} \label{eq:derivation commute}
 [\smallint f,\smallint h]=0 \iff \text{the derivations   } \{ f \, {}_{\lambda} \, \cdot \}\vert_{\lambda=0} \text{  
 and   } \{ h \, {}_{\lambda} \, \cdot \}\vert_{\lambda=0} \text{   of   } \mathcal{P} \text{   commute}. 
\end{equation}

\begin{defn} 
In this paper, we call integrable system an infinite
dimensional super-commutative Lie subalgebra of $\mathcal{P}\slash\partial\mathcal{P}$
or equivalently an infinite family of compatible Hamiltonian equations.
\end{defn}

A well-known method of constructing such an abelian subalgebra, or equivalently an infinite family of pairwise commuting derivations of $\mathcal{P}$, is the so-called \textit{Lenard-Magri scheme} \cite{Magri78}.
\begin{prop}[Lenard-Magri scheme]\label{prop:LM scheme}
Suppose that $\mathcal{P}$ is a PVsA endowed with two distinct  $\lambda$-brackets
$\{ \ {}_{\lambda} \ \}_{K}$ and $\{ \ {}_{\lambda}\ \}_{H}$. If there
exists a set of linearly independent even local functionals $\{ \int h_i| \, i\in\mathbb{Z}_+\}$
such that
\begin{equation} \label{eq:iteration}
[\, \smallint f\, ,\, \smallint h_{i+1}\, ]_{K}=[\, \smallint f\, ,\, \smallint h_i\, ]_{H}\,
\end{equation}
for any $f\in \mathcal{P}$ and $i\in \mathbb{Z}_+$, then 
\begin{equation} \label{eq:HIS-LM}
\frac{d\phi}{dt_i}=\{\, h_{i} \, {}_{\lambda}\, \phi\, \}_{H}\Big\vert_{\lambda=0}, \, \, i \in \mathbb{Z}_+
\end{equation}
is an integrable system. Moreover, all the local functionals $\int h_i$ are integrals of motion of the equations \eqref{eq:HIS-LM}.
\end{prop}

In the case of the affine PVA $ \mathcal{V}^k(\mathfrak{gl}_{n})$, there are two distinct $\lambda$-brackets $\{\ {}_\lambda \ \}_H$ and $\{\ {}_\lambda \ \}_K$ defined by 
\begin{equation}
    \{\, a\, {}_\lambda \,  b\, \}_H = [a,b]+k\lambda(a|b), \quad  \{\, a\, {}_\lambda \, b\, \}_K= (\mathbbm{1}_{n}|[a,b])
\end{equation}
for any $a,b\in \mathfrak{gl}_n$ and the $n\times n$ identity matrix $\mathbbm{1}_{n}$. Moreover one can find a sequence $\smallint h_i$ satisfying the conditions of Proposition \ref{prop:LM scheme} \cite{DS85}. As for the $\mathcal{W}$-superalgebra $\mathcal{W}^k(\mathfrak{g}, f),$ besides the  $\lambda$-bracket $\{\ {}_\lambda \ \}_H$ inherited from the affine PVsA bracket \eqref{eq: Affine PVA}, there is another $\lambda$-bracket induced from the bracket on the differential superalgebra $\mathcal{V}(\mathfrak{g}):$
\begin{equation} \label{eq:K-brakcet of W-alg}
    \{\, a\, {}_\lambda \, b\, \}_K = (s|[a,b]) \quad \text{ for } \quad a,b\in \mathfrak{g}
\end{equation}
where $s\in \text{ker}(\text{ad}\, \mathfrak{n})$. If $f+ z s \in \mathfrak{g}(\!(z^{-1})\!)$ is semisimple, integrable systems on $\mathcal{W}^k(\mathfrak{g},f)$ can be constructed using a super analogue of the Drinfeld-Sokolov construction (\cite{Suh18}, \cite{DSKV13}). 
Finally, let us come back to the Example \ref{ex:W(osp(1|2))}.

\begin{example}[super KdV]
    Let $\mathfrak{g}=\mathfrak{osp}(1|2)$ and $f$ be the even nilpotent element in Example \ref{ex:W(osp(1|2))}. Then $\mathcal{W}^k(\mathfrak{g},f)$ is endowed with two $\lambda$-brackets. The first $\{ \ {}_\lambda \ \}_H$, which is induced from $\mathcal{V}^k(\mathfrak{g})$, and the second $\{ \ {}_\lambda \ \}_K$ defined by \eqref{eq:K-brakcet of W-alg} with $s:=e$. Since $f+zs$ is semisimple, the methods of \cite{Suh18} can be applied and lead to the so-called {\it super KdV} integrable system:
    \begin{equation}\label{eq:sKdV}
    \arraycolsep=1.4pt\def\arraystretch{1.8}
     \left\{ 
     \begin{array}{l}
     \displaystyle \frac{d w_1}{dt}= 4 k^3 \partial^3 w_1 -6 k \partial(w_1)w_2 -3k w_1 \partial w_2,\\
     \displaystyle \frac{d w_2}{dt}= -k^3 \partial^3 w_2 -6k \partial(w_2) w_2 +12 k^2 w_2 \partial^{2}w_{1},
     \end{array}
     \right.
    \end{equation}
    where $w_1$ and $w_2$ are generators of $\mathcal{W}^k(\mathfrak{g},f)$ in Example \ref{ex:W(osp(1|2))}. Note that the system of equations \eqref{eq:sKdV} is equivalent to the derivation 
    \[ \frac{d\phi}{dt}= \big\{ \, w_2^2 +4k \partial(w_1)w_1\ {}_\lambda \ \phi \, \big\}_H \Big|_{\lambda=0}. \]
    
\end{example}

\newpage

\section{Adler-type operators associated with $\mathfrak{gl}(m|n)$} \label{Sec:ATO}

\subsection{Super Adler-type operators}
In this subsection, we introduce a super-analog of the so-called Adler-type operators
which were first defined in \cite{DSKV15}, motivated by \cite{Adler79}. Recall that we only consider associative, supercommutative and unital differential superalgebras with even derivations. 
A \textit{pseudo-differential operator} on a differential superalgebra $\mathcal{V}$
is an element in $\mathcal{V}(\!(\partial^{-1})\!)=\mathcal{V}\otimes\mathbb{C}(\!(\partial^{-1})\!)$. The space $\mathcal{V}(\!(\partial^{-1})\!)$ is a
superalgebra for the product defined by $\partial  v=v'+v\partial$ and 
\begin{alignat*}{1}
 &  \partial^{-1} v=\sum_{m\in\mathbb{Z}_{+}}(-1)^{m}v^{(m)}\partial^{-m-1},
\end{alignat*}
for $v\in\mathcal{V}$ and $v^{(m)}=\partial^{m}(v)$. 
The subspace $\mathcal{V}[\partial]$ is a differential
subalgebra of $\mathcal{V}(\!(\partial^{-1})\!)$ whose elements are called  \textit{differential
operators} on $\mathcal{V}$.

\vskip 2mm

For a nonzero pseudo-differential
operator $A(\partial)=\sum_{k\in \mathbb{Z}}a_{k}\partial^{k}$ on $\mathcal{V}$, 
the {\it order} of $A(\partial)$ is defined by 
$
\textrm{ord}(A(\partial))=N\in \mathbb{Z}$, where $N$ is the maximal integer such that $a_N\neq 0$.
An operator $A(\partial)$ can be uniquely decomposed as $$A(\partial)=A(\partial)_{+}+A(\partial)_{-}.$$ 
Here $A(\partial)_{+}=\sum_{k\in \mathbb{Z}_+}{\displaystyle a_{k}\partial^{k}}$
(resp. $A(\partial)_{-}=A(\partial)-A(\partial)_+$)
is called the \textit{differential part} (resp. \textit{integral part})
of $A(\partial)$. In addition, the \textit{residue} of the pseudo-differential operator $A(\partial)$ is defined by 
\[ \mathrm{Res}(A(\partial))= A_{-1} \in \mathcal{V}.\]
The \textit{symbol} of a pseudo-differential operator $A(\partial)$ is the formal Laurent series
$A(z)=\sum_{k\in \mathbb{Z}}a_{k}z^{k},$
for an even indeterminate $z$. The symbol of a product satisfies 
\[\label{symbol of product}
(A B)(z)=A(z+\partial)(B(z)).
\] 
In the RHS, we expand $(z+\partial)^{-1}$ as $\sum_{n\in \mathbb{Z}_+}(-1)^{n}z^{-n-1}\partial^{n}$ in such a way that only nonnegative powers of $\partial$ appear and act naturally on $B(z)$. Finally, the 
 \textit{adjoint} $A^{*}(\partial)$ of $A(\partial)$ is  the pseudo-differential operator
\begin{equation}\label{eq:adjoint}
  A^{*}(\partial)=\sum_{k\in \mathbb{Z}}(-\partial)^{k} a_{k}.
 \end{equation}

 Now we introduce a special family of pseudo-differential operators called \textit{Adler-type}, which is one of the main ingredients of this paper.
In the scalar case, our definition of Adler-type operator is identical to the nonsuper setting in \cite{DSKV16b} (this class of operators
first appeared in \cite{DSKV15}).

\begin{defn}
Let $\mathcal{V}$ be a differential superalgebra endowed with a $\lambda$-bracket
$\{\, {}_{\lambda}\, \}$. An even pseudo-differential operator $A(\partial)$
on $\mathcal{V}$ is called \textit{super Adler-type} (sATO) if the following identity:
\begin{equation}\label{eq:super Adler}
\{A(z){}_{\lambda}A(w)\}=A(w+\lambda+\partial)\iota_{z}(z-w-\lambda-\partial)^{-1}A^{*}(\lambda-z)-A(z)\iota_{z}(z-w-\lambda-\partial)^{-1}A(w),
\end{equation}
holds in $\mathcal{V}[\lambda](\!(z^{-1},w^{-1})\!)$ where $\iota_{z}(z-w-\lambda-\partial)^{-1}$  is obtained by taking the geometric expansion of $(z-w-\lambda-\partial)^{-1}$ for large $|z|$. 
In the RHS of \eqref{eq:super Adler}, we mean explicitly 
\[\Big(A(w+\lambda+\partial)\iota_{z}(z-w-\lambda-\partial)^{-1} \Big)(A^{*}(\lambda-z))-A(z) \Big(\iota_{z}(z-w-\lambda-\partial)^{-1} \Big)(A(w)).
\]
For the sake of simplicity, we will keep this slight abuse of notations in the rest of our paper. 
\end{defn}

We call such an operator sATO to emphasize that some of its coefficients have odd parity as opposed to an Adler-type operator(ATO), whose coefficients are all even. 
Let us first recall two elementary examples of ATOs.

\begin{example}
Let $\mathcal{V}=\mathbb{C}[q^{(n)}\big\vert n\in\mathbb{Z}_{+}]$ be the (even) differential algebra generated by $q$  and consider the $\lambda$-bracket  given by $\{q{}_{\lambda}q\}=\lambda$. Note that it defines a PVA structure on $\mathcal{V}$ by Theorem \ref{thm:PVA extension}. One can
check explicitly that $A(\partial)=\partial +q$ is an ATO since
\begin{align*}
    &A(w+\lambda+\partial)\iota_{z}(z-w-\lambda-\partial)^{-1}A^{*}(\lambda-z)\\
    & =\,(w+\lambda+\partial-z +z+q)\iota_{z}(z-w-\lambda-\partial)^{-1}(z-\lambda-\partial-w+w +q)(1)\\
   & =\,-(z-\lambda+ q)+(z+q) +(z+q)\iota_{z}(z-w-\lambda-\partial)^{-1}(w +q)\\
   & =\,  \{A(z){}_{\lambda}A(w)\}+A(z)\iota_{z}(z-w-\lambda-\partial)^{-1}A(w).
\end{align*}
\end{example}

\begin{example}\label{ex:Virasoro PVA}
Consider the differential operator $A(\partial)=\partial^2+u \partial+v.$ One can check that it is an ATO for the $\lambda$-bracket defined on the generators by
$$ \{ u \, {}_{\lambda} \, u \}= 2 \lambda,  \, \, \{ u \, {}_{\lambda} \, v \}=\lambda^2+u \lambda, \, \, \{ v \, {}_{\lambda} \, v \}=-\lambda^3+(u^2-2v+2u')\lambda+u''+uu'-v'.$$
In this PVA, there is a one-parameter family of Virasoro, or conformal, elements
$$\omega_{\alpha}=-v+\frac{1}{2}u^2+\alpha u', \, \, \alpha \in \mathbb{C} .$$
Indeed they satisfy the relations
\[\{ \omega_{\alpha} \, {}_{\lambda} \, \omega_{\alpha} \}=(\partial+2\lambda)\omega_{\alpha}+(-2\alpha^2+2\alpha-1)\lambda^3.\]
\end{example}
In order to connect sATOs with the Gelfand-Dickey brackets, we need the following lemma.
\begin{lem} \label{abc}
Let $A$ and $B$ be two differential operators on a differential superalgebra $\mathcal{V}$ of order at most $N$. Then for any $\lambda, z, w \in \mathbb{C}$ the three following expressions are equal : 
\begin{enumerate}[]
    \item \quad {\textup (i)} \  $ A(z)\iota_z(z-w-\lambda-\partial)^{-1}(B(w))-A(\partial+\lambda+w) \iota_z(z-w-\lambda-\partial)^{-1} (B^*(\lambda-z))$, \\
    \item \quad {\textup (ii)} \ $\big(A(z)(z-w-\lambda-\partial)^{-1}B(\partial+w)-A(\partial+\lambda+w) (z-w-\lambda-\partial)^{-1} B^*(\lambda-z) \big)(1)$, \\
    \item  \quad {\textup (iii)} \ $\displaystyle \sum_{i,j=0}^{N-1} z^{i}w^j\mathrm{ Res } \big( (A(\partial+\lambda) (\partial+\lambda)^{-i-1})_+ B(\partial) \partial^{-j-1}-A(\partial+\lambda) ((\partial+\lambda)^{-i-1}B(\partial))_+ \partial^{-j-1} \big)$.
\end{enumerate}
\end{lem}
\begin{proof}
We show first that $(i)=(iii)$.
If $(i,j)$ is not in $\{0,\cdots,N-1\}^{2}\subset \mathbb{Z}^2$, then we can easily see that 
$$\mathrm{ Res } \big( (A(\partial+\lambda) (\partial+\lambda)^{-i-1})_+ B(\partial) \partial^{-j-1}-A(\partial+\lambda) ((\partial+\lambda)^{-i-1}B(\partial))_+ \partial^{-j-1} \big)=0. $$
Hence we can rewrite $(iii)$ as a double infinite sum
\begin{equation} 
\begin{split}
    (iii) &= \sum_{i,j \in \mathbb{Z}} z^{i}w^j\mathrm{ Res } \big( (A(\partial+\lambda) (\partial+\lambda)^{-i-1})_+ B(\partial) \partial^{-j-1}-A(\partial+\lambda) ((\partial+\lambda)^{-i-1}B(\partial))_+ \partial^{-j-1} \big) \\
     &= \sum_{i \in \mathbb{Z}} \big( (A(\partial+\lambda) z^{i} (\partial+\lambda)^{-i-1})_+ B(\partial)-A(\partial+\lambda) ( z^{i} (\partial+\lambda)^{-i-1}B(\partial))_+  \big) (w) \\
     &= \sum_{i \in \mathbb{Z}} \big( (A(z) z^{i} (\partial+\lambda)^{-i-1})_+ B(\partial)-A(\partial+\lambda) ( z^{i} (\partial+\lambda)^{-i-1}B^*(\lambda-z))_+  \big) (w).
    \end{split}
\end{equation}
The third line follows from the fact that the formal delta distribution $\delta(z,w):=\sum_{i \in \mathbb{Z}} z^{i}w^{-i-1}$ satisfies
$$(z-\partial-\lambda) \delta(z,\partial+\lambda)= \delta(z,\partial+\lambda) (z-\partial-\lambda)=0.$$
Hence we have 
\begin{equation*}
    \begin{split}
        (iii) &= \sum_{i \in \mathbb{Z}}\big(A(z)(  z^{i} (\partial+\lambda)^{-i-1})_+ B(\partial)- A(\partial+\lambda)( z^{i} (\partial+\lambda)^{-i-1})_+B^*(\lambda-z)  \big) (w)\\
        &= \sum_{i \in \mathbb{Z}_+}\big(A(z)(   z^{-i-1} (\partial+\lambda)^{i}) B(\partial)- A(\partial+\lambda)( z^{-i-1} (\partial+\lambda)^{i})B^*(\lambda-z)  \big) (w)\\
        &=\sum_{i \in \mathbb{Z}_+} A(z)(   z^{-i-1} (\partial+w+\lambda)^{i}) (B(w))- A(\partial+w+\lambda)( z^{-i-1} (\partial+w+\lambda)^{i})(B^*(\lambda-z))  =(i).
    \end{split}
\end{equation*}
We now prove that $(ii)=(iii)$. 
\begin{equation*}
    \begin{split}
        (iii)&= \sum_{i \in \mathbb{Z}_+, j \in \mathbb{Z}} \mathrm{ Res } \big( (A(\partial+\lambda) z^{i}(\partial+\lambda)^{-i-1})_+ B(\partial) \partial^{-j-1}-A(\partial+\lambda) (z^{i}(\partial+\lambda)^{-i-1}B(\partial))_+ \partial^{-j-1} \big)w^j \\
        &= \sum_{i \in \mathbb{Z}_+}z^{i} \big( (A(\partial+\lambda) (\partial+\lambda)^{-i-1})_+ B(\partial)-A(\partial+\lambda) (z^{i}(\partial+\lambda)^{-i-1}B(\partial))_+  \big)(w) \\
        &= \bigg.\big( (A(\partial+\lambda) (\partial+\lambda-z)^{-1})_+ B(\partial) -A(\partial+\lambda) ((\partial+\lambda-z)^{-1}B(\partial))_+  \big)(w) \Big.\\
        &=  \big( A(z) (z-\partial-\lambda)^{-1} B(\partial) -A(\partial+\lambda)(z-\partial-\lambda)^{-1}B^*(\lambda-z)  \big)(w) \bigg. \\
        &= \Big. \big( A(z) (z-\partial-w-\lambda)^{-1} B(\partial+w) -A(\partial+\lambda+w)(z-\partial-w-\lambda)^{-1}B^*(\lambda-z)  \big)(1) = (ii) .\Big.
    \end{split}
\end{equation*}
To deduce the fourth line above, we have used the identities
\begin{equation*}
    \begin{split}
        & \Big.(A(\partial+\lambda) (\partial+\lambda-z)^{-1})_+=A(\partial+\lambda) (\partial+\lambda-z)^{-1}-A(z)(\partial+\lambda-z)^{-1}\;\; \text{and} \\
        & \Big.((\partial+\lambda-z)^{-1}B(\partial))_+ =(\partial+\lambda-z)^{-1}B(\partial)-(\partial+\lambda-z)^{-1}B^*(\lambda-z).
    \end{split}
\end{equation*}
\end{proof}

\subsection{Matrix super Adler-type operators} 
Let  $I$ be a finite set equipped with a parity map 
$p : I \mapsto \{0,1\}$, $i\mapsto \tilde{\imath}$. We say that an index $i \in I$ is even (resp. odd) if $\tilde{\imath} = 0$ (resp. $\tilde{\imath}=1$). We define the superalgebra $\mathfrak{gl}(I)$ as the $\mathbb{C}$-vector superspace generated by the elements $e_{ij}$ for $i,j \in I$  with the parity $p(e_{ij}) \equiv \tilde{\imath}+\tilde{\jmath}$ (mod 2) and relations $e_{ij} e_{kl} = \delta_{jk} e_{il}$. This superalgebra is, in particular, a Lie superalgebra for the bracket 
\begin{equation}
[e_{ij}, e_{kl}] = \delta_{jk}e_{il}- (-1)^{(\tilde{\imath}+\tilde{\jmath})(\tilde{k}+\tilde{l})} \delta_{il} e_{kj}.
\end{equation}
This Lie superalgebra is isomorphic to $\mathfrak{gl}(m|n)$ where $m=|I|-n$ is the number of even elements in $I$, but it is useful for us to keep this flexibility in the notations, as it will become clear later on.
Given any differential superalgebra $\mathcal{V}$, we can extend the superalgebra $\mathfrak{gl}(I)$ to $ \mathfrak{gl}(I) \otimes \mathcal{V}$ by letting for $a, b \in \mathfrak{gl}(I)$ and $v, w \in \mathcal{V}$,
\begin{equation} \label{eq:ordinary product}
(a \otimes v) \circ (b \otimes w) = (-1)^{\tilde{b}\tilde{v}} ab\otimes vw.
\end{equation} 
In addition, we define another product $\star$ on the space $ \mathfrak{gl}(I) \otimes \mathcal{V}$ by
\begin{equation} \label{eq:stars product}
(a \otimes v) \star (b \otimes w)  = (-1)^{\tilde{b}\tilde{v}+ \tilde{a}\tilde{b}} ab\otimes vw
\end{equation}
for $a, b \in \mathfrak{gl}(I)$ and $v, w \in \mathcal{V}$.
Note that if $a \otimes v$ is even, there is no sign in \eqref{eq:stars product} and both products are associative. We stress here that one of the main differences between the constructions in this paper and the analogue results in the nonsuper setting is the need to use these two products. 

Let $A(\partial)=\sum_{k\in \mathbb{Z}}a_{k}\partial^{k}\in\left(\mathfrak{gl}(I)\otimes\mathcal{V}\right)(\!(\partial^{-1})\!)$ be 
a matrix pseudo-differential operator over $\mathcal{V}$. By definition,
$A(\partial)$ is a square matrix whose $ij$-th entry $A_{ij}(\partial)$ is in $\mathcal{V}(\!(\partial^{-1})\!)$. A matrix pseudo-differential operator over $\mathcal{V}$
is called \textit{monic} of order $N$ if $\text{ord}(A_{ij}(\partial))\leq N$ for any $i,j\in I$ and $a_{N}=\ensuremath{\mathbbm{1}}_{I}=\sum_{i \in I} e_{ii}$.
We now define the super-analog of matrix-valued Adler-type operators. The nonsuper case of matrix Adler-type operator was introduced in \cite{DSKV15}.
\begin{defn} \label{def:ATO}
Let $\mathcal{V}$ be a differential superalgebra endowed with a $\lambda$-bracket
$\{ \, {}_{\lambda} \, \}$. An even pseudo-differential operator $A(\partial) \in \left(\mathfrak{gl}(I)\otimes\mathcal{V}\right)(\!(\partial^{-1})\!)$ is said to be a matrix \textit{super Adler-type operator (sATO)} on $\mathcal{V}$ if it satisfies the set of identities
\begin{equation} \label{mato}
\begin{aligned}
 \{A_{ij}(z){}_{\lambda}A_{hk}(w)\}=&(-1)^{\tilde{\imath}\tilde{\jmath}+\tilde{\imath}\tilde{h}+\tilde{\jmath}\tilde{h}}A_{hj}(w+\lambda+\partial)\iota_{z}(z-w-\lambda-\partial)^{-1}(A_{ik})^{*}(\lambda-z) \\
&-(-1)^{\tilde{\imath}\tilde{\jmath}+\tilde{\imath}\tilde{h}+\tilde{\jmath}\tilde{h}}A_{hj}(z)\iota_{z}(z-w-\lambda-\partial)^{-1}A_{ik}(w)
\end{aligned}
\end{equation}
for all $i,j,h,k\in I$. In particular, Definition \ref{def:ATO} recovers the Definition 2.1 \cite{DSKV16b} of matrix Adler-type operators if $I$ is a purely even index set.

\end{defn}

\begin{ex}\label{ex:Affine Adler}
Let us consider the affine Poisson vertex superalgebra $\mathcal{V}:=\mathcal{V}^{-1}(\mathfrak{gl}(I))$ of level $-1$ and the matrix pseudo-differential operator $A(\partial)\in (\mathfrak{gl}(I)\otimes \mathcal{V})(\!(\partial^{-1})\!)$ whose $ij$-th entry is $A_{ij}(\partial)= \delta_{ij}\partial +(-1)^{\tilde{\imath}}q_{ij}$.
One can check that 
\begin{equation}\label{affine lambda bracket}
  \begin{aligned}
    & \{(-1)^{\tilde{\imath}}q_{ij}\,{}_\lambda\,(-1)^{\tilde{h}}q_{hk}\}_{\mathrm{Aff}}  =(-1)^{\tilde{\imath}+\tilde{h}}(\delta_{hj}q_{ik}-(-1)^{(\tilde{\imath}+\tilde{\jmath})(\tilde{h}+\tilde{k})}\delta_{ik}q_{hj}-(-1)^{\tilde{\imath}}\delta_{hj}\delta_{ik}\lambda ) \\
    & \hskip 3cm = (-1)^{\tilde{\imath}\tilde{\jmath}+\tilde{\imath}\tilde{h}+\tilde{\jmath}\tilde{h}}\left( (-1)^{\tilde{\imath}}\delta_{jh}q_{ik}-(-1)^{\tilde{h}}\delta_{ik}q_{hj}
      -\delta_{hj}\delta_{ik}\lambda \right)\\
      & \hskip 3cm = (-1)^{\tilde{\imath}\tilde{\jmath}+\tilde{\imath}\tilde{h}+\tilde{\jmath}\tilde{h}} 
      \left((\delta_{hj}z+(-1)^{\tilde{h}}q_{hj})\iota_z(z-w-\lambda-\partial)^{-1}(\delta_{ik}w+(-1)^{\tilde{\imath}}q_{ik})\right. \\
     & \hskip 3cm - \left. (\delta_{hj}(w+\lambda+\partial)+(-1)^{\tilde{h}}q_{hj})\iota_z(z-w-\lambda-\partial)^{-1}(\delta_{ik}(z-\lambda)+(-1)^{\tilde{\imath}}q_{ik}) \right).\\
     \end{aligned}
\end{equation}
The first equality is the definition of  $\lambda$-bracket of affine PVsA while the second and third equality can be checked by direct computations.
Hence $A(\partial)$ is a matrix sATO on $\mathcal{V}^{-1}(\mathfrak{gl}(I)).$
\end{ex}

In Definition \ref{def:ATO}, we are given a differential superalgebra $\mathcal{V}$ together with a $\lambda$-bracket. In the following theorem, we show that the equality \eqref{mato} implies the skew-symmetry and Jacobi-identity for this $\lambda$-bracket restricted to the subalgebra of $\mathcal{V}$ generated by the coefficients of the matrix sATO.

\begin{thm} \label{Thm_gl:Skew-Jacobi,scalar}
 Let $A(\partial)=(A_{ij}(\partial))_{i,j\in I}$ be a matrix sATO on the differential superalgebra $\mathcal{V}$ endowed with the $\lambda$-bracket $\{\, {}_\lambda  \,\}$. For $a,b,c,d,e,f\in I$, we have
  \begin{enumerate}
    \item \textup{(skew-symmetry)} \quad $
    \{A_{ab}(z){}_\lambda A_{cd}(w)\}= -(-1)^{(\tilde{a}+\tilde{b})(\tilde{c}+\tilde{d})}\{A_{cd}(w){}_{-\lambda-\partial} A_{ab}(z)\},
    $
    \item \textup{(Jacobi-identity)}
    \begin{equation*}\label{eqn:Jacobi}
      \begin{aligned}
    & \{A_{ab}(z_1){}_\lambda \{A_{cd}(z_2){}_\mu A_{ef}(z_3)\}\}  \\
    & = \{\{A_{ab}(z_1){}_\lambda A_{cd}(z_2)\}_{\lambda+\mu}A_{ef}(z_3)\}\}+(-1)^{(\tilde{a}+\tilde{b})(\tilde{c}+\tilde{d})}\{A_{cd}(z_2){}_\mu \{ A_{ab}(z_1)_\lambda A_{ef}(z_3)\}\}.
      \end{aligned}
    \end{equation*}
  \end{enumerate}
  Hence if the entries of $A(\partial)$ are differentially algebraically independent, the differential superalgebra 
 $\mathcal{U}$ generated by these entries is a PVsA for the $\lambda$-bracket $\{\, {}_\lambda \, \}.$  We call such a PVsA an \textit{Adler-type PVsA}.
\end{thm}
\begin{proof}
  
  For any $u,v \in \mathcal{V}$ and $n \in \mathbb{Z}_{+}$, we use the notation
  \begin{equation} \label{eq:-lambda-partial}
      \big( u (\lambda+\partial)^n v \big)_{-\lambda - \partial}:= \big( (-\lambda-\partial)^n u \big ) v.
  \end{equation}
  From equation \eqref{mato}, we have 
    \begin{equation*}
      \begin{split}
         &(-1)^{\tilde{b}\tilde{d}+\tilde{a}\tilde{b}+\tilde{a}\tilde{d}}\{A_{cd}(w){}_{-\lambda-\partial} A_{ab}(z)\} \\
           &= \big (A_{ad}(w) \sum_{n \in \mathbb{Z}_+}\frac{(z+\lambda+\partial)^n}{w^{n+1}}A_{cb}(z)\big)_{-\lambda-\partial}-\big (A_{ad}(z+\partial+\lambda) \sum_{n \in \mathbb{Z}_+}\frac{(z+\lambda+\partial)^n}{w^{n+1}}A^*_{cb}(-w+\lambda) \big)_{-\lambda-\partial} \\
          & = \sum_{n \in \mathbb{Z}_+}\frac{(z-\lambda-\partial)^n}{w^{n+1}}( A_{ad}(w) )A_{cb}(z)- \sum_{n \in \mathbb{Z}_+}\frac{(z-\lambda-\partial)^n}{w^{n+1}}(A_{ad}^*(\lambda-z))A_{cb}(\lambda+\partial+w). 
          \end{split}
    \end{equation*}
    Hence
    \begin{equation*}
      \begin{split}
      &(-1)^{\tilde{a}\tilde{c}+\tilde{c}\tilde{d}+\tilde{a}\tilde{d}}\{A_{cd}(w){}_{-\lambda-\partial} A_{ab}(z)\} \\
          & =A_{cb}(z)\sum_{n \in \mathbb{Z}_+}\frac{(z-\lambda-\partial)^n}{w^{n+1}}( A_{ad}(w) )- A_{cb}(\lambda+\partial+w)\sum_{n \in \mathbb{Z}_+}\frac{(z-\lambda-\partial)^n}{w^{n+1}}(A_{ad}^*(\lambda-z))\\
           & =-A_{cb}(z)i_w(z-w-\lambda-\partial)^{-1}( A_{ad}(w) )+ A_{cb}(\lambda+\partial+w)i_w(z-w-\lambda-\partial)^{-1}(A_{ad}^*(\lambda-z)). 
      \end{split}
      \end{equation*}
    This concludes the proof of the skew-symmetry property, keeping in mind that in the defining equation of Adler-type operators the expansion $\iota_w(z-w-\lambda-\partial)^{-1}$ can be replaced by $\iota_z(z-w-\lambda-\partial)^{-1}$ due to the properties of the delta distribution and differential operators.
    Similarly, the proof of the Jacobi identity can be derived from Lemma 3.5 and Lemma 3.2 in \cite{DSKV15}.  Finally, by Theorem \ref{thm:PVA extension}, we conclude that the differential subalgebra $\mathcal{U}$ generated by the coefficients of $A$ is a PVsA, provided that its coefficients are differentially algebraically independent.
\end{proof}

\subsection{Matrix sATOs and Gelfand-Dickey brackets} \label{subsec:Generic}

Given an index set $I$ with parity map and a positive integer $N$, we consider the differential superalgebra 
\begin{equation} \label{eq:generic PVA}
 \mathcal{V}^N_{I}:= \mathbb{C}[u_{M,ab}^{(l)}|\,  M=0,1,\cdots,N-1, \, a,b\in I,\ \text{and} \ l \in \mathbb{Z}_+],
\end{equation}
where $\tilde{u}_{M,ab}^{(l)} = \tilde{a}+\tilde{b}$ for all $a, b \in I$ and $\partial(u_{M,ab}^{(l)})=u_{M,ab}^{(l+1)}$. We define the matrix differential operator
\begin{equation} \label{eq:generic}
  L(\partial)= \sum_{a\in I} e_{aa} \otimes \partial^N+ \sum_{M=0}^{N-1} \, \sum_{a,b\in I}e_{ab}\otimes u_{M,ab}\partial^M \, \in \, \mathfrak{gl}(I) \otimes \mathcal{V}_I^{N}(\!( \partial^{-1})\!)
\end{equation}
 whose coefficients are the generators of $\mathcal{V}^N_{I}$.
 
\begin{lem}\label{affbracket}
    There is a unique $\lambda$-bracket on the superalgebra $\mathcal{V}^N_{I}$ which lets $\mathcal{V}^N_{I}$ be the 
   Adler-type PVsA associated with $L(\partial)$. We call this $\lambda$-bracket  the \textit{generic bracket} associated to $I$ and $N$. 
    The superalgebra $\mathcal{V}^N_{I}$
    is a PVsA for the $\lambda$-bracket \eqref{mato} by Theorem \ref{Thm_gl:Skew-Jacobi,scalar}.
\end{lem}
\begin{proof}
    The coefficients of the operator are algebraically independent, hence we only need to check that the equation \eqref{mato} is well-defined, which is the case as the exponents of both $z$ and $w$ are bounded by $N-1$ in the RHS by Lemma \ref{abc}.
\end{proof}

The PVsA structure constructed above is none other than the lifting of the well-known quadratic Gelfand-Dickey Poisson bracket  \eqref{eq:adler bracket} on the space of functions on matrix pseudo-differential operators of degree $N$.
Let us first recall the definition of the variational derivative $\frac{\delta f}{\delta L}$ for an element $f \in \mathcal{V}^N_{I}$
\begin{equation}\frac{\delta f}{\delta L}=\sum_{a,b \in I}\sum_{k=0}^{N-1}(-1)^{\tilde{a}} e_{ba} \otimes \partial^{-k-1} \frac{\delta f}{\delta u_{k,ab}}. 
\end{equation}
This definition is justified by the property

\begin{equation} \label{universal}
    \begin{split}
       \frac{d}{d\epsilon} &\int f(L+\epsilon A) = \int A \star \frac{\delta f}{\delta L} = \sum_{a,b \in I}\sum_{k=0}^{N-1} \int A_{k,ab}\frac{\delta f}{\delta u_{k,ab}}
    \end{split}
\end{equation}
for any even matrix differential operator $A \in \mathfrak{gl}(I)\otimes \mathcal{V}^N_{I}$ and  $f \in \mathcal{V}^N_{I}$. Note that $\mathcal{V}^N_{I}$ is a superalgebra which is why the explicit form of $\frac{\delta f}{\delta L}$ is different from \eqref{eq:adler bracket}. However, in both cases the universal property \eqref{universal} is satisfied. The \textit{supertrace} is the linear map from $\mathfrak{gl}(I) \otimes \mathcal{V}_I^N$ to $\mathcal{V}_I^N$ defined by 
$$\text{str } (e_{ij} \otimes f) :=(-1)^{\tilde{\imath}} \delta_{ij} f $$
for all $i,j \in I$ and $f \in \mathcal{V}_I^N$.

\begin{prop} \label{prop:GD}
    The Gelfand-Dickey Lie bracket on $\mathcal{V}_I^{N}/ \partial \mathcal{V}_I^{N}$ coincides with the Lie bracket induced by the generic $\lambda$-bracket on $\mathcal{V}_I^{N}$. More precisely, for all $f, g \in \mathcal{V}_I^{N}$, we have

    \begin{equation}
       \int  \{f \, _{\lambda} \, g \} \Big\vert_{\lambda=0} =  \, \, \mathrm{Res} \,\,  \mathrm{str} \int (L \star \frac{\delta f}{\delta L})_+ \star L \star\frac{\delta g}{\delta L}-L \star(\frac{\delta f}{\delta L} \star L)_+  \star \frac{\delta g}{\delta L}.
    \end{equation}

\end{prop}
\begin{proof}
    First, it follows immediately from the super master formula \eqref{eq:master formula} that 
    \begin{equation}
      \int  \{f \, _{\lambda} \, g \} \Big\vert_{\lambda=0} =\sum_{a,b,c,d \in I}  \sum_{k,l=0}^{N-1}  (-1)^{(\tilde{f}+\tilde{a}+\tilde{b})(\tilde{c}+\tilde{d})}\int \Big( \{u_{k,ab} \, _{\partial} \, u_{l,cd} \}_{\to}\frac{\delta f}{\delta u_{k,ab}}\Big)\frac{\delta g}{\delta u_{l,cd}}.
    \end{equation}
    On the other hand, we have
    \begin{equation*}
        \begin{split}
           &\mathrm{Res} \,\,  \mathrm{str} \int (L \star \frac{\delta f}{\delta L})_+ \star L \star\frac{\delta g}{\delta L}-L \star(\frac{\delta f}{\delta L} \star L)_+  \star \frac{\delta g}{\delta L} \\
           &=  \mathrm{Res} \, \, \mathrm{str}  \int \sum_{k,l=0}^{n-1} \sum_{a,b,c,d, \in I}\Big(e_{cb} \otimes L_{cb} \star (-1)^{\tilde{a}} e_{ba} \otimes \partial^{-k-1} \frac{\delta f}{\delta u_{k,ab}} \Big)_+ \star e_{ad} \otimes L_{ad} \star (-1)^{\tilde{c}} e_{dc} \otimes \partial^{-l-1} \frac{\delta g}{\delta u_{l,cd}} \\
           &-  \mathrm{Res} \, \,\mathrm{str}\, \,   \int \sum_{k,l=0}^{n-1} \sum_{a,b,c,d, \in I}e_{cb} \otimes L_{cb} \star \Big((-1)^{\tilde{a}} e_{ba} \otimes \partial^{-k-1} \frac{\delta f}{\delta u_{k,ab}}  \star e_{ad} \otimes L_{ad} 
           \Big)_+\star (-1)^{\tilde{c}} e_{dc} \otimes \partial^{-l-1} \frac{\delta g}{\delta u_{l,cd}} \\
           &=  \mathrm{Res} \, \, \mathrm{str}  \int \sum_{k,l=0}^{n-1} \sum_{a,b,c,d, \in I}\Big((-1)^{\tilde{a}}e_{ca}  \otimes L_{cb}  \partial^{-k-1} \frac{\delta f}{\delta u_{k,ab}} \Big)_+ \star (-1)^{\tilde{c}} e_{ac} \otimes L_{ad}  \partial^{-l-1} \frac{\delta g}{\delta u_{l,cd}} \\
           &-  \mathrm{Res} \, \, \mathrm{str}  \int \sum_{k,l=0}^{n-1} \sum_{a,b,c,d, \in I}e_{cb} \otimes L_{cb} \star \Big((-1)^{\tilde{a}+\tilde{f}(\tilde{a}+\tilde{d})} e_{bd} \otimes \partial^{-k-1} \frac{\delta f}{\delta u_{k,ab}}   L_{ad} 
           \Big)_+\star (-1)^{\tilde{c}} e_{dc} \otimes \partial^{-l-1} \frac{\delta g}{\delta u_{l,cd}} \\
           &-   \int \sum_{k,l=0}^{n-1} \sum_{a,b,c,d, \in I} (-1)^{\tilde{a}+\tilde{c}+\tilde{f}(\tilde{a}+\tilde{c})} \mathrm{Res} \,  L_{cb}  \Big( \partial^{-k-1} \frac{\delta f}{\delta u_{k,ab}}   L_{ad} 
           \Big)_+ \partial^{-l-1} \frac{\delta g}{\delta u_{l,cd}}.
        \end{split}
    \end{equation*}
    Hence we need to check the following differential operator identities for all $a,b,c,d \in I$ and $k,l \in \{0,\cdots, N-1\}$ 
    \begin{equation*} 
(-1)^{\tilde{a}\tilde{b}+\tilde{a}\tilde{c}+\tilde{b}\tilde{c}} \{ u_{k,ab} {}_{\partial} \, u_{l,cd} \}_{\to}(F)= (-1)^{\tilde{F}(\tilde{a}+\tilde{d})}\mathrm{Res} \Big( \, \Big( L_{cb}  \partial^{-k-1} F \Big)_+L_{ad}  \partial^{-l-1} - L_{cb}  \Big(\partial^{-k-1} F L_{ad}\Big)_+  \partial^{-l-1} \Big),
    \end{equation*}
for all $F \in \mathcal{V}^N_{I}$. Taking the symbol of both sides yields
\begin{equation*} 
\begin{aligned}
& (-1)^{\tilde{a}\tilde{b}+\tilde{a}\tilde{c}+\tilde{b}\tilde{c}}  \{ u_{k,ab} {}_{\lambda} \, u_{l,cd} \} \\
& = \mathrm{Res} \Big( \, \Big( L_{cb}(\partial+\lambda)  (\partial+\lambda)^{-k-1} \Big)_+L_{ad}(\partial)  \partial^{-l-1} - L_{cb}(\partial+\lambda)  \Big((\partial+\lambda)^{-k-1}  L_{ad}(\partial)\Big)_+  \partial^{-l-1} \Big).
\end{aligned}
    \end{equation*}
    We are done since this equality holds by Lemma \ref{abc} and definition of the $\lambda$-bracket \eqref{mato}.
\end{proof}

\subsection{Second compatible bracket}  \label{subsec:second bracket}

For all $\epsilon \in \mathbb{C}$, there exists a unique $\lambda$-bracket $\{\, {}_{\lambda} \,\}_{\epsilon}$ on the differential superalgebra $\mathcal{V}^N_{I}$ such that $L^{\epsilon}(\partial):=L(\partial)+ \sum_{a\in I}e_{aa}\otimes \epsilon$ is sATO. Note that for any element $v$ in any PVsA, we have $\left \{ v {}_{\lambda} 1 \right \} = 0$ by the Leibniz rule. Hence for all $a,b,c,d \in I$, 
\begin{equation} \label{fff}
  \{L^{\epsilon}_{ab}(z){}_\lambda L^{\epsilon}_{cd}(w)\}_{\epsilon}= \{L_{ab}(z){}_\lambda L_{cd}(w)\}_{\epsilon}.
\end{equation}
It is clear that the contribution of $\epsilon^2$ is trivial in the RHS of equation \eqref{mato}. Therefore, we can decompose the $\lambda$-bracket $\{\, {}_{\lambda} \,\}_{\epsilon}$ as
\begin{equation}
  \{ \cdot\,{}_{\lambda} \,\cdot\}_{\epsilon}=\{\cdot\,{}_{\lambda} \,\cdot\}_{H}+\epsilon\,\{\cdot\,{}_{\lambda} \,\cdot\}_{K},
\end{equation}
where $\{\, _\lambda \,\}_{H}$ and $\{\, _\lambda \,\}_{K}$ define two compatible PVA structures on 
$\mathcal{V}^N_I$. The bracket $\{\, _\lambda \,\}_{H}$ is the generic bracket on $\mathcal{V}^N_I$ constructed in Section \ref{subsec:Generic}. As for the bracket $\{\, _\lambda \,\}_{K}$, its values on the generators of the differential superalgebra $\mathcal{V}^N_I$ are given by
\begin{equation}\label{3.13}
    \begin{split} (-1)^{\tilde{a}\tilde{b}+\tilde{a}\tilde{c}+\tilde{b}\tilde{c}}\{L_{ab}(z) \,_\lambda\, L_{cd}(w) \}_{K}&=
    \delta_{ad}(L_{cb}(z)\!-\!{L_{cb}}(w+\lambda+ \partial))\iota_{z}(z\!-\!w\!-\!\lambda-\partial)^{-1}\\
    &+\delta_{cb}\iota_{z}(z\!-\!w\!-\!\lambda\!-\!\partial)^{-1}(L_{ad}(w)\!-\!(L_{ad})^{*}(-z+\lambda)).
    \end{split}
\end{equation}
Note that, in the case of the affine PVsA corresponding to $N=1$, this $K$-bracket is trivial. It is a particular case of the following Lemma.
\begin{lem} \label{K=0}
  For all $a,b,c,d \in I$, we have
  \begin{enumerate}
   \item[$(a)$] {\makebox[3.5cm]{$\left\{ u_{N-1,ab} \;\,_\lambda\;\, L_{cd}(w) \right\}_{H}$} = $ (-1)^{\tilde{a}\tilde{b}+\tilde{a}\tilde{c}+\tilde{b}\tilde{c}}\left[\,\delta_{cb}L_{ad}(w)-\delta_{ad}L_{cb}(w+\lambda)\,\right]$},
    \item[$(b)$] {\makebox[3.5cm]{$\left\{ L_{ab}(z) \;\,_\lambda\;\, u_{N-1,cd} \right\}_{H}$} = $ (-1)^{\tilde{b}\tilde{c}+\tilde{b}\tilde{d}+\tilde{c}\tilde{d}}\left[\,\delta_{cb}(L_{ad})^{*}(-z+\lambda)-\delta_{ad}L_{cb}(z)\,\right]$},
    \item[$(c)$] {\makebox[3.5cm]{$\left\{ u_{N-1,ab} \;\,_\lambda\;\, u_{N-1,cd} \right\}_{H}$} = $(-1)^{\tilde{a}\tilde{b}+\tilde{a}\tilde{c}+\tilde{b}\tilde{c}} \left[\delta_{cb}u_{N-1,ad}-\delta_{ad}u_{N-1,cb}-\delta_{ad}\delta_{cb}N\lambda\,\right] $},
    \item[$(d)$] {\makebox[3.5cm]{$\left\{ u_{N-1,ab} \;\,_\lambda\;\, L_{cd}(w) \right\}_{K}$} = $\left\{ L_{ab}(z) \;\,_\lambda\;\, u_{N-1,cd} \right\}_{K}=0$}.
\end{enumerate}

\end{lem}

\begin{proof}
  We first check \text{(a)}.
  \begin{align*}
       (-1)^{\tilde{a}\tilde{b}+\tilde{a}\tilde{c}+\tilde{b}\tilde{c}} \left\{ u_{N-1,ab} \;\,_\lambda\;\, L_{cd}(w) \right\}_{H}
      =& (-1)^{\tilde{a}\tilde{b}+\tilde{a}\tilde{c}+\tilde{b}\tilde{c}} \, \,  \mathrm{Res}_{z}\left[\left\{ L_{ab}(z) \;\,_\lambda\;\, L_{cd}(w) \right\}_{H}z^{-N}\right]\\
      =&\text{Res}_{z}\!\left[L_{cb}(z)\iota_{z}(z\!-\!w\!-\!\lambda\!-\!\partial)^{-1}\!{L_{ad}}(w)z^{-N}\right] \\
      -&\text{Res}_{z}\!\left[L_{cb}(w\!+\!\lambda\!+\!\partial)\iota_{z}(z\!-\!w\!-\!\lambda\!-\!\partial)^{-1}\!(L_{ad})^{*}(-z\!+\!\lambda)z^{-N}\right]\\
      =&\delta_{cb}L_{ad}(w)-\delta_{ad} L_{cb}(w+\lambda+\partial)(1) \\
      =&\delta_{cb}L_{ad}(w)-\delta_{ad}L_{cb}(w+\lambda).
  \end{align*}
  Next, \text{(b)} follows from (a) by skew-symmetry:
  \begin{align*}
      \left\{ L_{ab}(z) \;\,_\lambda\;\, u_{N-1,cd} \right\}_{H}
      &=-(-1)^{(\tilde{a}+\tilde{b})(\tilde{c}+\tilde{d})}\!\left\{ u_{N-1,cd} \;\,_{ \,-\!\lambda-\partial}\;\, L_{ab}(z) \right\}_{H}\\
      &=-(-1)^{(\tilde{a}+\tilde{b})(\tilde{c}+\tilde{d})}(-1)^{\tilde{a}\tilde{c}+\tilde{a}\tilde{d}+\tilde{c}\tilde{d}}\left[\,\delta_{ad}L_{cb}(z)-\delta_{cb}L_{ad}(z+\lambda)\,\right]_{ -\lambda-\partial}\\
      &= (-1)^{\tilde{b}\tilde{c}+\tilde{b}\tilde{d}+\tilde{c}\tilde{d}}\left[\,\delta_{cb}(L_{ad})^{*}(-z\!+\!\lambda)-\delta_{ad}L_{cb}(z)\,\right].
  \end{align*}
  The third statement \text{(c)} follows from taking the coefficient of $w^{N-1}$ in both sides of \text{(a)}.
  Finally, we prove \text{(d)} as follows
  \begin{align*}
       (-1)^{\tilde{a}\tilde{b}+\tilde{a}\tilde{c}+\tilde{b}\tilde{c}} \left\{ u_{N-1,ab} \;\,_\lambda\;\, L_{cd}(w) \right\}_{K} 
      =& (-1)^{\tilde{a}\tilde{b}+\tilde{a}\tilde{c}+\tilde{b}\tilde{c}} \, \,\text{Res}_{z}\left[\left\{ L_{ab}(z) \;\,_\lambda\;\, L_{cd}(w) \right\}_{K}z^{-N}\right]\\
      =&\text{Res}_{z}\delta_{ad}(L_{cb}(z)-L_{cb}(w+\lambda+\partial))\iota_{z}(z\!-\!w\!-\!\lambda-\partial)^{-1}z^{-N}\\
      +&\text{Res}_{z}\delta_{cb}\iota_{z}(z\!-\!w\!-\!\lambda\!-\!\partial)^{-1}(L_{ad}(w)-(L_{ad})^{*}(-z+\lambda))z^{-N}\\
      =&\delta_{ad}\delta_{cb}-\delta_{cb}\delta_{ad}=0.
  \end{align*}
\end{proof}

Using the super master formula \eqref{eq:master formula} and equation \eqref{3.13}, one can show that the linear Gelfand-Dickey Lie bracket on $\mathcal{V}^N_I/ \partial \mathcal{V}^N_I$ is identical to the Lie bracket induced by $\{\cdot\, _\lambda \,\cdot\}_{K}$ following the lines of Section \ref{subsec:Generic}. Summarizing both results, we obtain the analogue formula to \eqref{eq:adler bracket}.
 Namely we have for all $f,g \in \mathcal{V}^N_I$
    \begin{equation}
    \begin{split}
        \int  \{f \, _{\lambda} \, g \}_{H} \Big\vert_{\lambda=0} = & \, \, \mathrm{Res} \,\,  \mathrm{str} \int (L \star \frac{\delta f}{\delta L})_+ \star L \star\frac{\delta g}{\delta L}-L \star(\frac{\delta f}{\delta L} \star L)_+  \star \frac{\delta g}{\delta L}, \\
       \int  \{f \, _{\lambda} \, g \}_{K} \Big\vert_{\lambda=0} =&  \, \, \mathrm{Res} \,\,  \mathrm{str} \int \Big(L \star \frac{\delta f}{\delta L}-L \star \frac{\delta f}{\delta L}\Big)_+  \star\frac{\delta g}{\delta L}.
       \end{split}
    \end{equation}

\newpage

\section{$\mathcal{W}$-superalgebras associated with Adler-type operators} \label{sec:W-algebra}

\subsection{
Quasi-determinants of matrix sATOs
}
In this subsection, we construct more examples of sATOs by taking quasi-determinants of matrix sATOs. 
For a detailed introduction to quasi-determinants, we refer to \cite{GGRW05}, \cite{DSKV16a} and \cite{DSKV18}. We recall that all matrix differential operators are assumed to be even.

Let  $I=I_{\bar{0}} \sqcup I_{\bar{1}}$ be a finite set with the parity map $p:I\rightarrow\{0,1\}$, where $I_{\bar{0}}:=p^{-1}(0)$ is the set of even indices and $I_{\bar{1}}:=p^{-1}(0)$ is the set of odd indices. We assume in the sequel that these finite index sets are ordered. For a differential superalgebra
$\mathcal{V}$, we recall the $\star$-product \eqref{eq:stars product} on the space of matrix pseudo-differential operators $\mathfrak{gl}(I)\otimes\mathcal{V}(\!(\partial^{-1})\!)$. If  pseudo-differential operators $A(\partial)=\sum_{i,j\in I} e_{ij}\otimes a_{ij}$ and $B(\partial)=\sum_{i,j\in I} e_{ij}\otimes b_{ij}$ are even in $(\mathfrak{gl}(I)\otimes\mathcal{V})_{\bar{0}}$, then
\begin{equation}
    A(\partial)\star B(\partial):=\displaystyle\sum_{i,j\in I}\displaystyle\sum_{s\in I} e_{ij}\otimes a_{is}b_{sj}.
\end{equation}
More generally, for finite index sets $I,J,K$ with parity, 
we define the $\star$-product between even matrix pseudo-differential operators of different sizes as follows 
\begin{align}\label{star product2}
(\mathrm{Mat}_{I\times J}\otimes\mathcal{V}(\!(\partial^{-1})\!))_{\bar{0}}\times(\mathrm{Mat}_{J \times K}\otimes\mathcal{V}(\!(\partial^{-1})\!))_{\bar{0}} & \rightarrow(\mathrm{Mat}_{I \times K}\otimes\mathcal{V}(\!(\partial^{-1})\!))_{\bar{0}}\\
\nonumber \left(A(\partial), B(\partial)\right) &\mapsto (A\star B)(\partial)
\end{align}
in the same way as in \eqref{eq:stars product},
where $\mathrm{Mat}_{I\times J}$ is the superspace spanned by the elements $e_{ij}$ for $i\in I$ and $j\in J$ with the parity $\tilde{\imath}+\tilde{\jmath}$ (mod 2).

We say that a matrix pseudo-differential operator $A(\partial)\in \mathfrak{gl}(I)\otimes \mathcal{V}(\!( \partial^{-1})\!),$ is {\it $\star$-invertible}  if there exists $A^{\mathrm{inv}}(\partial)\in \mathfrak{gl}(I)\otimes \mathcal{V}(\!( \partial^{-1})\!)$ such that $A(\partial)\star A^{\mathrm{inv}}(\partial)= A^{\mathrm{inv}}(\partial)\star A(\partial)= \mathbbm{1}_{I}.$ Suppose that $J$ and $K$ are  subsets of $I$ such that
\begin{equation} \label{eq:as-parity-index}
|J \cap I_{\bar{0}}|= |K\cap I_{\bar{0}}| \quad \text{ and } \quad |J \cap I_{\bar{1}}|= |K \cap I_{\bar{1}}| \ . 
\end{equation} 
Denote by $A(\partial)_{JK}\in \mathrm{Mat}_{J\times K} \otimes \mathcal{V}(\!(\partial^{-1})\!)$ the submatrix consisting of $jk$-entries of $A(\partial)$ for $j\in J$ and $k\in K.$ 
If there exists a matrix valued operator  $\big(A(\partial)_{JK}\big)^{\mathrm{inv}}\in \mathrm{Mat}_{K\times J}\otimes\mathcal{V}(\!(\partial^{-1})\!)$ satisfying  
\[ A(\partial)_{JK}\star \big(A(\partial)_{JK}\big)^{\mathrm{inv}} = \mathbbm{1}_J  \quad \text{ and } \quad  \big(A(\partial)_{JK}\big)^{\mathrm{inv}} \star A(\partial)_{JK} = \mathbbm{1}_K, \]
then we call $\big(A(\partial)_{JK}\big)^{\mathrm{inv}}$  the {\it $\star$-inverse} of $A(\partial)_{JK}$ and say that $A(\partial)_{JK}$
is {\it $\star$-invertible}.

\begin{defn} \label{def:qausi-det}
Let $A(\partial)\in \mathfrak{gl}(I)\otimes \mathcal{V}(\!( \partial^{-1})\!)$ be a $\star$-invertible matrix pseudo-differential and $J,K\subset I$ be subsets satisfying \eqref{eq:as-parity-index}. Suppose $A^{\mathrm{inv}}(\partial)_{KJ}$ is also $\star$-invertible. Then the $(J,K)$ {\it quasi-determinant} of $A(\partial)$ is 
\[ |A(\partial)|_{JK}:= \big( (A^{\mathrm{inv}}(\partial))_{KJ}\big)^{\mathrm{inv}}.\]
\end{defn}

For subsets $I$, $J$, $K$ in Definition \ref{def:qausi-det}, let $J^c=I \setminus J$ and $K^c=I \setminus K$. If the $(J,K)$ quasi-determinant of the matrix pseudo-differential operator $A(\partial)$ is well-defined and 
the submatrix $(A_{J^{c}K^{c}})(\partial)$ is $\star$-invertible, then we have
\begin{equation}\label{qdet,another form}
   |A(\partial)|_{JK}=A_{JK}(\partial)-A_{JK^{c}}(\partial)\star(A_{J^{c}K^{c}})^{\mathrm{inv}}(\partial)\star A_{J^{c}K}(\partial).
\end{equation}
See \cite{DSKV16a} for the proof of \eqref{qdet,another form}.
We also have the following lemma which directly follows by \eqref{mato}.

 \begin{lem} \label{lem:submatix-sato}
 Let $A(\partial)\in \mathfrak{gl}(I)\otimes \mathcal{V}(\!( \partial^{-1})\!)$ be a sATO. For two subsets $J$ and $K$ of $I$ satisfying \eqref{eq:as-parity-index}, we can identify $\mathrm{Mat}_{J\times K}$ with $\mathfrak{gl}(J)$. Moreover, under this identification   $\big(A(\partial)\big)_{JK}\in\mathrm{Mat}_{J\times K}\otimes \mathcal{V}(\!( \partial^{-1})\!) \cong \mathfrak{gl}(J)\otimes \mathcal{V}(\!( \partial^{-1})\!)$
 is also a sATO.
 \end{lem}

By Definition \ref{def:qausi-det} and Lemma \ref{lem:submatix-sato}, in order to show that the $(J,K)$ quasi-determinant of a sATO is still a sATO for $J$ and $K$ satisfying \eqref{eq:as-parity-index}, it is enough to show that the $\star$-inverse of a sATO is again a sATO. 

\begin{lem} \label{lem:star inverse}
Let $A(\partial)\in \mathfrak{gl}(I) \otimes \mathcal{V}(\!(\partial^{-1})\!)$ be a monic matrix pseudo-differential operator of order $N$.
There exists a unique $\star$-inverse of order $-N$.
\end{lem}

\begin{proof}
Let $A(\partial)=  \mathbbm{1}_I \partial^N+ \sum_{M<N} U_M \partial^M$ and write  $X:=\sum_{M<N} U_M \partial^{M-N}$. Then we have
$A(\partial)^{\mathrm{inv}}=\partial^{-N}\big(  
\mathbbm{1}_I-X+X^{\star 2}-X^{\star 3}\cdots
\big)$, where $X^{\star n}$ is the $n$-th power of $X$ with respect to the $\star$-product.
\end{proof}
\begin{lem}\label{Lemma:Inverse operator}
     Suppose that a differential superalgebra $\mathcal{V}$ is endowed with a $\lambda$-bracket $\left\{  \,{}_{\lambda}\, \right\}$ and  let $C(\partial)=(C_{ij}(\partial))_{i,j\in I} \in \mathfrak{gl}{(I)}\otimes\mathcal{V}(\!(\partial^{-1})\!)$ be a $\star$-invertible matrix pseudo-differential operator. 
    For $a \in  \mathcal{V}$, we have
   \begin{equation} \label{eq:super_lemma_first equality} 
  \begin{aligned}
  &\{a \,{}_{\lambda}\, (C^{\mathrm{inv}})_{ij}(z)\}\\
  &=-\displaystyle\sum_{r,t\in I}\displaystyle\sum_{n \in \mathbb{Z}}\displaystyle(-1)^{\tilde{a}(\tilde{\imath}+\tilde{r})}(C^{\mathrm{inv}})_{ir}(\lambda+z+\partial)\left\{ a \,{}_{\lambda}\, C_{rt;n} \right\}(z+\partial)^n(C^{\mathrm{inv}})_{tj}(z)
  \end{aligned}
  \end{equation}
  and
  \begin{equation}\label{eq:super_lemma_second equality} 
  \begin{aligned}
    &\{(C^{\mathrm{inv}})_{ij}(\lambda) \,{}_{\lambda+z}\, a\}\\
    &=-\displaystyle\sum_{r,t\in I}\displaystyle\sum_{n \in \mathbb{Z}}\displaystyle(-1)^{\tilde{a}(\tilde{\jmath}+\tilde{t})+(\tilde{\imath}+\tilde{r})(\tilde{a}+\tilde{r}+\tilde{\jmath})}\left\{ {C_{rt;n}} \,{}_{\lambda+z+\partial}\, a \right\}(\lambda+\partial)^n(C^{\mathrm{inv}})_{tj}(\lambda)
   (C^{\mathrm{inv}})^{*}_{ir}(z),
    \end{aligned}
    \end{equation}
    where $C_{ij}(\partial)=\sum_{n\in \mathbb{Z}}\,C_{ij;n}\partial^{n}$ and negative powers of $(x+\partial)$ are expanded using the geometric series with nonnegative powers of $\partial$.
\end{lem}
\begin{proof}
    This can be shown by direct calculations similar to the proof of Lemma 2.8 in \cite{LSS23}.
\end{proof}

\begin{prop}\label{prop:Properties of sATO}
    Let $\mathcal{V}$ be a differential superalgebra with a $\lambda$-bracket $\{\ {}_{\lambda}\ \}$ and  $A(\partial)\in  \mathfrak{gl}(I) \otimes\mathcal{V}(\!( \partial^{-1})\!)$ be a sATO on $\mathcal{V}$. If $A(\partial)$ is $\star$-invertible, then the $\star$-inverse $A(\partial)^{\mathrm{inv}}$ is also a sATO on $\mathcal{V}$ with respect to the opposite bracket $-\{\ {}_\lambda \ \}$. Hence a quasi-determinant of a sATO is a sATO.
\end{prop}
\begin{proof}
    The proposition follows from the proof of Proposition 3.7 \cite{DSKV16b} but with additional sign considerations. 

Using both statements in Lemma \ref{Lemma:Inverse operator}, we see that the $\lambda$-bracket between inverse operators is as follows
\begin{equation} \label{eq:prop4.5-1}
\begin{aligned}
    \left\{ (A^{\mathrm{inv}})_{ij}(z) \,_{\lambda}\, (A^{\mathrm{inv}})_{hk}(w) \right\}=&\sum_{p,q,s,t\in I}(-1)^{(\tilde{\imath}+\tilde{\jmath})(\tilde{h}+\tilde{s})}(-1)^{(\tilde{s}+\tilde{t})(\tilde{q}+\tilde{\jmath})+(\tilde{\imath}+\tilde{p})(\tilde{s}+\tilde{t}+\tilde{p}+\tilde{\jmath})}\\&(A^{\mathrm{inv}})_{hs}(w+\lambda+\partial)\left\{ A_{pq}(z+x_{1}) \,_{\lambda+x_{1}+x_{3}}\, A_{st}(w+x_{2}) \right\}\\
    &\big(\big\vert _{x_{1}=\partial} (A^{\mathrm{inv}})_{qj}(z)\big)\big(\big\vert _{x_{3}=\partial} \left(
(A^{\mathrm{inv}})_{ip}\right)^{*}(\lambda-z)\big) \big(\big\vert _{x_{2}=\partial} (A^{\mathrm{inv}})_{tk}(w)\big).
\end{aligned}
\end{equation}
In \eqref{eq:prop4.5-1}, the notation $x_1^k x_2^l \big(\big\vert _{x_{1}=\partial} a \big) \big(\big\vert _{x_{2}=\partial} b \big)$ represents $\partial^k(a) \partial^l(b)$. By the definition of sATO, we have 
\begin{align*}
\eqref{eq:prop4.5-1}=& \sum_{p,q,s,t\in I }(-1)^{(\tilde{\imath}+\tilde{\jmath})(\tilde{h}+\tilde{s})+(\tilde{s}+\tilde{t})(\tilde{q}+\tilde{\jmath})+(\tilde{\imath}+\tilde{p})(\tilde{s}+\tilde{t}+\tilde{p}+\tilde{\jmath})+\tilde{\imath}\tilde{p}+\tilde{p}+(\tilde{t}+\tilde{k})(\tilde{p}+\tilde{\imath})+\tilde{p}\tilde{q}+\tilde{p}\tilde{s}+\tilde{q}\tilde{s}}  (A^{\mathrm{inv}})_{hs}(w+\lambda+\partial) \\
&\quad\times \Bigl(A_{sq}(w\!+x_{1}\!+x_{2}\!+x_{3}\!+\lambda\!+\partial)\iota_{z}(z\!-w\!-x_{2}-x_{3}-\lambda\!-\partial)^{-1} (A_{pt})^{*}(\lambda+x_{3}-z)\Bigr.\\ 
&\quad \hskip 5cm  \Bigl.
    -A_{sq}(z+x_{1})\iota_{z}(z\!-w\!-\lambda\!-x_{2}-x_{3}-\partial)^{-1}A_{pt}(w+x_{2})\Bigr)\\
    &\quad \times \Big(\Big\vert _{x_{1}=\partial} (A^{\mathrm{inv}})_{qj}(z)\Big)\Big(\Big\vert _{x_{2}=\partial} (A^{\mathrm{inv}})_{tk}(w)\Big)\Big(\Big\vert _{x_{3}=\partial} \left(
(A^{*\mathrm{inv}})_{pi}\right)(\lambda-z)\Big)\\
= & \sum_{p,q,s,t\in I}(-1)^{\tilde{\imath}\tilde{h}+\tilde{\jmath}\tilde{h}+\tilde{t}\tilde{q}+\tilde{t}\tilde{\jmath}+\tilde{\imath}\tilde{\jmath}+\tilde{p}\tilde{\jmath}+\tilde{k}\tilde{p}+\tilde{k}\tilde{\imath}+\tilde{p}\tilde{q}+\tilde{p}\tilde{t}+\tilde{t}+(\tilde{\imath}+\tilde{p})(\tilde{t}+\tilde{k}+\tilde{q}+\tilde{\jmath})} \\
& \quad  \times (A^{\mathrm{inv}})_{hs}(w+\lambda+\partial)A_{sq}(w\!+x_{1}\!+x_{2}\!+x_{3}\!+\lambda\!+\partial) \iota_{z}(z\!-w\!-x_{2}-x_{3}-\lambda\!-\partial)^{-1}\\
& \times (A^{*}_{tp})(\lambda+x_{3}-z) \Big(\Big\vert _{x_{3}=\partial} 
(A^{*\mathrm{inv}})_{pi}(\lambda-z)\Big)\Big(\Big\vert _{x_{1}=\partial} (A^{\mathrm{inv}})_{qj}(z)\Big)\Big(\Big\vert _{x_{2}=\partial} (A^{\mathrm{inv}})_{tk}(w)\Big)\\
- &\sum_{p,q,s,t\in I}(-1)^{\tilde{\imath}\tilde{h}+\tilde{\jmath}\tilde{h}+\tilde{t}\tilde{q}+\tilde{t}\tilde{\jmath}+\tilde{\imath}\tilde{\jmath}+\tilde{p}\tilde{\jmath}+\tilde{k}\tilde{p}+\tilde{k}\tilde{\imath}+\tilde{p}\tilde{q}+(\tilde{p}+\tilde{t})(\tilde{q}+\tilde{\jmath})}\\
&\times (A^{\mathrm{inv}})_{hs}(w+\lambda+\partial)A_{sq}(z+x_{1})\Big(\Big\vert _{x_{1}=\partial} (A^{\mathrm{inv}})_{qj}(z)\Big)\iota_{z}(z\!-w\!-\lambda\!-x_{2}-x_{3}-\partial)^{-1}\\
&\times A_{pt}(w+x_{2})\Big(\Big\vert _{x_{2}=\partial} (A^{\mathrm{inv}})_{tk}(w)\Big)\Big(\Big\vert _{x_{3}=\partial} \left(
(A^{*\mathrm{inv}})_{pi}\right)^{*}(\lambda-z)\Big)\\
= &\sum_{q,t\in I}(-1)^{\tilde{\imath}\tilde{h}+\tilde{\jmath}\tilde{h}+\tilde{t}\tilde{q}+\tilde{t}\tilde{\jmath}+\tilde{t}+\tilde{\imath}\tilde{t}+\tilde{\imath}\tilde{q}}\delta_{hq}(A^{\mathrm{inv}})_{qj}(z) \iota_{z}(z\!-w\!-\lambda\!-\partial)^{-1}\delta_{ti}(A^{\mathrm{inv}})_{tk}(w)\\
- & \sum_{p,s\in I}(-1)^{\tilde{\imath}\tilde{h}+\tilde{\jmath}\tilde{h}+\tilde{\imath}\tilde{\jmath}+\tilde{k}\tilde{p}+\tilde{k}\tilde{\imath}}(A^{\mathrm{inv}})_{hs}(w+\lambda+\partial)\delta_{sj} \iota_{z}(z\!-w\!-\lambda\!-\partial)^{-1}\delta_{pk}(A^{*\mathrm{inv}})_{pi}(\lambda-z)\\
= &(-1)^{\tilde{\imath}\tilde{\jmath}+\tilde{\imath}\tilde{h}+\tilde{\jmath}\tilde{h}}\big( (A^{\mathrm{inv}})_{hj}(z) \iota_{z}(z\!-w\!-\lambda\!-\partial)^{-1}(A^{\mathrm{inv}})_{ik}(w)\big.\\
\big.&\hspace{2cm}-(A^{\mathrm{inv}})_{hj}(w+\lambda+\partial) \iota_{z}(z\!-w\!-\lambda\!-\partial)^{-1}\left(
(A^{\mathrm{inv}})_{ik}\right)^{*}(\lambda-z)\big).
\end{align*}
This completes the proof. 
\end{proof}

\subsection{Rectangular $\mathcal{W}$-superalgebras and sATOs} \label{sec:4.3}
\indent In this subsection we fix $m,n,N\in \mathbb{Z}_+$ and $N \geq 2$.
We explain how to construct the rectangular $\mathcal{W}$-superalgebra $\mathcal{W}(\mathfrak{gl}(Nm|Nn),f)$ within the theory of sATOs where $f$ is the $N \times (m|n)$ rectangular nilpotent in $\mathfrak{gl}(Nm|Nn)$ as defined in Example \ref{ex: rectangular W-super}.

Consider
 the index set $\Pi=\{1, 2, \cdots, N(m+n)\}$ with the parity map $p:\Pi\rightarrow \{0,1\}$ defined by
\begin{equation}\label{eq:index_parity}
\tilde{\imath}:=p(i)=
\begin{cases} 
\,0 \qquad&\text{ if }\ 0<i\leq m \,(\mathrm{mod}\;m\!+\!n),\\
\,1 \qquad&\text{ if }\ m<i\leq m+n\,(\mathrm{mod}\;m\!+\!n) .\\
\end{cases}
\end{equation}
Then $\mathfrak{gl}(\Pi)\cong \mathfrak{gl}(Nm|Nn)$ and 
an element in $\mathfrak{gl}(\Pi)$ is presented by a matrix \eqref{ex:gl(ml|nl)} in Example \ref{ex: rectangular W-super}. In other words,  $\mathfrak{gl}(\Pi)= \mathfrak{gl}_{N} \otimes \mathfrak{gl}(m|n)$. 
Let us consider the sATO $A(\partial) \in \mathfrak{gl}(\Pi)\otimes \mathcal{V}\big(\mathfrak{gl}(\Pi)\big)(\!( \partial^{-1})\!)$ of degree one as in Example \ref{ex:Affine Adler}, i.e., the $ij$-th entry for $i,j\in \Pi$ is   
\begin{equation}\label{sATO of Affine PVsA}
 A_{ij}(\partial)=\delta_{ij}\partial+(-1)^{\tilde{\imath}}q_{ij}.
\end{equation}
As in Example \ref{ex: rectangular W-super}, we can decompose $A(\partial)$ into $N^2$ submatrix-valued operators 
$\big(A_{[uv]}(\partial)\big)_{1\leq u,v\leq N}$, each of which is isomorphic to a $\mathcal{V}(\mathfrak{gl}(\Pi))$-valued $(m|n)\!\times\!(m|n)$ matrix pseudo-differential operator. This can be described by the following picture:
\begin{equation} \label{eq:A(partial)}
A(\partial)=\left[
\begin{array}{cccc}
A_{[11]}(\partial) & A_{[12]}(\partial) & \cdots & A_{[1N]}(\partial) \\
A_{[21]}(\partial) &A_{[22]}(\partial) & \cdots & A_{[2N]}(\partial) \\
\vdots & \vdots & \ddots
& \vdots  \\
A_{[N1]}(\partial) & A_{[N2]}(\partial) & \cdots & A_{[NN]}(\partial) \\
\end{array}
\right].
\end{equation}
Denote $q_{\frac{[uv]}{(ij)}}:=q_{(u-1)(m+n)+i,(v-1)(m+n)+j}$ so that 
\begin{equation} \label{eq:A_uv}
A_{[uv]}(\partial):=\left(A_{\frac{[uv]}{(ij)}}(\partial)\right)_{i,j\in \{1,\cdots,m+n\}}\;\text{for} \quad \;A_{\frac{[uv]}{(ij)}}(\partial):=\delta_{uv}\delta_{ij}\partial+(-1)^{\tilde{\imath}}q_{\frac{[uv]}{(ij)}}
\end{equation}

and 
\[
\mathcal{V}(\mathfrak{gl}(\Pi))\cong \mathbb{C}[q_{\frac{[uv]}{(ab)}}^{(k)}|1\leq u,v\leq N,\, 1\leq a,b\leq m+n \;\text{and}\; k\in \mathbb{Z}_+].
\]
Let us fix index subsets $I, J$ as follows  
\[
I=\{(N-1)(m+n)+1,(N-1)(m+n)+2, \cdots,N(m+n)\}\subset \Pi \quad\textrm{and}\;\; J=\{1,2,\cdots, m+n\}\subset \Pi.
\] 

\vskip 3mm 

 Let $\mathfrak{g}=\mathfrak{gl}(\Pi)$ and  recall the definitions of $f$ in \eqref{ex:gl(ml|nl)-f},  $\mathfrak{p}$ in \eqref{eq:Lie subalgebra} and 
the differential superalgebra homomorphism $\rho:\mathcal{V}(\mathfrak{g})\rightarrow \mathcal{V}(\mathfrak{p})$ defined by $\rho(a)=\pi_{\mathfrak{p}}(a)+(f|a)$ for $a\in\mathfrak{g}.$ We also denote by $\rho$ its extension to $\mathfrak{g} \otimes \mathcal{V}(\mathfrak{g})(\!(\partial^{-1})\!)$ such that 
\[ \rho(a \otimes q(\partial)):= a \otimes  \rho(q(\partial))\]
for $a\in \mathfrak{g}$ and $q(\partial)\in \mathcal{V}(\mathfrak{g})(\!(\partial^{-1})\!).$ We are now ready to define our candidate matrix differential operator 
\begin{equation}\label{eq:L}
L(\partial)=(-1)^{N-1}|\rho(A(\partial))|_{IJ}=(-1)^{N-1}|\mathbbm{1}_{(m|n)}\partial+f^{\bot}+\displaystyle\sum_{q_{ij}\in\mathfrak{g}_{\leq 0}}(-1)^{\tilde{\imath}}e_{ij}\otimes q_{ij}|_{IJ}
\end{equation}
whose coefficients will be generators of the rectangular $\mathcal{W}$-superalgebra $\mathcal{W}(\mathfrak{gl}(Nm|Nn),f)$, where 
\[{f^{\bot}}=\sum_{u=1}^{N-1}\left(e_{\frac{[u,u+1]}{(11)}}+e_{\frac{[u,u+1]}{(22)}}+\cdots+e_{\frac{[u,u+1]}{(m+n,m+n)}}\right)=
\left[
\begin{array}{ccccc}
0 &\mathbbm{1}_{(m|n)} &0  & \cdots & 0  \\
 0& 0 & \mathbbm{1}_{(m|n)}& \cdots & 0 \\
0  &  0 & 0& \cdots  & 0 \\
\vdots & \vdots & \vdots& \ddots & \vdots  \\
0 & 0 & 0 & \cdots&0 \\
\end{array}\right].\]
 Remark that the diagonal element 
\begin{equation}\label{diagonal h}
    h=\sum_{u=1}^{N}(N-2u+1)\left(e_{\frac{[uu]}{(11)}}+e_{\frac{[uu]}{(22)}}+\cdots+e_{\frac{[uu]}{(m+n,m+n)}}\right)
\end{equation}
gives a $\frac{\mathbb{Z}}{2}$-grading of $\mathfrak{gl}(\Pi)$ which is needed to define the $\mathcal{W}$-superalgebra $\mathcal{W}(\mathfrak{gl}(Nm|Nn),f)$. 

\vskip 3mm

Since both $\rho (A(\partial))$ and $\rho(A(\partial))_{I^{c}J^{c}}$ are $\star$-invertible, we can express the quasi-determinant formula \eqref{qdet,another form} using the $\star$-product and block matrix differential operators $A_{[uv]}$:
\begin{equation*}\label{block matrix multiplication}
    |\rho(A(\partial))|_{IJ}=\left[\begin{array}{c}
A_{[N1]} 
\end{array}\right]
-\left[\begin{array}{cccc}
A_{[N2]} & \cdots & A_{[NN]} \\
\end{array}\right]\star
\left[\begin{array}{cccc}
\mathbbm{1}_{(m|n)} & 0 & \cdots & 0 \\
A_{[22]} & \mathbbm{1}_{(m|n)} & \cdots & 0 \\
\vdots & \vdots & \ddots & \vdots  \\
A_{[N-1,2]} & A_{[N-1,3]} & \cdots & \mathbbm{1}_{(m|n)} \\
\end{array}\right]^{\mathrm{inv}}\star
\left[\begin{array}{cccc}
A_{[11]}  \\
A_{[21]}  \\
\vdots   \\
A_{[N-1,1]} \\
\end{array}\right].
\end{equation*}
The matrix differential operator $L(\partial)$ is monic
of degree $N$.
Each component of the symbol $L(z)$ can be written as a product of symbols of scalar differential operators $A_{\frac{[uv]}{(st)}}(z)$:  

\begin{align}\label{operator L(z)}
L_{ij}(z) & =(-1)^{N-1}\sum_{k=0}^{N-1}{\displaystyle \sum_{\tiny\begin{array}{c}
(i_{1},\cdots,i_{k})\,\text{s.t.}\\
l>i_{1}>\cdots>i_{k}>0
\end{array}}}\!{\displaystyle \sum_{\tiny\begin{aligned}\begin{array}{c}
s_{1,}\cdots s_{k}=1\end{array}\end{aligned}
}^{m+n}\!}(-1)^{k}A_{\frac{[N,i_{1}+1]}{(i,s_{1})}} A_{\frac{[i_{1},i_{2}+1]}{(s_{1}s_{2})}}\cdots A_{\frac{[i_{k},1]}{(s_{k},j)}}(z),
\end{align}
for $i,j\in J=\{1,2,\cdots, m+n\}.$

\begin{lem} \label{lem:W-algebra}
Let $u\in \{1,2,\cdots, N-1\}$ and $a,b\in J$. Then the following equalities hold: 
    \begin{enumerate}
\item[$(a)$]  For $\alpha>u+1$, 
\begin{center}
    $\rho \Big(\big\{q_{\frac{[u,u+1]}{(ab)}}\, {}_{\lambda}\, A_{\frac{[\alpha,u+1]}{(\beta b)}}A_{\frac{[u,u]}{(ba)}}(z)-  A_{\frac{[\alpha,u]}{(\beta a)}}(z)\big\}\Big)=0$.
\end{center}

\item [$(b)$]For $\beta \in J \setminus \{b\}$,  
\vskip 2mm
\begin{center}
    $\rho\Big( \big\{q_{\frac{[u,u+1]}{(ab)}}\, {}_{\lambda}\, A_{\frac{[u+1,u+1]}{(\beta b)}}A_{\frac{[u,u]}{(ba)}}(z)- A_{\frac{[u+1,u]}{(\beta a)}}(z)\big\}\Big)=0$.
\end{center}

\item[$(c)$]  For $\beta \in J \setminus \{a\}$, 
\vskip 2mm
\begin{center}
    $\rho\Big(\big\{q_{\frac{[u,u+1]}{(ab)}}\, {}_{\lambda} \,  A_{\frac{[u+1,u+1]}{(ba)}}A_{\frac{[u,u]}{(a\beta)}}(z)- A_{\frac{[u+1,u]}{(b\beta)}} 
    (z)\big\}\Big)=0$.\vspace{0.2cm}
\end{center}

\item[$(d)$] For $a,b\in J$ with $a\neq b$,
\vskip 2mm
\begin{center}
    $\rho\Big(\big\{q_{\frac{[u,u+1]}{(ab)}}\, {}_{\lambda} \, 
    A_{\frac{[u+1,u+1]}{(bb)}}A_{\frac{[u,u]}{(ba)}}(z)+A_{\frac{[u+1,u+1]}{(ba)}}A_{\frac{[u,u]}{(aa)}}(z)-A_{\frac{[u+1,u]}{(ba)}}(z)\big\}\Big)=0$.\vspace{0.2cm}
\end{center}

\item[$(e)$]  For $\alpha<u$, 
\vskip 2mm
\begin{center}
    $\rho\Big(\big\{q_{\frac{[u,u+1]}{(ab)}}\, {}_{\lambda}\,  A_{\frac{[u+1,u+1]}{(ba)}}A_{\frac{[u,\alpha]}{(a\beta)}}(z)- A_{\frac{[u+1,\alpha]}{(b\beta)}}(z)\big\}\Big)=0.$
\end{center}
\end{enumerate}
\end{lem}

\begin{proof}
Here, we give the full proof of (a), (d) and (e). (b) and (c) can be proved similarly.

The following computation shows (a): 
\begin{align*}
    &\rho \Big(\big\{q_{\frac{[u,u+1]}{(ab)}}\, {}_{\lambda}\, A_{\frac{[\alpha,u+1]}{(\beta b)}}A_{\frac{[u,u]}{(ba)}}(z)\big\}\Big)
    \\
= &\rho\Big(\big\{q_{\frac{[u,u+1]}{(ab)}}{}_{\lambda}  q_{\frac{[\alpha,u+1]}{(\beta b)}}(\delta_{ab}z+\!(-1)^{p(b)}\!q_{\frac{[u,u]}{(ba)}}\big)\big\}\Big) =\rho\Big(\{q_{\frac{[u,u+1]}{(ab)}}{}_{\lambda} q_{\frac{[\alpha,u+1]}{(\beta b)}}q_{\frac{[u,u]}{(ba)}}\} \Big) \\
= & -\rho\Big((-1)^{(a+b)(\beta+a)}q_{\frac{[\alpha,u+1]}{(\beta b)}}\Big) =\rho\Big(\big\{q_{\frac{[u,u+1]}{(ab)}}{}_{\lambda} q_{\frac{[\alpha u]}{(\beta a)}}\big\}\Big)= \rho \Big(\big\{q_{\frac{[u,u+1]}{(ab)}}\, {}_{\lambda}\,A_{\frac{[\alpha,u]}{(\beta a)}}(z)\big\}\Big),
\end{align*}
where the third equality holds by the fact $\rho\big(q_{\frac{[u,u+1]}{(a,b)}}\big)=(-1)^{\tilde{a}}\delta_{ab}$.
\vskip 2mm
(d) can be obtained in the following way:  
\begin{align*}
 & \rho\Big(\big\{q_{\frac{[u,u+1]}{(ab)}}{}_{\lambda}A_{\frac{[u+1,u+1]}{(bb)}}A_{\frac{[u,u]}{(ba)}}(z)+\!A_{\frac{[u+1,u+1]}{(ba)}}A_{\frac{[u,u]}{(aa)}}(z)\big\}\Big)\\
 =& \rho\Big(\big\{q_{\frac{[u,u+1]}{(ab)}}{}_{\lambda}(-1)^{b}(q_{\frac{[u+1,u+1]}{(ba)}}+q_{\frac{[u,u]}{(ba)}})z+(-1)^{b}q_{\frac{[u,u]}{(ba)}}'+q_{\frac{[u+1,u+1]}{(bb)}}q_{\frac{[u,u]}{(ba)}} +(-1)^{a+b}q_{\frac{[u+1,u+1]}{(ba)}}q_{\frac{[u,u]}{(aa)}}\big\}\Big)\\
 =& \rho\Big(\big(-(-1)^{a}q_{\frac{[u,u+1]}{(bb)}}+(-1)^{b}q_{\frac{[u,u+1]}{(aa)}}\big)z-(\lambda+\partial)(-1)^{a}q_{\frac{[u,u+1]}{(bb)}}\\
 & +\big(q_{\frac{[u,u+1]}{(ab)}}q_{\frac{[u,u]}{(ba)}}-(-1)^{a+b}q_{\frac{[u+1,u+1]}{(bb)}}q_{\frac{[u,u+1]}{(bb)}}\big)+(-1)^{a+b}\big(q_{\frac{[u,u+1]}{(aa)}}q_{\frac{[u,u]}{(aa)}}-(-1)^{a+b}q_{\frac{[u+1,u+1]}{(ba)}}q_{\frac{[u,u+1]}{(ab)}}\big)\Big)\\
 =& (-1)^{b}\Big( q_{\frac{[u,u]}{(aa)}}-(-1)^{a+b}q_{\frac{[u+1,u+1]}{(bb)}}-(-1)^{a}\lambda\Big) \\
 = & \rho\Big(\big\{q_{\frac{[u,u+1]}{(ab)}}{}_{\lambda}A_{\frac{[u+1,u]}{(ba)}}\big\}\Big).
\end{align*}
\vskip 2mm
(e) is proved as follows: 
\begin{align*}
&\rho\Big(\big\{q_{\frac{[u,u+1]}{(ab)}}{}_{\lambda} A_{\frac{[u+1,u+1]}{(ba)}} A_{\frac{[u,\alpha]}{(a\beta)}}(z)\big\}\Big)\\ & =\rho\Big(\big\{q_{\frac{[u,u+1]}{(ab)}}{}_{\lambda}\delta_{ab}q_{\frac{[u,\alpha]}{(a\beta)}}'+\delta_{ab}q_{\frac{[u,\alpha]}{(a\beta)}}\partial+\!(-1)^{a+b}\!q_{\frac{[u+1,u+1]}{(ba)}}q_{\frac{[u,\alpha]}{(a\beta)}})\big\}\Big)\\
 & =\rho\Big(\big\{q_{\frac{[u,u+1]}{(ab)}}{}_{\lambda}\!(-1)^{a+b}\!q_{\frac{[u+1,u+1]}{(ba)}}q_{\frac{[u,\alpha]}{(a\beta)}})\big\}\Big)\\
 & =\rho\Big((-1)^{a+b}q_{\frac{[u,u+1]}{(aa)}}q_{\frac{[u,\alpha]}{(a\beta)}}\Big) =\rho\Big(\big\{q_{\frac{[u,u+1]}{(ab)}}{}_{\lambda} A_{\frac{[u+1,\alpha]}{(b\beta)}}\big\}\Big).
\end{align*}
\end{proof}

\begin{prop}\label{prop:in W}
The coefficients of $L(\partial)$ in \eqref{eq:L} are elements in the $\mathcal{W}$-superalgebra $\mathcal{W}(\mathfrak{gl}(Nm|Nn),f)$ associated with the $N\times (m|n)$ rectangular nilpotent element $f$.
\end{prop}
\begin{proof}
Recall that $\mathfrak{gl}(\Pi)\cong\mathfrak{gl}(Nm|Nn)$ and hence we can consider $\mathcal{W}(\mathfrak{gl}(Nm|Nn),f)$ as a subalgebra of $\mathcal{V}(\mathfrak{gl}(\Pi))$. 

First we check that the coefficients of $L(z)$ lie in $\mathcal{V}(\mathfrak{p})$. 
It is clear since  $q_{\frac{[uv]}{(ij)}}\in \mathfrak{gl}(\Pi)_{v-u}$. Next we should check that $\rho\{\nu{}_{\lambda}L_{ij}(z)\}=0$
for all $\nu\in\mathfrak{n}=\mathfrak{gl}(\Pi)_{\geq1}$ and $i,j\in  J$. It is enough to show that 
 \[\label{claim of prop:in W}
\rho\Big(\big\{q_{\frac{[u,u+1]}{(ab)}}{}_{\lambda}L_{ij}(z)\big\}\Big)=0,
\]
for $\nu=q_{\frac{[u,u+1]}{(ab)}}\in \mathfrak{gl}(\Pi)_{1}$ and  $a,b\in J$ and $u\in \{1,2, \cdots, N-1\}$ by the Jacobi identity and since $[\mathfrak{g}_{i} ,\ \mathfrak{g}_{j}]\subseteq\mathfrak{g}_{i+j}$.  To show \eqref{claim of prop:in W}, first we fix some $a, b$ and $u$.

\vskip 1mm

We can split the terms in \eqref{operator L(z)} into five types which depend on $a,b$ and $u$:  
\begin{enumerate}
\item (Type I) The terms in \eqref{operator L(z)}  containing $A_{\frac{[\alpha,u+1]}{(\beta b)}}A_{\frac{[uu]}{(ba)}}$
for some $\alpha \geq u+1$ and $\beta \in J$,
\item (Type II) The terms in \eqref{operator L(z)}  containing $A_{\frac{[u+1,u+1]}{(ba)}}A_{\frac{[u\alpha]}{(a\beta)}}$
for some $\alpha \leq u$ and $\beta \in J$,
\item (Type III) The terms in \eqref{operator L(z)}  containing 
$A_{\frac{[\alpha u]}{(\beta a)}}$
for some $\alpha \geq u+1$ and $\beta \in J$,
\item (Type IV) The terms in \eqref{operator L(z)}  containing $A_{\frac{[u+1,\alpha]}{(b\beta)}}$
for some $\alpha \leq u$ and $\beta \in J$,
\item (Type V) The other terms in \eqref{operator L(z)}. 
\end{enumerate}
Note that for a term $Z$ of Type V, we have  $\rho\big(\{q_{\frac{[u,u+1]}{(ab)}}{}_{\lambda}Z \}\big)=0$. Hence we focus on the terms of Type I -- Type IV. Let us divide each of Type I -- Type IV into three sub-types. For example,
\begin{center}
     Type I = Type I$_{(\alpha>u+1)}\;\sqcup $ Type I$_{(\alpha=u+1,\beta \neq b)} \;\sqcup $  Type I$_{(\alpha=u+1,\beta=b)},$
\end{center}
where Type I$_{(\alpha>u+1)}$ consists of the terms in Type I such that $\alpha>u+1$. Type I$_{(\alpha=u+1,\beta \neq b)}$ and Type I$_{(\alpha=u+1,\beta=b)}$ can be defined similarly. 
Now we list the following facts:

\begin{enumerate}[(i)]
\item  $\text{Type I}_{(\alpha>u+1)} \sqcup   \text{Type III}_{(\alpha>u+1)}$ : 
In $L_{ij}(z)$, all terms with  $A_{\frac{[\alpha,u+1]}{(\beta b)}}A_{\frac{[uu]}{(ba)}}$ (i.e. $\text{Type I}_{(\alpha>u+1)}$ terms) can be written as 
$XA_{\frac{[\alpha,u+1]}{(\beta b)}}A_{\frac{[uu]}{(ba)}}Y(z)$ for some $X(\partial),Y(\partial) \in \mathcal{V}(\mathfrak{p})[\partial].$ Similarly, with the same $X$ and $Y$, the terms with  $A_{\frac{[u,u+1]}{(ab)}}$ (i.e. $\text{Type III}_{(\alpha>u+1)}$ terms) in $L_{ij}(z)$ has the form $-XA_{\frac{[u,u+1]}{(ab)}}Y(z)$. 
Note that $\{q_{\frac{[u,u+1]}{(ab)}} \, {}_{\lambda}\, X(z)\}=0$ and $ \{q_{\frac{[u,u+1]}{(ab)}}\, {}_{\lambda}\, Y(z)\}=0$ and hence, by Lemma \ref{lem:W-algebra} (a), we have

\begin{equation} \label{eq:W-key-1}
    \rho \Big(\big\{q_{\frac{[u,u+1]}{(ab)}}\, {}_{\lambda}\, X A_{\frac{[\alpha,u+1]}{(\beta b)}}A_{\frac{[u,u]}{(ba)}}Y(z)- X A_{\frac{[\alpha,u]}{(\beta a)}}Y(z)\big\}\Big)=0.
\end{equation}

\item $\text{Type I}_{(\alpha=u+1,\beta \neq b)}\sqcup\text{Type III}_{(\alpha=u+1,\beta \neq b)}$ : 
By the same reason as (i), the sum of $\text{Type I}_{(\alpha=u+1,\beta \neq b)}$   and  $\text{Type III}_{(\alpha=u+1,\beta \neq b)}$ terms in $L_{ij}(z)$ has the form of 
$\big(X A_{\frac{[u+1,u+1]}{(\beta b)}}A_{\frac{[u,u]}{(ba)}}Y(z)-X A_{\frac{[u+1,u]}{(\beta a)}}Y(z)\big)$ for some $X(\partial),Y(\partial)\in \mathcal{V}(\mathfrak{p})[\partial]$ such that $\{q_{\frac{[u,u+1]}{(ab)}} \, {}_{\lambda}\, X(z)\}= \{q_{\frac{[u,u+1]}{(ab)}}\, {}_{\lambda}\, Y(z)\}=0$. Moreover, by  Lemma \ref{lem:W-algebra} (b), we have

\begin{equation} \label{eq:W-key-2}
    \rho\Big( \big\{q_{\frac{[u,u+1]}{(ab)}}\, {}_{\lambda}\, X A_{\frac{[u+1,u+1]}{(\beta b)}}A_{\frac{[u,u]}{(ba)}}Y(z)-X A_{\frac{[u+1,u]}{(\beta a)}}Y(z)\big\}\Big)=0.
\end{equation}

\item $\text{Type II}_{(\alpha=u,\beta \neq a)}\sqcup \text{Type IV}_{(\alpha=u,\beta \neq a)}$ : 
The sum of all terms of
the given types in $L_{ij}(z)$ has the form $\big(X A_{\frac{[u+1,u+1]}{(ba)}}A_{\frac{[u,u]}{(a\beta)}}Y(z)- X A_{\frac{[u+1,u]}{(b\beta)}}
    Y(z)\big)$ and,  by  Lemma  \ref{lem:W-algebra} (c), we have 
\begin{equation}\label{eq:W-key-3}
    \rho\Big(\big\{q_{\frac{[u,u+1]}{(ab)}}\, {}_{\lambda} \, X A_{\frac{[u+1,u+1]}{(ba)}}A_{\frac{[u,u]}{(a\beta)}}Y(z)- X A_{\frac{[u+1,u]}{(b\beta)}} 
    Y(z)\big\}\Big)=0.
\end{equation}

\item $\text{Type I}_{(\alpha=u+1,\beta=b)}\sqcup \text{Type II}_{(\alpha=u,\beta=a)}\sqcup \text{Type III}_{(\alpha=u+1,\beta=b)}$ with $a\neq b$ : 
The sum of all terms of the given types in $L_{ij}(z)$ has the form $\big(X
    A_{\frac{[u+1,u+1]}{(bb)}}A_{\frac{[u,u]}{(ba)}}Y(z)+XA_{\frac{[u+1,u+1]}{(ba)}}A_{\frac{[u,u]}{(aa)}}Y(z)-XA_{\frac{[u+1,u]}{(ba)}}Y(z)\big)$ and, by  Lemma \ref{lem:W-algebra} (d), we have
\begin{equation} \label{eq:W-key-4}
    \rho\Big(\big\{q_{\frac{[u,u+1]}{(ab)}}\, {}_{\lambda} \, X
    A_{\frac{[u+1,u+1]}{(bb)}}A_{\frac{[u,u]}{(ba)}}Y(z)+XA_{\frac{[u+1,u+1]}{(ba)}}A_{\frac{[u,u]}{(aa)}}Y(z)-XA_{\frac{[u+1,u]}{(ba)}}Y(z)\big\}\Big)=0.
\end{equation}

\item  $\text{Type II}_{(\alpha<u)}\sqcup \text{Type IV}_{(\alpha<u)}$ :
The sum of all terms of the given types in $L_{ij}(z)$ has the form $\big( X A_{\frac{[u+1,u+1]}{(ba)}}A_{\frac{[u,\alpha]}{(a\beta)}}Y(z)-X A_{\frac{[u+1,\alpha]}{(b\beta)}}Y(z)\big)$ and, by Lemma \ref{lem:W-algebra} (e), we have
\begin{equation} \label{eq:W-key-5}
    \rho\Big(\big\{q_{\frac{[u,u+1]}{(ab)}}\, {}_{\lambda}\, X A_{\frac{[u+1,u+1]}{(ba)}}A_{\frac{[u,\alpha]}{(a\beta)}}Y(z)-X A_{\frac{[u+1,\alpha]}{(b\beta)}}Y(z)\big\}\Big)=0.
\end{equation}
\end{enumerate} 
By adding \eqref{eq:W-key-1}--\eqref{eq:W-key-5}, we get $\rho\Big(\big\{ \{q_{\frac{[u,u+1]}{(ab)}}\, {}_{\lambda}\, L_{ij}(z)\big\} \Big)=0$.
\end{proof}

Using Proposition \ref{prop:in W}, we obtain the following theorems which are the main results of this section. 

\begin{thm} \label{thm:W-algebra and sATO-ver2}
The coefficients of $L(\partial)$ in \eqref{eq:L} freely generate the $\mathcal{W}$-superalgebra $\mathcal{W}(\mathfrak{gl}(Nm|Nn),f)$ associated with the $N\times (m|n)$ rectangular nilpotent element $f$.
\end{thm}

\begin{proof}
Let us name the coefficients of \eqref{operator L(z)} by
\[\label{eq:generators w_{ij:k}}
L_{ij}(\partial)=\sum_{k=0}^{N}w_{ij;k}\partial^{k}.
\] By Proposition \ref{prop:in W},  each coefficient $w_{ij;k}$ of $L_{ij}(z)$ lies in $\mathcal{W}(\mathfrak{gl}(Nm|Nn),f)$. Hence it is enough to show that $\left\{w_{ij;k}\right\}_{\tiny\substack{1\leq i,j\leq m+n,\\0\leq k\leq N-1}}$  freely generates $\mathcal{W}(\mathfrak{gl}(Nm|Nn),f)$. To this end, we use
Proposition \ref{prop:free generating set of W-algebra}. In other words, we aim to show that (i) $w_{ij;k}$ is homogeneous with respect to the conformal grading $\Delta$ in \eqref{eq:conformal weight} and that (ii) the linear part of each $w_{ij;k}$ spans $\mathfrak{g}^f$ modulo total derivatives. \\

\noindent
(i) We can extend the conformal grading on $\mathcal{V}_{I}^{\,l}$ to $\mathcal{V}_{I}^{\,l}(\!(\partial^{-1})\!)$ by letting $\Delta_{\partial}=1$. Then, from \eqref{eq:A_uv}, we have
\[ \Delta_{A_{\frac{[uv]}{(ij)}}(\partial)}=1+u-v,\]  for $u,v\in \{1,2,\cdots, N\}$ with $u\geq v$ and $i,j\in \{1,2, \cdots,m+n\}$. Hence for  $k\in \{0,\cdots, N-1\}$, 
\begin{equation} \label{eq:conf_A}
\Delta_{A_{\frac{[N,i_{1}+1]}{(is_{1})}}A_{\frac{[i_{1},i_{2}+1]}{(s_1,s_2)}}\cdots A_{\frac{[i_{k},1]}{(s_{k},j)}}}=(N-i_1)+ (i_1-i_2) +\cdots + (i_k-1+1)=N.
\end{equation}
Comparing \eqref{operator L(z)} and \eqref{eq:conf_A}, we get $\Delta_{L_{ij}(\partial)}=N$. In addition, we have  $\Delta_{w_{ij;k}}=l-k$ for $w_{ij;k}$ in \eqref{eq:generators w_{ij:k}}. \\

\noindent 
(ii) From \eqref{operator L(z)} we get the explicit formula: 
\begin{equation}\label{eq:w{ij,k}}
w_{ij;k}=\sum_{r=k}^{N-1}\sum_{i_{1},\cdots,i_{r}}\sum_{s_{1},\cdots,s_{r}=1}^{m+n}\text{Res}\,A_{\frac{[N,i_{1}+1]}{(is_{1})}}A_{\frac{[i_{1},i_{2}+1]}{(s_1,s_2)}}\cdots A_{\frac{[i_{r},1]}{(s_{r},j)}}(\partial)\partial^{-k-1}\in S(\mathbb{C}[\partial]\otimes\mathfrak{p}).
\end{equation}
Let us pick the summand in \eqref{eq:w{ij,k}} which yields the linear part of $w_{ij;k}$, that is 
\[
f_{ij;k}:=\sum_{i_{1},\cdots,i_{k}}\sum_{s_{1},\cdots,s_{k}=1}^{m+n}\text{Res}\,A_{\frac{[N,i_{1}+1]}{(is_{1})}}A_{\frac{[i_{1},i_{2}+1]}{(s_1,s_2)}}\cdots A_{\frac{[i_{k},1]}{(s_{k},j)}}(\partial)\partial^{-k-1}=(-1)^{k}\sum_{h=0}^{k}q_{\frac{[N+h-k,h+1]}{(ij)}}\in\mathfrak{p}.
\]
  One can check that $f_{ij;k}$ for $i,j \in \{1,2, \cdots, m+n\}$ and $k\in \{0,1, \cdots, N-1\}$ forms a basis of $\mathfrak{g}^{f}$ and 
satisfies
\eqref{eq:property of generator of W-algebra}:
\[
w_{ij;k}-f_{ij;k}\in\partial(\mathbb{C}[\partial]\otimes \mathfrak{p}) \oplus \bigoplus_{m\geq 2}(\mathbb{C}[\partial]\otimes \mathfrak{p})^{\otimes m}.
\]
Since we proved $(i)$ and $(ii)$, we can conclude that $\left\{w_{ij;k}\right\}_{\tiny\substack{1\leq i,j\leq m+n,\\0\leq k\leq N-1}}$ freely generates $\mathcal{W}(\mathfrak{gl}(Nm|Nn),f)$ by Proposition \ref{prop:free generating set of W-algebra}.
\end{proof}

\begin{thm} \label{thm:W-algebra and sATO-ver1}
    The matrix differential operator $L(\partial)$ in \eqref{eq:L} is a sATO of order $N\geq 2$ on  $\mathcal{W}(\mathfrak{gl}(Nm|Nn),f)$. In other words, $\mathcal{W}(\mathfrak{gl}(Nm|Nn),f)$ is the Adler-type PVsA associated with $L(\partial)$.
\end{thm}

\begin{proof}
Note that the coefficients of $L^{\mathrm{inv}}(z)$ are also in the $\mathcal{W}$-superalgebra $\mathcal{W}(\mathfrak{gl}(Nm|Nn),f)$ since they are in the differential superalgebra generated by coefficients of the monic operator $L(\partial)$. 
Let us denote the $\lambda$-bracket \eqref{eq: W-bracket} on $\mathcal{W}(\mathfrak{gl}(Nm|Nn),f)$ by $\{\ {}_\lambda \ \}_{\mathcal{W}}$ and the  affine bracket  \eqref{eq: Affine PVA} defined on $\mathcal{V}(\mathfrak{g}(\Pi))$  by  $\{\ {}_\lambda \ \}_{\mathrm{Aff}}$. We aim to
show that $L^{\mathrm{inv}}(\partial)$ is a sATO on $\mathcal{W}(\mathfrak{gl}(Nm|Nn),f)$ with respect to the opposite bracket $-\{ \ {}_\lambda \ \}_{\mathcal{W}}.$  Recall that the matrix operator $A(\partial)$  defined in \eqref{eq:A(partial)} is 
 a sATO for the affine bracket. Hence, the following equalities hold:
\begin{align*}
    &-\{(L^{\mathrm{inv}})_{ij}(z)\,{}_{\lambda}\,(L^{\mathrm{inv}})_{hk}(w)\}_{\mathcal{W}}
   =-\rho\{(A^\mathrm{inv})_{\frac{[l1]}{(ij)}}(z) \,{}_{\lambda}\,(A^\mathrm{inv})_{\frac{[l1]}{(hk)}}(z)\}_{\mathrm{Aff}}\\
    &=(-1)^{\tilde{\imath}\tilde{\jmath}+\tilde{\imath}\tilde{h}+\tilde{\jmath}\tilde{h}}\rho\Big( (A^{\mathrm{inv}})_{\frac{[l1]}{(hj)}}(w+\lambda+\partial) \iota_{z}(z\!-w\!-\lambda\!-\partial)^{-1}(
(A^{\mathrm{inv}})_{\frac{[l1]}{(ik)}})^{\!*}(\lambda-z)\Big.\\
\Big.&\hspace{6cm}-(A^{\mathrm{inv}})_{\frac{[l1]}{(hj)}}(z) \iota_{z}(z\!-w\!-\lambda\!-\partial)^{-1}(A^{\mathrm{inv}})_{\frac{[l1]}{(ik)}}(w)\Big)\\
&=(-1)^{\tilde{\imath}\tilde{\jmath}+\tilde{\imath}\tilde{h}+\tilde{\jmath}\tilde{h}}\Big( (L^{\mathrm{inv}})_{hj}(w+\lambda+\partial) \iota_{z}(z\!-w\!-\lambda\!-\partial)^{-1}(
(L^{\mathrm{inv}})_{ik})^{\!*}(\lambda-z)\Big.\\
\Big.&\hspace{6cm}-(L^{\mathrm{inv}})_{hj}(z) \iota_{z}(z\!-w\!-\lambda\!-\partial)^{-1}(L^{\mathrm{inv}})_{ik}(w)\Big)
\end{align*}
 Here, the first equality follows from \eqref{eq:W-bracket;property} and the second equality holds by  Proposition \ref{prop:Properties of sATO}. 
Hence the pseudo-differential operator $L^{\mathrm{inv}}(\partial)$ is a sATO with respect to the negative $\lambda$-bracket $-\{\ {}_{\lambda}\ \}_{\mathcal{W}}$ on the $\mathcal{W}$-superalgebra. Again, by Proposition \ref{prop:Properties of sATO}, its inverse $L(\partial)$ is a sATO on $\mathcal{W}(\mathfrak{gl}(Nm|Nn), f)$ with respect to $\{\ {}_{\lambda}\ \}_{\mathcal{W}}$.
\end{proof}

We can summarize the previous two theorems as follows.

\begin{thm} \label{summary}
Let $\mathcal{V}_{\Pi}^N$ be the Adler-type PVsA introduced in \eqref{eq:generic PVA}, where $\Pi$ is the index set \eqref{eq:index_parity}. Then 
  \begin{equation}
     \mathcal{V}_{\Pi}^N \simeq \mathcal{W}(\mathfrak{gl}(Nm|Nn),f)
 \end{equation}
 for the $N\times (m|n)$ rectangular nilpotent $f$.
\end{thm}
\begin{proof}
 By Theorem \ref{thm:W-algebra and sATO-ver2}, we have an isomorphism of differential superalgebras. Moreover, this isomorphism preserves the $\lambda$-brackets in both sides by Theorem  \ref{thm:W-algebra and sATO-ver1}.  
\end{proof}

\vskip 5mm

\subsection{Examples}

In this section, we describe the  rectangular $\mathcal{W}$-superalgebras $\mathcal{W}(\mathfrak{gl}(Nm|Nn),f)$ for the $N\times (m|n)$ rectangular nilpotent element $f$ when $N=2$ or $(N,m,n)=(3,2,1).$

\begin{ex}\label{ex:rectangular case with l=2}
Let us consider the rectangular $\mathcal{W}$-superalgebra $\mathcal{W}(\mathfrak{gl}(2m|2n),f)$ associated with the $2\times (m|n)$ type nilpotent $f.$
 The index set $\Pi=\{1, 2, \cdots, 2m+2n\}=I\sqcup J$ where $J=\{1, 2, \cdots, m+n\}$ and $I= \Pi\setminus J$. The parity on $\Pi$ is defined as in \eqref{eq:index_parity}. From the operator $A(\partial)$  in \eqref{sATO of Affine PVsA}, we obtain the sATO $L(\partial)$ in \eqref{eq:L}:
\[
L(\partial):=\left[\begin{array}{c}
A_{[22]} \\
\end{array}\right]\star
\left[\begin{array}{c}
\mathbbm{1}_{(m|n)} 
\end{array}\right]^{\mathrm{inv}}\star
\left[\begin{array}{c}
A_{[11]}  \\
\end{array}\right]-\left[\begin{array}{c}
A_{[21]} 
\end{array}\right]=A_{[22]}\star A_{[11]}-A_{[21]}.
\]
Then each entry of $L(\partial)$ has the order at most two. More precisely, 
\[
L(\partial)=\left(\delta_{ij}\partial^2+v_{ij}\partial+w_{ij}\right)_{1\leq i,j \leq m+n},
\]
where
\begin{equation}\label{generators in l=2}
\begin{aligned}
& v_{ij}=(-1)^{\tilde{\imath}}(q_{ij}+q_{i+m+n,j+m+n}),\\
&w_{ij}=(-1)^{\tilde{\imath}}q_{ij}'-(-1)^{\tilde{\imath}}q_{i+m+n,j}+{\displaystyle \sum_{k\in I}}(-1)^{\tilde{\imath}+\tilde{k}}q_{i+m+n,k+m+n}q_{kj}.
\end{aligned}
\end{equation}
Using the identity \eqref{eq:super Adler}, we can compute the $\lambda$-bracket between the coefficients in $L(\partial)$ which generate the algebra $\mathcal{W}(\mathfrak{gl}(2m|2n),f)$:
\begin{equation}\label{lambda relations in l=2}
    \begin{aligned}
    &\{v_{ij}\:_{\lambda}\:v_{hk}\}=(-1)^{\tilde{\imath}\tilde{\jmath}+\tilde{\imath}\tilde{h}+\tilde{\jmath}\tilde{h}}\left(\delta_{hj}v_{ik}-\delta_{ik}v_{hj}-2\delta_{hj}\delta_{ik}\lambda\right),\\
     &\{v_{ij}\:_{\lambda}\:w_{hk}\}=-(-1)^{\tilde{\imath}\tilde{\jmath}+\tilde{\imath}\tilde{h}+\tilde{\jmath}\tilde{h}}\delta_{ik}(v_{hj}\lambda+w_{hj})+(-1)^{\tilde{\jmath}}\delta_{hj}w_{ik}-(-1)^{\tilde{\jmath}}\delta_{jh}\delta_{ik}\lambda^{2},\\
     &\{w_{ij}\:_{\lambda}\:w_{hk}\}=
(-1)^{\tilde{\imath}\tilde{\jmath}+\tilde{\imath}\tilde{h}+\tilde{\jmath}\tilde{h}}\big((\lambda+\partial)\delta_{hj}w_{ik}+v_{hj}w_{ik}+\delta_{ik}\delta_{hj}\lambda^{3}+\delta_{ik}v_{hj}\lambda^{2}+\delta_{ik}w_{hj}\lambda\\
&\hspace{7cm}-\delta_{hj}(\lambda+\partial)^{2}v_{ik}-v_{hj}(\lambda+\partial)v_{ik}-w_{hj}v_{ik}\big).
\end{aligned}
\end{equation}

\end{ex}

\begin{example}
Let us consider the rectangular $\mathcal{W}$-superalgebra $\mathcal{W}(\mathfrak{gl}(6|3),f)$ associated with the $3\times (2|1)$ rectangular nilpotent $f.$
 The parity of the index set  $\Pi=\{1,2,\cdots, 9\}$ is defined as in \eqref{eq:index_parity}. From the operator $A(\partial)\in\mathfrak{gl}(\Pi)\otimes\mathcal{V}(\mathfrak{gl}(\Pi))(\!( \partial^{-1})\!)$  in \eqref{sATO of Affine PVsA}, we obtain the sATO $L(\partial)\in \mathfrak{gl}(2|1)\otimes\mathcal{V}(\mathfrak{gl}(\Pi))(\!(\partial^{-1})\!)$ in \eqref{eq:L} by direct computations. For example, the (1,1) entry of $L(\partial)$ is 
\begin{equation}\label{l=3, example, L_{11}}
\begin{aligned}
L_{11}(\partial)= & q_{\frac{[31]}{(11)}}-q_{\frac{[32]}{(11)}}(\partial+q_{\frac{[11]}{(11)}})-q_{\frac{[32]}{(12)}}q_{\frac{[11]}{(21)}}-q_{\frac{[32]}{(13)}}(-q_{\frac{[11]}{(31)}})-(\partial+q_{\frac{[33]}{(11)}})q_{\frac{[21]}{(11)}}\\
-&q_{\frac{[33]}{(12)}}q_{\frac{[21]}{(21)}}-q_{\frac{[33]}{(13)}}(-q_{\frac{[21]}{(31)}})+(\partial+q_{\frac{[33]}{(11)}})(\partial+q_{\frac{[22]}{(11)}})(\partial+q_{\frac{[11]}{(11)}})\\
+ &(\partial+q_{\frac{[33]}{(11)}})q_{\frac{[22]}{(12)}}q_{\frac{[11]}{(21)}}+(\partial+q_{\frac{[33]}{(11)}})q_{\frac{[22]}{(13)}}(-q_{\frac{[11]}{(31)}})\\
+ & q_{\frac{[33]}{(12)}}q_{\frac{[22]}{(21)}}(\partial+q_{\frac{[11]}{(11)}})+q_{\frac{[33]}{(12)}}(\partial+q_{\frac{[22]}{(22)}})q_{\frac{[11]}{(21)}}+q_{\frac{[33]}{(12)}}q_{\frac{[22]}{(23)}}(-q_{\frac{[11]}{(31)}})\\
+ & q_{\frac{[33]}{(13)}}(-q_{\frac{[22]}{(31)}})(\partial+q_{\frac{[11]}{(11)}})+q_{\frac{[33]}{(13)}}(-q_{\frac{[22]}{(32)}})q_{\frac{[11]}{(21)}}+q_{\frac{[33]}{(13)}}(\partial-q_{\frac{[22]}{(33)}})(-q_{\frac{[11]}{(31)}}),
\end{aligned}
\end{equation}
where $q_{\frac{[uv]}{(ab)}}=q_{(u-1)(m+n)+a,(v-1)(m+n)+b}$
for $1\leq u,v\leq 3$ and $1\leq a,b\leq 3$. 
By symbolizing the terms in RHS of \eqref{l=3, example, L_{11}}, we get 
\[L_{11}(z)=z^{3}+w_{11;2}z^{2}+w_{11;1}z+w_{11;0},\]
where
\begin{align*}
w_{11;2}&=q_{11}+q_{44}+q_{77},\\
w_{11;1}&=-q_{74}-q_{41}+2q_{11}'+q_{44}'\\
&+q_{44}q_{11}+q_{77}q_{11}+q_{77}q_{44}+q_{45}q_{21}-q_{46}q_{31}+q_{78}q_{54}+q_{78}q_{21}-q_{79}q_{64}-q_{79}q_{31},\\
w_{11;0}&=q_{11}''+q_{71}-q_{41}'+(q_{44}q_{11})'+q_{77}q_{11}'-q_{74}q_{11}+q_{75}q_{21}+q_{76}q_{31}-q_{77}q_{41}+q_{78}q_{51}+q_{79}q_{61}\\
&+(q_{45}q_{21})'-(q_{46}q_{31})'+q_{78}q_{21}'-q_{79}q_{31}'+q_{77}q_{44}q_{11}+q_{77}q_{45}q_{21}-q_{77}q_{46}q_{31}\\
&+q_{78}q_{54}q_{11}+q_{78}q_{55}q_{21}-q_{78}q_{56}q_{31}-q_{79}q_{64}q_{11}-q_{79}q_{65}q_{21}+q_{79}q_{66}q_{31}. 
\end{align*}
The three elements $w_{11;2}$, $w_{11;1}$, $w_{11;0}$ are in $\mathcal{W}(\mathfrak{gl}(6|3),f).$
\end{example}

\newpage
\section{Integrable systems and rectangular $\mathcal{W}$-superalgebras}
\label{Sec:integrable}
In this section, we construct integrable systems on the  rectangular $\mathcal{W}$-superalgebras $\mathcal{W}(\mathfrak{gl}(Nm|Nn),f)$.
By Theorem \ref{thm:W-algebra and sATO-ver2}, it amounts to define an integrable system on a PVsA defined by a generic sATO (see Section $3$). While the previous section was making use of the $\star$-product, we will now need to use the first product $\circ$ \eqref{eq:ordinary product}.

\subsection{Fractional powers of sATO}
\label{subsec:fractional}
Let $\mathcal{V}$ be a PVsA endowed with a $\lambda$-bracket $\{ \, _{\lambda} \,  \}$ and $I$ be a finite index set with parity map.

\begin{lem}\label{prop:fractional power, minus one power}
Let $A(\partial) \in \mathfrak{gl}(I) \otimes \mathcal{V} (\!(\partial^{-1})\!)$ be a monic matrix pseudo-differential operator of
order $N\geq1$.
\begin{enumerate}
\item[$(a)$] There exists a unique matrix pseudo-differential operator $A(\partial)^{-1}$ such that $A(\partial)^{-1}\circ A(\partial)=A(\partial) \circ A(\partial)^{-1}= \mathbbm{1}_I$, which is monic of order $-N$.
 \item[$(b)$]  There exists a unique matrix pseudo-differential operator $A^{\frac{1}{N}}(\partial)$ such that
$(A^{\frac{1}{N}})^{\circ N}=A$ which is monic of order
one.
\end{enumerate}
Here, we emphasize that matrix pseudo-differential operators $A(\partial)^{-1}$ and   $A^{\frac{1}{N}}$ are defined with respect to the $\circ$-product \eqref{eq:ordinary product}, not the $\star$-product.
\end{lem}
\begin{proof}
   (a) For the existence and uniqueness of the inverse, one can just apply the proof of Lemma \ref{lem:star inverse}.\\
   (b) Since the leading term is the identity matrix, the proof reduces to the scalar setting in which case the statement is clear. The idea is as follows. The coefficient $V_0$ in the operator 
    \[B(\partial)= \partial+ \sum_{k<0} V_k \partial^k\in \mathfrak{gl}(I)\otimes \mathcal{V}(\!(\partial^{-1})\!)\]
    satisfying $B(\partial)^N=A(\partial)$ can be obtained by comparing the coefficients of $\partial^{N-1}$ in the both sides. Inductively, we get $V_k$ for negative integers $k$.
\end{proof}

For $i,j \in I$, the $ij$-th entry of $A \circ B(z)$ is  
 $\left( A \circ B \right)_{ij}(z) = \sum_{t \in I}(-1)^{(\tilde{\imath}+\tilde{t})(\tilde{\jmath}+\tilde{t})} A_{it}(z+\partial)B_{tj}(z)$.
Hence, for any $ v \in \mathcal{V}$, by the Leibniz rule, we have
\begin{equation} \label{eq:bracket-fraction}
\begin{aligned}
\big\{ \big( A \circ B \big)_{ij}(z) \;\,_\lambda\;\, v \big\}
= &\displaystyle \sum_{t \in I} (-1)^{(\tilde{t}+\tilde{\jmath})(\tilde{v}+\tilde{t}+\tilde{\imath})} \left\{ A_{it}(z+\partial)\;\, {_{\lambda+\partial} }\;\,{v} \right\}_{\rightarrow}  B_{tj}(z)
\\&+\displaystyle \sum_{t \in I} (-1)^{\tilde{v}(\tilde{\imath}+\tilde{t})} \left\{ B_{tj}(z)\;\,_{\lambda+\partial}\;\, v \right\}_{\rightarrow}  (A^{*})_{ti}(-z+\lambda)
 \end{aligned}
 \end{equation}
 where the negative powers of $z+\partial$ are expanded for large $|z|$.

\begin{lem} \label{lem:fraction of L}
    Let $A(\partial)$ be a monic matrix pseudo-differential operator on $\mathcal{V}$ of order $N\,>\,0$. Then for all $a,b  \in I$ and for all $k \in \mathbb{Z}_+$, we have
        \begin{equation}
        \mathrm{Res}_z \big\{ {\mathrm{str} A^{\frac{k}{N}}(z)} \, _\lambda \, A_{ab}(w) \big\}\bigr|_{\lambda=0} = \frac{k}{N}\displaystyle\sum_{c,d \in I}
        (-1)^{(\tilde{a}+\tilde{b})(\tilde{c}+\tilde{d})+\tilde{d}}\, \mathrm{Res}_{z} \left\{ {A_{cd}(z+\partial) \, _\partial  \, A_{ab}(w) }\right\} (A^{\frac{k}{N}-1}_{dc}(z)).
         \end{equation}
        \end{lem}
\begin{proof}
By induction on $k$, it can be checked using the sesquilinearity and Leibniz rule axioms that 
    \begin{equation}\label{eqqq}
    \begin{split}
        \big\{ {\text{str} A^{\frac{k}{N}}(z)} \;\,_\lambda\;\, A_{ab}(w) \big\} =\displaystyle\sum_{i,s,t \in I} \displaystyle\sum_{l=1}^k (-1)^{\tilde{t}}(-1)^{(\tilde{t}+\tilde{s})(\tilde{s}+\tilde{\imath})}(-1)^{(\tilde{t}+\tilde{\imath})(1+\tilde{a}+\tilde{b})}(-1)^{(\tilde{s}+\tilde{\imath})(\tilde{a}+\tilde{b})} \\
        \times \big\{ A^{\frac{1}{N}}_{ts}(z+x) \;\,_{\lambda+x+y}\;\, A_{ab}(w) \big\} \,\big( \big\vert _{x=\partial} A^{\frac{k-l}{N}}_{si}(z)\big)\big( \big\vert _{y=\partial}(A^{*\frac{l-1}{N}})_{ti}(-z+\lambda) \big).
         \end{split}
    \end{equation}
We simplify the signs in \eqref{eqqq} as follows
\begin{align*}
    (-1)^{\tilde{t}}(-1)^{(\tilde{t}+\tilde{s})(\tilde{s}+\tilde{\imath})}(-1)^{(\tilde{t}+\tilde{\imath})(1+\tilde{a}+\tilde{b})}(-1)^{(\tilde{s}+\tilde{\imath})(\tilde{a}+\tilde{b})}=(-1)^{(\tilde{s}+\tilde{t})\tilde{\imath}+\tilde{s}\tilde{t}+\tilde{\imath}+(\tilde{s}+\tilde{t})(\tilde{a}+\tilde{b})+\tilde{s}}.
\end{align*}
Next, we take the residue and evaluate at $\lambda=0$ to get 
\begin{equation}\label{eqq}
\begin{split}
   &\text{Res}_z \big\{ {\text{str} A^{\frac{k}{N}}(z)} \;\,_\lambda\;\, A_{ab}(w) \big\}\bigr|_{\lambda=0}\\
    &= \text{Res}_z \displaystyle\sum_{i,s,t \in I} \displaystyle\sum_{l=1}^k \big\{ A^{\frac{1}{N}}_{ts}(z\!+x\!+y+\!\lambda) \;\,_{\lambda+x+y}\;\, A_{ab}(w) \big\} 
    \big( \big\vert _{x=\partial} A^{\frac{k-l}{N}}_{si}(z\!+\lambda\!+y)\big) \\
    & \hskip 3cm \times \big( \big\vert _{y=\partial} A^{\frac{l-1}{N}}_{it}(z) \big) \big \vert_{\lambda=0} \times (-1)^{(\tilde{s}+\tilde{t})\tilde{\imath}+\tilde{s}\tilde{t}+\tilde{\imath}} (-1)^{(\tilde{s}+\tilde{t})(\tilde{a}+\tilde{b})+\tilde{s}} \\
   &= k \, \text{Res}_z \displaystyle\sum_{s,t \in I} \big\{ A^{\frac{1}{N}}_{ts}(z+\partial) \;\,_{\partial}\;\, A_{ab}(w) \big\} ( A^{\frac{k-1}{N}}_{st}(z)) \times (-1)^{(\tilde{s}+\tilde{t})(\tilde{a}+\tilde{b})+\tilde{s}}.
   \end{split}
\end{equation}
Here, the first equality follows from \eqref{eq:bracket-fraction} and 
\begin{equation}
    \text{Res}_{z}\,B_{ij}(z)(C^{\,*})_{hk}(-z+\lambda)=\text{Res}_{z}\,B_{ij}(z+\lambda+\partial)C_{kh}(z).
\end{equation}
Recall that $A(z+x)$ is expanded using the geometric expansion for large $|z|$. For any $c,d \in I$, one gets 
\begin{align}
    \big\{ {A^{\frac{N}{N}}_{cd}(z)} \;\,_\lambda\;\, A_{ab}(w) \big\} =\displaystyle\sum_{s,t \in I}\displaystyle\sum_{l=1}^N (-1)^{(\tilde{a}+\tilde{b})(\tilde{s}+\tilde{d})}(-1)^{(\tilde{c}+\tilde{t})(\tilde{t}+\tilde{s}+\tilde{a}+\tilde{b}+\tilde{s}+\tilde{d})}(-1)^{(\tilde{c}+\tilde{t})(\tilde{t}+\tilde{d})+(\tilde{t}+\tilde{s})(\tilde{s}+\tilde{d})} \label{eq:cd-ab} \\
    \times \big\{ A^{\frac{1}{N}}_{ts}(z+x) \;\,_{\lambda+x+y}\;\, A_{ab}(w) \big\} \,\big( \big\vert _{x=\partial} A^{\frac{N-l}{N}}_{sd}(z)\big)\big( \big\vert _{y=\partial}(A^{*\frac{l-1}{N}})_{tc}(-z+\lambda) \big),
\end{align}
by the same process as we get \eqref{eqqq}.
For the simplicity of notation, let us denote the sign in \eqref{eq:cd-ab} by \[\eta=(-1)^{(\tilde{a}+\tilde{b})(\tilde{s}+\tilde{d})}(-1)^{(\tilde{c}+\tilde{t})(\tilde{t}+\tilde{s}+\tilde{a}+\tilde{b}+\tilde{s}+\tilde{d})}(-1)^{(\tilde{c}+\tilde{t})(\tilde{t}+\tilde{d})+(\tilde{t}+\tilde{s})(\tilde{s}+\tilde{d})}.\]
 After replacing $z$ by  $z+\partial$ and $\lambda$ by $\lambda +\partial$, applying both sides of our equation to $A^{\frac{k}{N}-1}_{dc}(z)$ and summing over all pairs of indices $(c,d)\in I\times I$, we obtain
\begin{equation}\label{eqqqq}
\begin{split}
    & \displaystyle \sum_{c,d \in I} \big\{ A_{cd}(z\!+\!x) \;\,_{\lambda+\partial}\;\, A_{ab}(w) \big\}\big(\big \vert_{x=\partial} A^{\frac{k}{N}-1}_{dc}(z)\big)\\
    &=\displaystyle\sum_{c,d \in I}\displaystyle\sum_{s,t \in I}\displaystyle\sum_{l=1}^N \big\{ A^{\frac{1}{N}}_{ts}(z\!+\!x) \;\,_{\lambda+x+y}\;\, A_{ab}(w) \big\} \,\big( \big\vert _{x=\partial} A^{\frac{N-l}{N}}_{sd}(z+x)\big)\big( \big\vert _{y=\partial}(A^{*\frac{l-1}{N}})_{tc}(-z+\lambda) \big) \\
    & \hskip 3cm \times \big( \big\vert _{x=\partial} A^{\frac{k}{N}-1}_{dc}(z)\big) \times {\eta}\\
    &=\displaystyle\sum_{c,d \in I}\displaystyle\sum_{s,t \in I} \displaystyle\sum_{l=1}^N \big\{ A^{\frac{1}{N}}_{ts}(z\!+\!x) \;\,_{\lambda+x+y}\;\, A_{ab}(w) \big\} \,\big( \big\vert _{x=\partial} A^{\frac{N-l}{N}}_{sd}(z+\partial)A^{\frac{k}{N}-1}_{dc}(z)\big) \\
    &\hskip 3cm \times \big( \big\vert _{y=\partial}(A^{*\frac{l-1}{N}})_{tc}(-z+\lambda) \big) \times {\eta} \cdot (-1)^{(\tilde{t}+\tilde{c})(\tilde{c}+\tilde{d})}\\
    &=\displaystyle\sum_{c \in I}\displaystyle\sum_{s,t \in I}\displaystyle\sum_{l=1}^N \big\{ A^{\frac{1}{N}}_{ts}(z\!+\!x) \;\,_{\lambda+x+y}\;\, A_{ab}(w) \big\} \,\big( \big\vert _{x=\partial} A^{\frac{k-l}{N}}_{sc}(z) \big)\big( \big\vert _{y=\partial}(A^{*\frac{l-1}{N}})_{tc}(-z+\lambda) \big) \\ 
    &\hskip 3cm \times {\eta} \cdot (-1)^{(\tilde{t}+\tilde{c})(\tilde{c}+\tilde{d})}(-1)^{(\tilde{s}+\tilde{d})(\tilde{d}+\tilde{c})}.
    \end{split}
\end{equation}
The sign in the last equation of \eqref{eqqqq} is simplified as follows : 
\begin{align*}
    {\eta}\cdot(-1)^{(\tilde{t}+\tilde{c})(\tilde{c}+\tilde{d})}(-1)^{(\tilde{s}+\tilde{d})(\tilde{d}+\tilde{c})}=(-1)^{(\tilde{s}+\tilde{t})(\tilde{a}+\tilde{b})+\tilde{s}}  (-1)^{(\tilde{a}+\tilde{b})(\tilde{c}+\tilde{d})+\tilde{d}} (-1)^{(\tilde{s}+\tilde{c})(\tilde{t}+\tilde{c})}.
\end{align*} 
Taking the residue and evaluating $\lambda$ at $0$, we get 
\begin{equation}
\begin{split}
    &\displaystyle \sum_{c,d \in I} \text{Res}_{z}\left\{ A_{cd}(z\!+\!\partial) \;\,_{\partial}\;\, A_{ab}{w} \right\} (A^{\frac{k}{N}-1}_{dc}(z)) \times (-1)^{(\tilde{a}+\tilde{b})(\tilde{c}+\tilde{d})+\tilde{d}}\\
    &= \text{Res}_{z}\displaystyle\sum_{c \in I}\displaystyle\sum_{s,t \in I}\displaystyle\sum_{l=1}^N \big\{ A^{\frac{1}{N}}_{ts}(z\!+\!x) \;\,_{x+y}\;\, A_{ab}(w) \big\} \,\big( \big\vert _{x=\partial} A^{\frac{k-l}{N}}_{sc}(z) \big)\big( \big\vert _{y=\partial}(A^{*\frac{l-1}{N}})_{tc}(-z) \big) \\ 
    &\hskip 3cm \times (-1)^{(\tilde{s}+\tilde{t})(\tilde{a}+\tilde{b})+\tilde{s}}  (-1)^{(\tilde{s}+\tilde{c})(\tilde{t}+\tilde{c})}
    \\
    &= \text{Res}_{z}\displaystyle\sum_{c \in I}\displaystyle\sum_{s,t \in I}\displaystyle\sum_{l=1}^N \big\{ A^{\frac{1}{N}}_{ts}(z\!+\!x) \;\,_{x+y}\;\, A_{ab}(w) \big\} \,\big( \big\vert _{x=\partial} A^{\frac{k-l}{N}}_{sc}(z+y) \big)\big( \big\vert _{y=\partial}(A^{\frac{l-1}{N}})_{ct}(z) \big) \\ 
    &\hskip 3cm \times (-1)^{(\tilde{s}+\tilde{t})(\tilde{a}+\tilde{b})+\tilde{s}}  (-1)^{(\tilde{s}+\tilde{c})(\tilde{t}+\tilde{c})}
    \\
    &=N\,\text{Res}_{z}\displaystyle\sum_{s,t \in I} \big\{ A^{\frac{1}{N}}_{ts}(z\!+\!\partial) \;\,_{\partial}\;\, A_{ab}(w) \big\} \, (A^{\frac{k-1}{N}}_{st}(z)) \times (-1)^{(\tilde{s}+\tilde{t})(\tilde{a}+\tilde{b})+\tilde{s}}.
    \end{split}
    \end{equation}
We conclude the proof by comparing the last equation with \eqref{eqq}.
\end{proof}

\subsection{integrable hierarchies on a rectangular $\mathcal{W}$-superalgebra}
While lemmas in Section  \ref{subsec:fractional} are stated for arbitrary choices of monic matrix pseudo-differential operator $A$,  we specialize now to the differential superalgebra $\mathcal{V}:=\mathcal{V}_I^{N}$ generated by the coefficients of the generic sATO
\begin{equation*}
   L(\partial) :=\partial^N+ \sum_{ M=0}^{N-1} \, \sum_{a,b\in I}e_{ab}\otimes u_{M,ab}\partial^M,
\end{equation*}
which is isomorphic to a rectangular $\mathcal{W}$-superalgebra
by Theorem \ref{thm:W-algebra and sATO-ver2}.

For all positive integer $k$, let us define the differential polynomial 
\begin{equation} \label{eq:Hamiltonian}
    h_k:=\frac{N}{k}\text{Res}\;\text{str}L^{\frac{k}{N}}(\partial).
\end{equation}
Then the variational derivative of $h_k$ can be written in terms of the coefficients of $L(\partial)$ as follows.
\begin{lem}
    Let $i\in \mathbb{Z}_+$ such that $i<N$. For $a,b \in I$ and a positive integer $k$, we have
    \begin{equation}\label{variational derivative lemma}
        \frac{\delta h_{k}}{\delta u_{i,ab}}=\mathrm{Res}_{z}(z+\partial)^{-i-1}(L^{\frac{k}{N}-1})_{ba}(z).
    \end{equation}
\end{lem}
\begin{proof} 
    By Lemma \ref{lem:fraction of L}, 
    \begin{equation}
        \mathrm{Res}_z \big\{ {\mathrm{str} L^{\frac{k}{N}}(z)} \;\,_\lambda\;\, L_{ab}(w) \big\}\big\vert_{\lambda=0} = \frac{k}{N}\displaystyle\sum_{c,d \in I} (-1)^{(\tilde{a}+\tilde{b})(\tilde{c}+\tilde{d})+\tilde{d}}
        \text{Res}_{z} \left\{ {L_{cd}(z+\partial) \;\,_\partial\;\, L_{ab}(w) }\right\}(L^{\frac{k}{N}-1})_{dc}(z)).
        \end{equation}
    Taking the coefficients of $w^{-j-1}$ in both sides, we get
    \begin{align} 
    \left\{ h_k \;\,_\lambda\;\, u_{j;ab} \right\}\Big\vert_{\lambda=0}&= \displaystyle\sum_{c,d \in I} 
    (-1)^{(\tilde{a}+\tilde{b})(\tilde{c}+\tilde{d})+\tilde{d}}\,\text{Res}_{z} \left\{ {L_{cd}(z+\partial) \;\,_\partial\;\, u_{j;ab} }\right\} ((L^{\frac{k}{N}-1})_{dc}(z)) \\
    &=\displaystyle\sum_{c,d \in I} \displaystyle\sum_{i=0}^{N-1}(-1)^{(\tilde{a}+\tilde{b})(\tilde{c}+\tilde{d})}\left\{ u_{i,cd}\; {}_{\partial}\; u_{j,ab} \right\}\,\text{Res}_{z}(z+\partial)^{-i-1}((L^{\frac{k}{N}-1})_{dc}(z)).
\end{align}
On the other hand, using the master formula \eqref{eq:master formula} and by definition of the variational derivative, we obtain 
\begin{align}
    \left\{ h_k \;\,_\lambda\;\, u_{j;ab} \right\}\big\vert_{\lambda=0}&=\displaystyle\sum_{i=0}^{N-1}\displaystyle\sum_{c,d \in I}(-1)^{(\tilde{a}+\tilde{b})(\tilde{c}+\tilde{d})}\left\{ u_{i,cd}\; {}_{\lambda+ \partial}\; u_{j,ab} \right\}(-\lambda-\partial)^{m}\frac{\partial h_{k}}{\partial u_{i,cd}^{(m)}}\Bigg\vert_{\lambda=0}\\
    &=\displaystyle\sum_{i=0}^{N-1}\displaystyle\sum_{c,d \in I}(-1)^{(\tilde{a}+\tilde{b})(\tilde{c}+\tilde{d})}\left\{ u_{i,cd}\; {}_{\partial}\; u_{j,ab} \right\}\Big(\frac{\delta h_{k}}{\delta u_{i,cd}}\Big).
\end{align}
The equation \eqref{variational derivative lemma} follows from coefficient comparison between the above two formulas, since the  constant matrix differential operator $(\left\{ u_{i,cd}\; {}_{\partial}\; u_{j,ab} \right\})_{i,j \in \{0,1,\cdots, N-1\} ; a,b,c,d \in I}$ is arbitrary.
\end{proof}

Recall from Section \ref{subsec:Generic} that $L(\partial)$ is even and that we have constructed two compatible PVsA structures on the algebra $\mathcal{V}$. The first one, denoted by $\{ \, _{\lambda} \,  \}_{H}$, is defined via the Adler-type identity for $L(\partial)$. The second bracket $\{ \, _{\lambda} \,  \}_{K}$ defined in Section \ref{subsec:second bracket} 
is obtained from $L^{\epsilon}(\partial)$, which is a deformation of $L(\partial)$.

\begin{lem} \label{1.5.}
  For every $a,b \in I$ and a positive integer $k$, we have  
  \begin{align*}
  (a) & \  \big\{ h_k \;\,_\lambda\;\, L_{ab}(w) \big\}_{H}\big\vert_{\lambda=0}=\displaystyle\sum_{c \in I}(-1)^{\tilde{c}(\tilde{a}+\tilde{b}+1)+\tilde{a}\tilde{b}} \left( (L^{\frac{k}{N}})_{ac}(w+\partial)_{+}L_{cb}(w)-L_{ac}(w+\partial)(L^{\frac{k}{N}})_{cb}(w)_{+} \right), \\
  (b) & \ \big\{ h_k \;\,_\lambda\;\, L_{ab}(w) \big\}_{K}\big\vert_{\lambda=0}=\displaystyle\sum_{c \in I} (-1)^{\tilde{c}(\tilde{a}+\tilde{b}+1)+\tilde{a}\tilde{b}} \big( (L^{\frac{k}{N}-1})_{ac}(w+\partial)_{+}L_{cb}(w)-L_{ac}(w+\partial)(L^{\frac{k}{N}-1})_{cb}(w)_{+} \big).
  \end{align*}
\end{lem}

\begin{proof}
  (a) The first bracket is defined on the generators of $\mathcal{V}$ by the super Adler-type identity:
  \begin{align*}
      &(-1)^{\tilde{a}\tilde{c}+\tilde{a}\tilde{d}+\tilde{c}\tilde{d}} \big\{ L_{ab}(z) \;_\lambda\; L_{cd}(w) \big\}_{H} \\
      &=\left[ L_{cb}(z)\iota_{z}(z\!-\!w\!-\!\lambda\!-\!\partial)^{-1}{L_{ad}}(w)-L_{cb}(w+\lambda+\partial)\iota_{z}(z\!-\!w\!-\!\lambda\!-\!\partial)^{-1}(L_{ad})^{*}(-z+\lambda)\right].  
  \end{align*}
  Applying Lemma \ref{lem:fraction of L}, we get 
     \begin{align*} 
      &\left\{ h_k \;\,_\lambda\;\, L_{ab}(w) \right\}_{H}\Big\vert_{\lambda=0} \\
      &=\displaystyle\sum_{c,d \in I}\text{Res}_{z}\big( L_{ad}(z+x){\iota_{z}}(z\!-\!w\!-\!\partial)^{-1}{L_{cb}}(w)-{L_{ad}}(w+x)\iota_{z}(z\!-\!w\!-\!\partial)^{-1}(L_{cb})^{*}(-z)\big) \\ 
      & \hskip 1cm \times \big( \big\vert _{x=\partial}{L^{\frac{k}{N}-1}}_{dc}(z) \big) \times \beta\\
      &=\displaystyle\sum_{c,d \in I} {\beta} \cdot (-1)^{(\tilde{d}+\tilde{c})(\tilde{c}+\tilde{b})}\times \big\{ \text{Res}_{z}\,L_{ad}(z+\partial)(L^{\frac{k}{N}-1})_{dc}(z)\iota_{z}(z\!-\!w\!-\!\partial)^{-1}\big\}{L_{cb}}(w) \\
      &\hskip 1cm -\displaystyle\sum_{c,d \in I} {\beta} \cdot (-1)^{(\tilde{d}+\tilde{c})(\tilde{c}+\tilde{b})}\times L_{ad}(w+\partial)\big\{\text{Res}_{z}\,(L^{\frac{k}{N}-1})_{dc}(z)\iota_{z}(z\!-\!w\!-\!\partial)^{-1}{({L}_{cb})^{*}}(-z)\big\} \\
      &=\displaystyle\sum_{c \in I} \text{Res}_{z} \left(L_{ac}(z) \iota_{z}(z\!-\!w\!-\!\partial)^{-1} \right)\, \,L_{cb}(w) \times (-1)^{\tilde{a}\tilde{c}+\tilde{b}\tilde{c}+\tilde{c}+\tilde{a}\tilde{b}}\\
      & \hskip 1cm -\displaystyle\sum_{d \in I} L_{ad}(w\!+\!\partial)\,(L^{\frac{k}{N}})_{db}(w)_{\textbf{+}} \times (-1)^{\tilde{a}\tilde{d}+\tilde{b}\tilde{d}+\tilde{d}+\tilde{a}\tilde{b}} \\
      &=\displaystyle\sum_{c \in I} (L^{\frac{k}{N}})_{ac}(w\!+\!\partial)_{\textbf{+}}\,L_{cb}(w) \times (-1)^{\tilde{a}\tilde{c}+\tilde{b}\tilde{c}+\tilde{c}+\tilde{a}\tilde{b}}-\displaystyle\sum_{d \in I} L_{ad}(w\!+\!\partial)\,(L^{\frac{k}{N}})_{db}(w)_{\textbf{+}} \times (-1)^{\tilde{a}\tilde{d}+\tilde{b}\tilde{d}+\tilde{d}+\tilde{a}\tilde{b}} \\
      &=\displaystyle\sum_{c \in I} \big\{ (L^{\frac{k}{N}})_{ac}(w\!+\!\partial)_{\textbf{+}}\,L_{cb}(w)-L_{ac}(w\!+\!\partial)\,(L^{\frac{k}{N}})_{cb}(w)_{\textbf{+}}\big\} \times (-1)^{\tilde{c}(\tilde{a}+\tilde{b}+1)+\tilde{a}\tilde{b}},
  \end{align*}
  where $\beta={(-1)^{\tilde{c}\tilde{a}+\tilde{c}\tilde{b}+\tilde{a}\tilde{b}} \cdot (-1)^{(\tilde{a}+\tilde{b})(\tilde{c}+\tilde{d})+\tilde{d}}}$.
In details, the sign simplification reads
\begin{align*}
  \beta \cdot(-1)^{(d+c)(c+b)}&=(-1)^{(\tilde{a}+\tilde{d})(\tilde{d}+\tilde{c})} \ (-1)^{\tilde{a}\tilde{c}+\tilde{b}\tilde{c}+\tilde{c}+\tilde{a}\tilde{b}}\\
  &\overset{\mathrm{or}}{=}(-1)^{(\tilde{d}+\tilde{c})(\tilde{c}+\tilde{b})} \ (-1)^{\tilde{a}\tilde{d}+\tilde{b}\tilde{d}+\tilde{d}+\tilde{a}\tilde{b}}.
\end{align*} 
\noindent 
(b) Recall that the second PVsA bracket is defined on the generators of $\mathcal{V}$ by 
\begin{align*}
  \left\{ L_{cd}(z) \,_\lambda\, L_{ab}(w) \right\}_{K}&= (-1)^{\tilde{a}\tilde{c}+\tilde{b}\tilde{c}+\tilde{a}\tilde{b}}\delta_{bc}(L_{ad}(z)\!-\!{L_{ad}}(w\!+\!\lambda)) \iota_{z}(z\!-\!w\!-\!\lambda)^{-1} \\
  &-(-1)^{\tilde{a}\tilde{c}+\tilde{b}\tilde{c}+\tilde{a}\tilde{b}}\delta_{ad}\iota_{z}(z\!-\!w\!-\!\lambda\!-\!\partial)^{-1}(L_{cb}(w)\!-\!(L_{cb})^{*}(-z\!+\!\lambda)).
\end{align*}
By Lemma 5.2, we have
\begin{align*}
  \left\{ h_k \;\,_\lambda\;\, L_{ab}(w) \right\}_{K}\big\vert_{\lambda=0}= 
  \displaystyle\sum_{c,d \in I} & \zeta\, \ \!\text{Res}_{z} \left[ \delta_{cb}{\iota_{z}}(z\!-\!w)^{-1}\!\left({L_{ad}}(z\!+\!\partial)-{L_{ad}}(w\!+\!\partial)\right) \right]  (L^{\frac{k}{N}-1})_{dc}(z) \\
  + & \zeta\, \ \!\text{Res}_{z} \left[\delta_{ad}\iota_{z}(z\!-\!w\!-\!\partial)^{-1}\!\left({L_{cb}(w)-(L_{cb})^{*}(-z)}\right) \right]  (L^{\frac{k}{N}-1})_{dc}(z),
\end{align*}
where $\zeta=(-1)^{\tilde{c}\tilde{a}+\tilde{c}\tilde{b}+\tilde{a}\tilde{b}}(-1)^{(\tilde{a}+\tilde{b})(\tilde{c}+\tilde{d})+\tilde{d}}$. Hence
\begin{align*}
  \left\{ h_k \;\,_\lambda\;\, L_{ab}(w) \right\}_{K}\big\vert_{\lambda=0} 
  &=\displaystyle\sum_{c,d \in I}{\zeta} \cdot \delta_{cb} \text{Res}_z \left[\iota_{z}(z\!-\!w)^{-1}L_{ad}(z+\partial)(L^{\frac{k}{N}-1})_{dc}(z)\right]\\  
  & -  \displaystyle\sum_{c,d \in I} {\zeta} \cdot \delta_{cb} L_{ad}(w+\partial)\text{Res}_z \left[\iota_{z}(z\!-\!w)^{-1}(L^{\frac{k}{N}-1})_{dc}(z)\right] \\
  & +  \displaystyle\sum_{c,d \in I} {\zeta} \cdot \delta_{ad}\cdot (-1)^{(\tilde{d}+\tilde{c})(\tilde{c}+\tilde{b})} \left[ \text{Res}_{z}\, (L^{\frac{k}{N}-1})_{dc}(z)\iota_{z}(z\!-\!w\!-\!\partial)^{-1}\right]L_{cb}(w) \\
  & -  \displaystyle\sum_{c,d \in I} {\zeta} \cdot \delta_{ad} \cdot (-1)^{(\tilde{d}+\tilde{c})(\tilde{c}+\tilde{b})}\left[\text{Res}_{z}\,(L^{\frac{k}{N}-1})_{dc}(z)\iota_{z}(z\!-\!w\!-\!\partial)^{-1}{({L}_{cb})^{*}}(-z)\right] \\
  &=(L^{\frac{k}{N}})_{ab}(w)_{+} -\displaystyle\sum_{d \in I} (-1)^{\tilde{d}(\tilde{a}+\tilde{b}+1)+\tilde{a}\tilde{b}} L_{ad}(w\!+\!\partial)(L^{\frac{k}{N}-1})_{db}(w)_{+} \\
  & +  \displaystyle\sum_{c \in I} (-1)^{\tilde{c}\,(\tilde{a}+\tilde{b}+1)+\tilde{a}\tilde{b}}(L^{\frac{k}{N}-1})_{ac}(w\!+\!\partial)_{+}L_{cb}(w) -(L^{\frac{k}{N}})_{ab}(w)_{+} \\
  &=\displaystyle\sum_{c \in I} (-1)^{\tilde{c}\,(\tilde{a}+\tilde{b}+1)+\tilde{a}\tilde{b}}\left( (L^{\frac{k}{N}-1})_{ac}(w+\partial)_{+}L_{cb}(w)-L_{ac}(w+\partial)(L^{\frac{k}{N}-1})_{cb}(w)_{+} \right).
   \end{align*}
\end{proof}

We now state two crucial facts that will enable us to generate integrable systems on $\mathcal{V}$ using the Lenard-Magri scheme (see Proposition \ref{prop:LM scheme}).

\begin{lem} \label{lem:LM}
  For every positive integer $k$ and $u\in \mathcal{V}$, we have the Lenard-Magri recursion
  \begin{align} \label{14}
      \left\{ h_k \;\,_\lambda\;\, u \right\}_{H}\big\vert_{\lambda=0}=\left\{ h_{k+N} \;\,_\lambda\;\, u \right\}_{K}\big\vert_{\lambda=0}.
  \end{align}
\end{lem}

\begin{proof}
It is a direct consequence of Lemma \ref{1.5.} (a) and (b). 
\end{proof}

\begin{lem} \label{lem:LM-initial}
  For every $k \in \left\{1, \cdots, N\right\}$ and $u \in \mathcal{V}$, we have 
  \begin{align} \label{15}
      \left\{ h_k \;\,_\lambda\;\, u \right\}_{K}\big\vert_{\lambda=0}=0.
  \end{align}
\end{lem}

\begin{proof}
  It is enough to show this for $u=L(w)$, since $L(w)$ consists of generators of $\mathcal{V}$. For $ k \in \{1,2, \cdots, N-1\}$, we have $L^{\frac{k}{N}-1}(w)_{+}=0$, and therefore equation \eqref{15} holds by Lemma \ref{1.5.} (b).  Moreover in the case of $k=N$, we have  $\left\{ h_N \;\,_\lambda\;\, L(w)\right\}_{K}\big\vert_{\lambda=0}=L(w)-L(w+\partial) \cdot 1=0$.
\end{proof}

\newpage

\begin{prop} \label{thm:integragle system}
Let $d/dt_k$ be the Hamiltonian derivation of $\mathcal{V}$
\begin{equation} \label{eq:Ham-eq}
\frac{dv}{dt_k}:=  \left\{ h_k \;\,_\lambda\;\, v \right\}_{H}\big\vert_{\lambda=0}, \, \, v \in \mathcal{V}
\end{equation}
associated with the Hamiltonian $h_k$ in \eqref{eq:Hamiltonian}.
\begin{enumerate}
\item[$(a)$]  The equation \eqref{eq:Ham-eq} is bihamiltonian. More precisely, we have 
\begin{equation}
    \frac{dv}{dt_k}:=  \left\{ h_k \;\,_\lambda\;\, v \right\}_{H}\big\vert_{\lambda=0}=\left\{ h_{k+N} \;\,_\lambda\;\, v \right\}_{K}\big\vert_{\lambda=0}.
\end{equation}
Hence it is an integrable system by the Lenard-Magri scheme.
\item[$(b)$] The derivations $d/dt_k$ pairwise commute. In other words, we have 
\[ [\smallint h_k \, , \, \smallint h_{k'} ]_H =0 \]
for all positive integers $k$ and $k'$. Hence local functionals $\int h_{k'}$ are all  integral of motions of \eqref{eq:Ham-eq}.
\item[$(c)$] The equation \eqref{eq:Ham-eq} is equivalent to the Lax equation:
\begin{equation}\label{eq:Lax equation}
      \displaystyle\frac{dL}{dt_{k}}=({L}^{\frac{k}{N}})_{+} \circ  L- L \circ ({L}^{\frac{k}{N}})_{+}\, . \!
  \end{equation} 
\end{enumerate}

\end{prop}
\begin{proof} (a) directly follows from  Lemma  \ref{lem:LM}. Recall that
$$  [ \smallint h_k \;, \smallint h_m ]_{H}:= \smallint \left\{ \, h_k \, {}_\lambda \, h_m \, \right\}_H \big\vert_{\lambda=0}$$
hence (b) also follows from Lemma  \ref{lem:LM} and \ref{lem:LM-initial}. Finally, (c) follows from the following computations:
\begin{equation*}
\begin{split}
[({L}^{\frac{k}{N}})_{+}, L\,]_{ab}(w) &= \big(({L}^{\frac{k}{N}})_{+}(\partial+w) \circ  L(w)\big)_{ab}- \big(L(\partial+w) \circ ({L}^{\frac{k}{N}})_{+}(w)\big)_{ab}\\
&=\sum_{c \in I}(-1)^{(\tilde{a}+\tilde{c})(\tilde{b}+\tilde{c})}\big(({L}^{\frac{k}{N}})_{ac \, +}(\partial+w) L(w)_{cb}- L_{ac}(\partial+w)({L}^{\frac{k}{N}})_{cb \, +}(w)\big)\\
&= \left\{ h_k \;\,_\lambda\;\, L_{ab}(w) \right\}_{H}\big\vert_{\lambda=0}.
\end{split}
\end{equation*}
\end{proof}

\begin{thm} \label{thm:main-2}
    Let $m,n,N$ be positive integers. For  $N\geq 2$, let $f$ be the $N\times (m|n)$ rectangular nilpotent element in $\mathfrak{gl}(Nm|Nm).$ Then the  Hamiltonian equation \eqref{eq:Ham-eq}
is an integrable bihamiltonian system on the $\mathcal{W}$-superalgebra $\mathcal{W}(\mathfrak{gl}(Nm|Nn),f)$. In addition, every local functional $\int h_{k'}$ for a positive integer $k$ is its integral of motion.
\end{thm}
\begin{proof}
    It is a direct consequence of Theorem 
    \ref{summary} and Proposition \ref{thm:integragle system}.
\end{proof}

Note that by Lemma \ref{K=0} (d) and Proposition \ref{thm:integragle system}, these systems of equations can be reduced to the the sub PVsA of $\mathcal{V}$ generated by the elements $u_{k,ab}$ for $k \leq N-2$. Indeed, the generators $u_{N-1,ab}$ are all constants of the system. The first bracket $\{ \, {}_{\lambda} \, \}_{H}$ can also be reduced to this subalgebra via the so-called \textit{Dirac reductions}. This reduced system is the specialization of noncommutative KdV \cite{OS98} to the algebra $(\mathfrak{gl}(I) \otimes \mathcal{V}_I^N, \circ)$.

\begin{ex}
We construct an integrable hierarchy on the $\mathcal{W}$-superalgebra $\mathcal{W}(\mathfrak{gl}(2|2),f)$ associated with the rectangular nilpotent $f$ which corresponds to the partition $2\times(1|1)$. Consider the generic super Adler-type operator $L(\partial)=\mathbbm{1}_{(1|1)}\partial^2+V\partial+W\in \mathcal{V}_{(1|1)}^2$, where 
      \begin{equation*}
      V=\begin{bmatrix} 
          v_{11} & v_{12}\\
          v_{21} & v_{22}
         \end{bmatrix},\quad
     W=\begin{bmatrix} 
         w_{11} & w_{12}\\
         w_{21} & w_{22}
         \end{bmatrix}\\.
  \end{equation*}
The PVsA $\mathcal{V}_{(1|1)}^2$ is isomorphic to the $\mathcal{W}$-superalgebra $\mathcal{W}(\mathfrak{gl}(2|2),f)$ by Theorem 
    \ref{summary}. 
    
    Proposition \ref{thm:integragle system} provides  an integrable system on this PVsA whose conserved densities are given by
  \[h_k=\frac{2}{k} \, \, \mathrm{Res}_{z}\;\mathrm{str}L^{\frac{k}{2}}(z),\]
  for $k=1,3,5\cdots$. When $k=1$, the equation \eqref{eq:Lax equation} is $$\frac{dW}{dt_{1}}= W'-\frac{1}{2}V''-\frac{1}{2}V \circ V'+\frac{1}{2}(V \circ W - W \circ V), \, \, \, \,  \frac{dV}{dt_1}=0.$$
  As we observed in the remark preceding this example, the four generators in $V$ are constant for all the derivations $d/dt_{2s+1}$ where $s\in \mathbb{Z}_{+}$. We can hence reduce the integrable system to the sub PVsA generated by $W$ and obtain a specialization of the noncommutative KdV hierarchy. In particular the equation corresponding to $k=3$ in this reduced hierarchy is  
\[
\frac{dW}{dt_{3}}=\frac{1}{4}W'''+\frac{3}{4}W \circ W'+ \frac{3}{4} W' \circ W,
\]
which is equivalent to the following system of differential equations
      \[\left\{\begin{array}{lr}
      \displaystyle\frac{dw_{11}}{dt_{3}}=\frac{1}{4}w_{11}'''+\frac{3}{4}(w_{11}w_{11}'-w_{12}w_{21}'), \\ \\
      \displaystyle\frac{dw_{12}}{dt_{1}}=\frac{1}{4}w_{12}'''+\frac{3}{4}(w_{11}w_{12}'+w_{12}w_{22}'),\\ \\ 
      \displaystyle\frac{dw_{21}}{dt_{1}}=\frac{1}{4}w_{21}'''+\frac{3}{4}(w_{21}w_{11}'+w_{22}w_{21}'),\\ \\
      \displaystyle\frac{dw_{22}}{dt_{1}}=\frac{1}{4}w_{22}'''+\frac{3}{4}(-w_{21}w_{12}'+w_{22}w_{22}').
      \end{array}\right.
    \] 
\end{ex}

\end{document}